\documentclass[journal,draftcls,onecolumn,12pt,twoside]{IEEEtran}
\IEEEoverridecommandlockouts

\usepackage{amsmath, amssymb, cite}
\makeatletter
\renewenvironment{subequations}{%
  \refstepcounter{equation}%
  \protected@edef\theparentequation{\theequation}%
  \setcounter{parentequation}{\value{equation}}%
  \setcounter{equation}{0}%
  \def\theequation{\theparentequation\roman{equation}}%
}{%
  \setcounter{equation}{\value{parentequation}}%
}
\makeatother
\allowdisplaybreaks
\usepackage{amsthm,bm,bbm}
\usepackage{mathtools}
\usepackage{multirow,color,amsfonts}
\usepackage{tabulary}
\usepackage{subfigure}
\usepackage{graphicx}
\usepackage{setspace}
\usepackage{enumerate}
\usepackage{algorithm}
\usepackage{algpseudocode}
\usepackage{comment}
\usepackage{pdfpages}
\usepackage{hyperref,url}

\usepackage{cancel}

\usepackage[size=small]{caption}

\usepackage[normalem]{ulem}
\usepackage{xcolor}

\let\oldsout\sout
\renewcommand{\sout}[1]{\textcolor{red}{\oldsout{#1}}}
  \DeclareMathAlphabet\mathbfcal{OMS}{cmsy}{b}{n}

\newcommand\independent{\protect\mathpalette{\protect\independenT}{\perp}}
\def\independenT#1#2{\mathrel{\rlap{$#1#2$}\mkern2mu{#1#2}}}

\newtheorem{theo}{Theorem}
\newtheorem{defi}{Definition}
\newtheorem{assu}{Assumption}
\newtheorem*{assu*}{Assumption}
\newtheorem{prop}{Proposition}

\newtheorem{cor}{Corollary}

\newtheorem{lem}{Lemma}
\newtheorem{ex}{Example}

\newcommand{\Rf}{R_{\rm f}}
\newcommand{\RSWsum}{R_{\rm SW}^{\Sigma}}
\newcommand{\RAHsum}{R_{\rm AH}^{\Sigma}}
\newcommand{\RAHrfdsum}{R_{\rm AH,rfd}^{\Sigma}}
\newcommand{\RKMsum}{R_{\rm KM}^{\Sigma}}
\newcommand{\RKMalternativesum}{R_{\rm KM,alt}^{\Sigma}}
\newcommand{\RKMsymsum}{R_{\rm KM,sym}^{\Sigma}}
\newcommand{\RKMsymrfdsum}{R_{\rm KM,sym,rfd}^{\Sigma}}
\newcommand{\RHKsum}{R_{\rm HK}^{\Sigma}}

\newcommand{\RKMsumrecursiveinnerproduct}{R_{\rm KM, recursive}^{\Sigma}}
\newcommand{\RKMsumrecursiveinnerproductsymmetric}{R_{\rm KM, recursive-sym.}^{\Sigma}}
\newcommand{\RKMsumnestedinnerproductsymmetric}{R_{\rm KM, nes.-sym.}^{\Sigma}}

\newcommand{\RSsum}{R_{\rm S}^{\Sigma}}
\newcommand{\RKMORsum}{R_{\rm KM-OR}^{\Sigma}}
\newcommand{\RKMORsumB}{R_{\rm KM-OR \mid {\bf B}}^{\Sigma}}

\newcommand{\pol}{\mbox{PolyDot}}
\newcommand{\mat}{\mbox{MatDot}}
\newcommand{\p}{\mbox{Polynomial}}

\usepackage{enumitem}

\def\compactify{\itemsep=0pt \topsep=0pt \partopsep=0pt \parsep=0pt}
\let\latexusecounter=\usecounter

\begin{document}

\title{Structured Codes for Distributed \\ Matrix Multiplication}
\author{Derya Malak 
\thanks{The author is with the Commun. Systems Dept., EURECOM, Biot Sophia Antipolis, 06904 France (derya.malak@eurecom.fr).}
\thanks{A preliminary version of this work was presented in part at the 2024 Int. Symp. Inf. Theory, Athens, Greece~\cite{malak2024diststruc}.}
\thanks{This research was partially supported by European Research Council ERC-StG Project SENSIBILITÉ under Grant 101077361, by the Huawei France-Funded Chair Toward Future Wireless Networks, and by the program ``PEPR Networks of the Future" of France 2030. \hfill Manuscript last revised: \today.}}

\maketitle

\begin{abstract}
Our work addresses the well-known open problem of distributed computing of bilinear functions of two correlated sources ${\bf A}$ and ${\bf B}$. In a setting with two nodes, with the first node having access to ${\bf A}$ and the second to ${\bf B}$, we establish bounds on the optimal sum rate that allows a receiver to compute an important class of non-linear functions, and in particular bilinear functions, including dot products $\langle {\bf A},{\bf B}\rangle$, and general matrix products ${\bf A}^{\intercal}{\bf B}$ over finite fields. The bounds are tight for large field sizes, for which case we can derive the exact fundamental performance limits for all problem dimensions and a large class of sources. Our achievability scheme involves the design of non-linear transformations of ${\bf A}$ and ${\bf B}$, carefully calibrated to work synergistically with the structured linear encoding scheme by K\"orner and Marton. The subsequent converses derived here, calibrate the Han-Kobayashi approach and the strong converse of Ahlswede-G{\'a}cs-K\"orner to yield relatively tight converses on the sum rate. We exhibit unbounded compression gains over Slepian-Wolf coding, depending on the source correlations. In the end, this work characterizes the fundamental limits of distributed computing for a crucial class of functions, while succinctly capturing the inherent computation structures and source correlations. 
\end{abstract}

\begin{IEEEkeywords}
Distributed computation, source coding for compression, structured linear coding, distributed dot-product computation, and distributed matrix multiplication. 
\end{IEEEkeywords}

\section{Introduction}
\label{sec:intro}

Basic functions like matrix multiplication, currently constitute the bulk of computational load in scientific computing, as they are omnipresent in applications that include convolution~\cite{strang2023introduction}, large linear transforms, Fourier transforms, quantum computing~\cite{buhrman2001quantum}, as well as in applications of machine learning such as linear regression, least squares modeling~\cite{strang2023introduction}, and many more. The unprecedented intensity of such computational loads often brings to the fore the necessity for massive parallelization techniques, and we are now witnessing the deployment of massive distributed computing systems, geared at tackling complex distributed computing tasks.

It is the case though that to be successfully deployed, distributed computing requires an intense exchange of information among the participating nodes. In most scenarios, including matrix multiplication, it is evident that to meaningfully parallelize across multiple workers, one must maintain a reduced communication load, which is now considered as a main bottleneck of parallel processing. The need for minimizing this load is clear and evident, and this is a need that has motivated several of the noteworthy parallel computing techniques, such as in~\cite{yu2017polynomial,dutta2019optimal,tandon2017gradient,raviv2020gradient,halbawi2018improving,soleymani2021analog,YuRavSoAve2018} to mention just a fraction, many of which have been designed and tested with success.

Motivated by the above, we here study the communication cost of distributed computing, and we do so for a prevalent class of non-linear functions, namely of bilinear functions. As suggested above, this setting finds itself at the core of various technological fields in edge and cloud computing~\cite{shamsi2013data} and machine learning~\cite{10129894}, and again, as suggested, this is a setting that entails considerable communication overheads as well as an intriguingly intertwined relationship between communication and computational parallelization. This bottleneck has been studied in seminal works such as~\cite{korner1979encode,han1987dichotomy,nazer2007computation,lim2019towards,lalitha2013linear}, focusing often on the linear function case. 
Our work focuses on the classical problem of distributed computing of bilinear functions of two correlated sources, placing emphasis on dot and matrix products, while also capturing an important element of modern large data; the strong structural correlations of this data that serves as computing input. In a context similar to~\cite{SlepWolf1973,korner1979encode,han1987dichotomy} where the receiver wishes to compute bilinear functions of the distributed sources, our aim is to establish bounds on the optimal sum rate (the minimum amount of information) that allows a receiver to compute these functions.

The technical contributions of this work are summarized in the following.

\subsection{Main Contributions of this Work}
\label{sec:contributions}

\begin{itemize}
\item {\bf New structured source codes for distributed computing of dot and matrix products.} We devise an encoding framework for computing the product of two distributed correlated source variables ${\bf A}$ and ${\bf B}$ (vectors or matrices), over finite fields. To this end, our achievability scheme involves the novel design of {\emph{non-linear transformations}} for long sequences of ${\bf A}$ and ${\bf B}$ drawn i.i.d. across realizations, according to some joint probability distribution. These transformations are then carefully calibrated to {\emph{work synergistically with the structured linear encoding scheme by K\"orner and Marton}}~\cite{korner1979encode} as well as the {\emph{more general scheme by Ahlswede and Han}}~\cite{ahlswede1983source} for computing a class of bilinear functions, all with a vanishing error probability.

Our achievability results include constructions for distributed computing of dot products (Corollary~\ref{cor:KW_sum_rate_for_inner_product}), matrix products that are symmetric  (Propositions~\ref{prop:KW_sum_rate_for_symmetric_matrix_product} and~\ref{prop:KW_sum_rate_for_inner_product+vs_SW_sum_rate}, and Theorem~\ref{theo:achievability_symmetric_matrix_products}), and general square matrix products (Proposition~\ref{prop:KW_sum_rate_for_general_matrix_product}, and Theorem~\ref{theo:achievability_square_matrix_products}). 
We also explore a hybrid encoding scheme (Proposition~\ref{prop:KM-OR_sum_rate_for_matrix_vector_product}), as well as consider recursive and nested applications of the dot product   (Propositions~\ref{prop:recursive_inner_product_KM}-\ref{prop:achievability_results_symmetric_matrices_nested_KM_q_2}) for distributed matrix multiplication.  
Our schemes are flexible, allowing the receiver to recover ${\bf A}^{\intercal}{\bf B}$ without imposing structural constraints on ${\bf A}$ and ${\bf B}$.

\item {\bf Achievable compression gains.} 
Contrasting the sum rates of our structured source codes (see Propositions~\ref{prop:KW_sum_rate_for_symmetric_matrix_product}-\ref{prop:KM-OR_sum_rate_for_matrix_vector_product} and Theorems~\ref{theo:achievability_symmetric_matrix_products} and~\ref{theo:achievability_square_matrix_products}) with the state-of-the-art codes (e.g.,~\cite{SlepWolf1973,krithivasan2011distributed,pradhan2020algebraic}) reveals significant gains when computing dot products and matrix products (of distributed sources ${\bf A}$ and ${\bf B}$). Our schemes carefully harness the structure of the source data, and the corresponding sum-rate gains are naturally more pronounced in the presence of stronger correlations (see Examples~\ref{ex:innerproduct_length_m_binary}-\ref{ex:general_matrix_q3}, and see also Figure~\ref{fig:innerproducts_general_comparison} which relates to Propositions~\ref{prop:KW_sum_rate_for_symmetric_matrix_product} and~\ref{prop:KM-OR_sum_rate_for_matrix_vector_product}).

\item {\bf Converse results.} Leveraging the Ahlswede-G{\'a}cs-K\"orner approach in~\cite{ahlswede1976bounds} allows us here to derive a strong converse for the square matrix product setting, as $q\to\infty$ (Theorem~\ref{theo:strong_converse_general_matrix_product}). Furthermore, calibrating the Han-Kobayashi approach in~\cite{han1987dichotomy} yields a relatively tight lower bound on the sum rate for all $q\geq 2$ (Theorem~\ref{theo:converse_general_matrix_product_Han_Kobayashi}). In addition to the main converse bounds, for independently and uniformly drawn source matrices ${\bf A}$ and ${\bf B}$ when $q\to\infty$, we also derive a matching strong converse to the achievability Theorem~\ref{theo:achievability_square_matrix_products} for the case of square matrix products  (Corollary~\ref{cor:achievable_rate_lem:entropy_general_matrix_product}). Finally, for the cases of symmetric and square matrix products over $\mathbb{F}_2$, we upper bound the optimality gaps of our design (see Propositions~\ref{prop:multiplicative_gain_for_binary_symmetric_matrix_products} and~\ref{prop:multiplicative_gain_for_binary_square_matrix_products} respectively).

\item Our schemes have some additional properties, like balancing the computation load across the nodes, as well as offering some security advantages. We briefly discuss these in Section~\ref{sec:conclusions}.
\end{itemize}

\subsection{Related Work and Connections of Our Work to the State of the Art}
\label{sec:related_work}

Below, we detail two main approaches used in distributed coding for computation: unstructured coding-based approaches (also known as random binning, relying on hashing functions), and structured coding-based approaches that are more directly geared towards leveraging the structure of the computation task.

\paragraph{Unstructured coding for computing} 
\label{sec:coding_for_computing}

Most of the early approaches use the idea of {\emph{random binning}} for lossless source coding. 
Focusing on the case of two statistically dependent, finite alphabet source variables $X_1$ and $X_2$ separately observed by two transmitters, the seminal work by Slepian and Wolf~\cite{SlepWolf1973} provided an unstructured coding technique for the asymptotically lossless compression of i.i.d. sequences $X_1^n=\{X_{1i}\}_{i=1}^n$ and $X_2^n=\{X_{2i}\}_{i=1}^n$, and established the well-known result that $\RSWsum=H_q(X_1,X_2)$ is the optimal rate for jointly recovering $(X_1^n,X_2^n)$. 
In some cases though, this sum rate can be significantly reduced when the network's task is to compute a function of the sources rather than to communicate the sources themselves. 
Taking a step towards distributed computing, Yamamoto derived the minimum rate (guaranteeing vanishing probability of error) at which a source has to compress $X_1^n$ for distributed computing of a general function $f(X_1^n,X_2^n)$ with side information $X_2^n$ at the receiver (cf.~\cite{yamamoto1982wyner}). Additional related work can also be found in~\cite{Kor73}, where the authors exploit characteristic graphs to derive lower bounds on perfect hashing. Drawing on~\cite{Kor73}, the graph-theoretic approaches in~\cite{OR01,malak2022fractional,malak2024multi}, as well as those in~\cite{lenz2023function}, have addressed various computing scenarios.

\paragraph{Structured coding for computing} 
\label{sec:struc_coding}
Structured coding approaches, on the other hand, deviate from the random coding approach, and instead entail correlated binning of sources and the use of algebraic codes. 
K\"orner and Marton (cf.~\cite{korner1979encode}) devised a {\emph{structured linear encoding}} strategy for distributed computing the modulo-two sum $X_1\oplus_2 X_2$ of two i.i.d. binary source sequences $(X_1^n, X_2^n)$, where in particular $(X_{1i},X_{2i})$ has the same joint distribution as $(X_1,X_2)$, for all $i\in[n]$. Their technique, based on the method of Elias~\cite{elias1955coding}, constructs linear codes that achieve an asymptotically vanishing probability of error at a derived minimum sum rate of $2H(X_1\oplus_2 X_2)$, when the joint distribution of $X_1$ and $X_2$ is symmetric (see also Definition~\ref{def:DSBS}). 
Furthermore, the interesting work in~\cite{lalitha2013linear} provided subspace-based lossless linear computation schemes using nested codes, where these schemes have been shown to generalize those in~\cite{korner1979encode}, as well as have been proven to be sum-rate optimal for a class of source distributions. 
Additionally, for the binary modulo-two sum problem, Ahlswede and Han also derived --- by combining the source coding technique from~\cite{berger1977multiterminal} with the method of Elias~\cite{elias1955coding} --- an achievable rate region for general binary sources~\cite[Theorem~10]{ahlswede1983source}, which contains the regions derived in~\cite{SlepWolf1973} and~\cite{korner1979encode}, and which is generally larger than the convex hull of both.

Furthermore, Han and Kobayashi generalized the structured encoding strategy in~\cite{korner1979encode} to the setting of computing $X_1\oplus_q X_2$ given $q$-ary sources $X_1$ and $X_2$, with $\oplus_q$ denoting addition over $\mathbb{F}_q$~\cite{han1987dichotomy}. Motivated by the problem of compressing non-additive functions, in~\cite{han1987dichotomy}, the authors also identified key function features that differentiate the Slepian-Wolf and K\"orner-Marton regions, providing conditions under which computing a general bivariate function $f(X_1^n,X_2^n)=\{f(X_{1i},X_{2i})\}_{i=1}^n$ requires a rate lower than $\RSWsum$. Along related lines, Ahlswede and Csiszár demonstrated that computing most binary-valued or componentwise functions of distributed source sequences $(X_1^n,X_2^n)$ necessitates transmission rates from separate encoders that are nearly as high as those needed for full reconstruction of $(X_1^n,X_2^n)$~\cite{ahlswede1981get}. Furthermore, this same work also revealed that for a class of functions --- like for example, computing the joint type (joint composition) of $X_1^n$ and $X_2^n$, or computing their Hamming distance or its parity --- then determining $f(X_1^n, X_2^n)$ in the knowledge of $X_2^n$, typically required the encoder of $X_1$ to use a rate comparable to that needed for fully reproducing $X_1^n$ itself. Additionally, it was also revealed that, given a distortion criterion, the problem of an exact characterization of the achievable rate region for $f(X_1^n, X_2^n)$ --- excluding componentwise functions --- may be as hard a problem as determining the achievable rate region for reproducing $X_1^n$ and $X_2^n$. In distributed computation of non-linear functions of $(X_1,X_2)$, identifying an injective mapping from the target function to a representation $X_1 \oplus_q X_2$, defined over a sufficiently large prime field $\mathbb{F}_q$~\cite{krithivasan2011distributed,pradhan2020algebraic,sohail2022unified}, known as {\emph{function embedding}}, followed by {\emph{structured binning}}, may yield rate savings over~\cite{SlepWolf1973}.

As our interest lies in distributed matrix multiplication, we proceed to provide some of the state of art related to this broad problem.

\paragraph{Distributed matrix multiplication and codes}
\label{sec:distributed_matrix_multiplication} 
Coded matrix multiplication recasts matrix multiplication tasks into a computation channel. Numerous strategies have been developed to enhance distributed coded matrix multiplication to reduce download costs and mitigate stragglers, such as Short-Dot~\cite{dutta2016short}, $\p$~\cite{yu2017polynomial}, $\mat$~\cite{lopez2022secure}, and $\pol$ codes~\cite{dutta2019optimal,yu2020straggler,yang2018secure}, all over finite fields.  
These approaches split source matrices into submatrices via linear transformations and transform matrix multiplication into inner or outer product computations. Distributing subtasks across worker nodes enables efficient, linearly separable processing of rows and columns of matrices. For example, $\mat$~\cite{lopez2022secure} and $\pol$ codes~\cite{dutta2019optimal} reduce communication costs and improve security, while $\p$ codes~\cite{yu2017polynomial}, as well as generalizations using algebraic function fields~\cite{fidalgo2024distributed}, namely finite field extensions of fields of rational functions~\cite{villa2006topics}, are commonly used to mitigate stragglers. Recent research focuses on achieving even greater reductions in communication costs (e.g.,~\cite{jia2021capacity, wan2021distributed, 9478901, fawzi2022discovering,  aliasgari2020private}). 
However, these approaches are not sum-rate optimal, even in the absence of stragglers. Furthermore, in the absence of stragglers, for $\p$ coding approaches, the user can directly recover the source matrices from subtasks.

{\emph{Our work builds on the foundational principles of structured codes to develop distributed matrix multiplication techniques over finite fields. We demonstrate significant compression savings for matrix product computations compared to~\cite{SlepWolf1973}, achieved by applying structured encoding to carefully designed non-linear mappings of the distributed source matrices.  
In addition to improved performance, our approach operates over smaller field sizes than those required in~\cite{krithivasan2011distributed,pradhan2020algebraic,sohail2022unified}.}} {\emph{This use of structured coding idea in the context of distributed matrix multiplication will prove pivotal in capturing source correlations and computation structures jointly, thus well capturing the matrix multiplication problem.}}

\subsection{Organization}
\label{sec:organization} 
Section~\ref{sec:system_model_problem_statement} formalizes the distributed matrix multiplication setting. Section~\ref{sec:achievability} presents our achievable coding schemes and corresponding rates for computing dot products, symmetric, and square matrix products. It also explores recursive and nested implementations of dot products for computing general matrix products.  Section~\ref{sec:converses} details our converse bounds and characterizes the optimality gaps of our schemes, while Section~\ref{sec:conclusions} concludes the paper. Throughout the paper, we present various examples to assist the reader in better understanding the results.

{\bf Notation.} 
We use regular type for random variables and boldface for vectors and matrices over the finite field $\mathbb{F}_q$. Logarithms base $2$ and $q>2$ are denoted by $\log$ and $\log_q$, respectively, and we write $\exp(\cdot)=q^{(\cdot)}$. 
We use $\oplus_q$ and $\ominus_q$ to denote addition and subtraction over the finite field $\mathbb{F}_q$, respectively, where $x \ominus_q y\triangleq x\oplus_q (-y)$. When $q$ is prime, $\oplus_q$ coincides with modulo-$q$ addition; for $q=p^m$, $m>1$, it denotes the standard field addition in $\mathbb{F}_{p^m}$. 
For a random variable $X$ with PMF $P_X$, its entropy in binary and $q$-ary units is $H(X)$ and $H_q(X)=H(X)/\log_2 q$, respectively. Likewise, for $(X_1,X_2)$ with joint PMF $P_{X_1,X_2}$, the joint and conditional entropies are $H_q(X_1,X_2)$ and $H_q(X_1 \mid X_2)$, respectively. The acronym i.i.d. stands for independent and identically distributed, and $X_1\independent X_2$ is used to describe statistical independence between $X_1$ and $X_2$. $\mathbb{P}(A)$ is the probability of an event $A$. For a binomial variable $X\sim {\rm Bin}(l,p)$ with $l\in\mathbb{N}$ and $p\in [0,1]$, the complementary cumulative distribution function is given by $\bar{F}(m;l,p)=\sum\nolimits_{i=m}^l {l\choose i}p^i(1-p)^{l-i}$. When $l=1$, then $X$ is a Bernoulli variable, denoted $X\sim {\rm Bern}(p)$, and $h(p)$ denotes the binary entropy function.

We denote by $[l]$ the set $\{1,\dots,l\}$, for $l\in\mathbb{Z}^+$, and by $[l_1,\ l_2]$ the set $\{l_1,\dots,l_2\}$ for $l_1,\  l_2\in\mathbb{Z}^+$ such that $l_1\leq l_2$. Given a random matrix ${\bf X}=\left(x_{ij}\right)_{i\in [m],\;j\in [l]}\in\mathbb{F}_q^{m\times l}$, its $i$-th row, $j$-th column, and transpose are given by ${\bf X}(i,:)$, ${\bf X}(:,j)$, ${\bf X}^{\intercal}$, respectively. Alternatively, ${\bf x}=\left(x_{i}\right)_{i\in [m]}\in\mathbb{F}_q^{m\times 1}$ (or $\mathbb{F}_q^m$) and ${\bf x}=\big(\left(x_{j}\right)_{j\in [l]}\big)^{\intercal}\in\mathbb{F}_q^{1\times l}$ denote column and row vectors, respectively. For a given ${\bf x}\in\mathbb{F}_q^{1\times l}$, for $i,\ j\in\mathbb{Z}^+$, and $i\leq j$, then ${\bf x}(i:j)\triangleq[x_i ,\ x_{i+1} ,\ \hdots ,\ x_j]$, and similarly for a column vector. The vertical concatenation of ${\bf A}\in\mathbb{F}_q^{m_1\times l}$ and ${\bf B}\in\mathbb{F}_q^{m_2\times l}$ is denoted by $[{\bf A}; {\bf B}]\in\mathbb{F}_q^{(m_1+m_2)\times l}$. 
The notations ${\bm 1}_{m\times l}$ and ${\bm 0}_{m\times l}$ denote $m\times l$ matrices of all ones and all zeros, respectively. We write $X^n \triangleq \{X_i\}_{i=1}^n = (X_1, X_2, \dots, X_n)$, and use both $[X_1; X_2; \dots; X_n]$ and $(X_1, X_2, \dots, X_n)$ to denote a sequence of i.i.d.\ realizations of $X$; the intended meaning will be clear from context. We extend this notation to matrices by defining ${\bf X}^n$, with ${\bf X}^n(i,j)$ representing the length-$n$ sequence of realizations of the $(i,j)$-th component ${\bf X}(i,j)\in\mathbb{F}_q$.

\section{System Model and Problem Statement}
\label{sec:system_model_problem_statement}

We consider a distributed scenario involving two sources with separate encoders, and a receiver. The distributed sources separately observe realizations of statistically dependent (i.e., correlated) matrix variables ${\bf A}=(a_{ij})_{i\in [m],\ j\in [l]}\in\mathbb{F}_q^{m\times l}$ and ${\bf B}=(b_{ij})_{i\in [m],\ j\in [l]}\in\mathbb{F}_q^{m\times l}$, respectively. In other words, Source $1$ observes ${\bf A}$ and Source $2$ observes ${\bf B}$, respectively. 
The receiver aims to compute $\mathbfcal{D}=(d_{ij})_{i,\ j\in[l]}={\bf A}^{\intercal}{\bf B}\in\mathbb{F}_q^{l\times l}$.

We take a non-real-time approach that relies on accumulating length-$n$ sequences of potentially correlated source matrix realizations. Specifically, the distributed sources are block-encoded with blocklength $n$. We assume statistically dependent finite alphabet two source sequences ${\bf A}^n=({\bf A}(1),{\bf A}(2),\dots,{\bf A}(n))$ and ${\bf B}^n=({\bf B}(1),{\bf B}(2),\dots,{\bf B}(n))$ corresponding to length-$n$ i.i.d. realizations of ${\bf A}$ and ${\bf B}$, respectively. We have the following additional assumption.

\begin{assu}[Memoryless and i.i.d. observations]
\label{assump:memoryless_iid}
The sequence of pairs $\{({\bf A}(i),{\bf B}(i))\}_{i=1}^n$ is i.i.d. according to some joint distribution $P_{{\bf A},{\bf B}}$, i.e., 
\[
P_{{\bf A}^n,{\bf B}^n}({\bf A}^n,{\bf B}^n)= \prod_{i=1}^n P_{{\bf A},{\bf B}}({\bf A}(i),{\bf B}(i)) \ .
\]
\end{assu}

Two distributed sources separately observe length-$n$ memoryless and i.i.d. realizations (Assumption~\ref{assump:memoryless_iid})  ${\bf A}^n=({\bf A}(1),{\bf A}(2),\dots,{\bf A}(n))$ and ${\bf B}^n=({\bf B}(1),{\bf B}(2),\dots,{\bf B}(n))$, respectively, and then encode their respective realizations independently. The two sources can perform general, and possibly non-linear, componentwise functions $g_1({\bf A}^n)=\{g_1({\bf A}(i))\}_{i=1}^n={\bf X}_1^n$ and $g_2({\bf B}^n)=\{g_2({\bf B}(i))\}_{i=1}^n={\bf X}_2^n$, where each component ${\bf X}_1(j)$ and ${\bf X}_2(j)$ lies in $\mathbb{F}_q$. As we will see later, in our setting, the encoders will apply the {\emph{structured linear coding}} technique of K\"orner-Marton in~\cite{korner1979encode} to the non-linear transformations ${\bf X}_1^n\in\mathbfcal{X}_1^n$ and ${\bf X}_2^n\in\mathbfcal{X}_2^n$, respectively.   
The separate encoders devise mappings $f_1: \mathbfcal{X}_1^n\to \mathcal{R}_{f_1}$ and $f_2: \mathbfcal{X}_2^n\to \mathcal{R}_{f_2}$, with ranges $\mathcal{R}_{f_1}$ and $\mathcal{R}_{f_2}$, respectively. The encoder outputs are then transmitted over {\emph{noiseless channels} to the common receiver. As we will also see, the receiver will add the received codewords,
\begin{align}
\label{Zn_modulo_q_sum}
{\bf Z}^n={\bf X}_1^n\oplus_q{\bf X}_2^n
\end{align}
before proceeding to recover the sequence ${\bf \mathbfcal{D}}^n=({\bf \mathbfcal{D}}(1),{\bf \mathbfcal{D}}(2),\dots,{\bf \mathbfcal{D}}(n))$ of desired matrix products, with ${\bf \mathbfcal{D}}(i)={\bf A}(i)^{\intercal}{\bf B}(i)\in\mathbb{F}_q^{l\times l}$ for all $i\in[n]$, with a small probability of error, as we will clarify later on (see Proposition~\ref{prop:KW_sum_rate_for_symmetric_matrix_product} and its proof).

\begin{defi}[An $(n,\epsilon)$-coding scheme~\cite{korner1979encode}]
\label{def:eps_n_coding}
The pair of functions $(f_1,f_2)$ is called an $(n,\epsilon)$-coding scheme if there exists a function $\phi: \mathcal{R}_{f_1}\times \mathcal{R}_{f_2} \to {\bf \mathbfcal{Z}}^n$ such that letting 
\begin{align}
{\bf \hat{Z}}^n \triangleq \phi(f_1({\bf X}_1^n),f_2({\bf X}_2^n))  
\end{align}
achieves an arbitrarily small probability of error $\mathbb{P}({\bf \hat{Z}}^n \neq {\bf Z}^n)<\epsilon$.  
\end{defi}

In this work, we devise $(n,\epsilon)$-coding schemes (see Section~\ref{sec:achievability}) that approximate $\mathbfcal{D}^n$, with accuracy $1-\epsilon$. 
The corresponding two-component source network with memoryless and i.i.d. observations is called the {\emph{general matrix multiplication source network}}, and is shown in Figure~\ref{fig:from_KM_to_matrices}.
\begin{figure}[t!]
\centering
\includegraphics[width=0.65\textwidth]{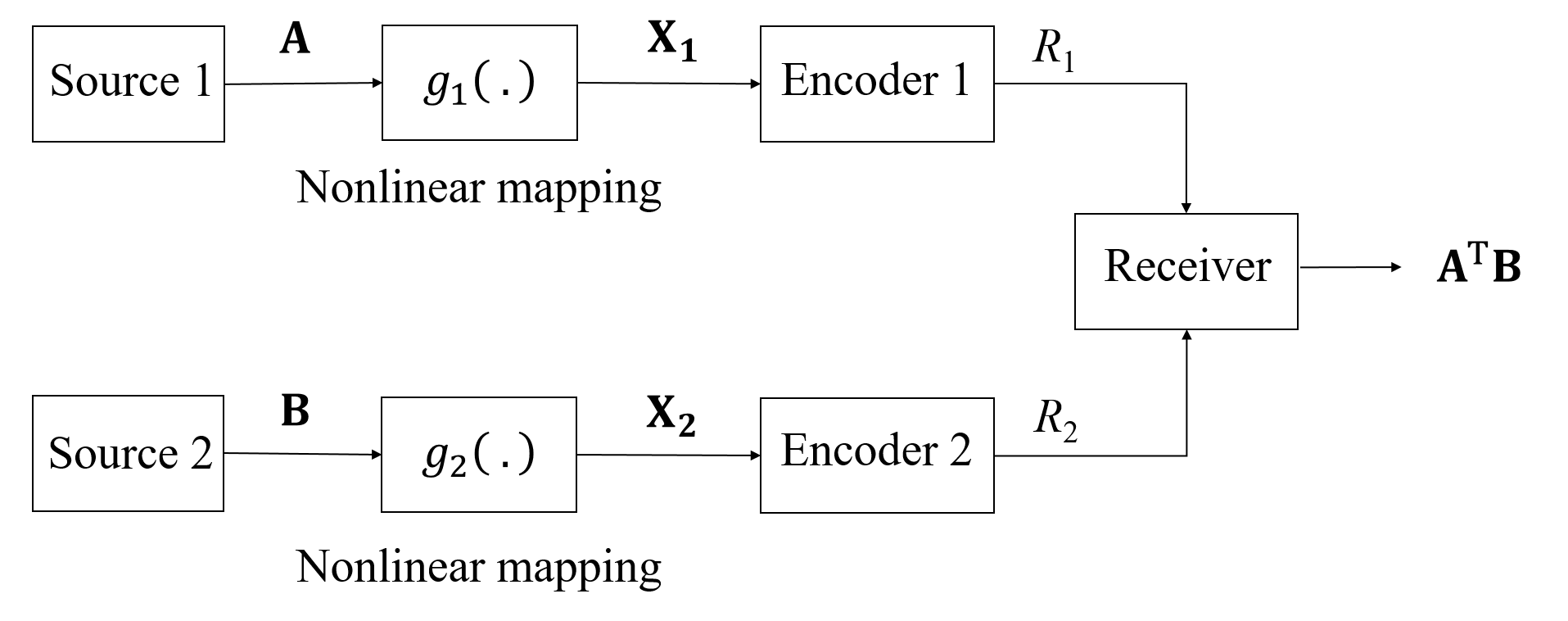}
\caption{A general matrix multiplication source network for distributed computation of $\mathbfcal{D}={\bf A}^{\intercal}{\bf B}$.}
\label{fig:from_KM_to_matrices}
\end{figure}

A special case of our setting is when the encoder sequences $X_1^n$ and $X_2^n$ are formed by i.i.d. realizations of distributed sources $X_1\in\mathbb{F}_2$ and $X_2\in\mathbb{F}_2$, respectively, where $(X_{1i},X_{2i})$ has the same joint distribution as $(X_1,X_2)$. For this case, K\"orner-Marton in~\cite{korner1979encode} have designed $(n,\epsilon)$-coding schemes for computing $X_1\oplus_2 X_2$ of a doubly symmetric binary source (DSBS) pair $(X_1, X_2 )$, and for any $\epsilon\in(0,1)$ at rates $R_1=R_2=H(X_1\oplus_2 X_2)$, which achieve the optimum.

\begin{defi}[Doubly symmetric binary source (DSBS)]
\label{def:DSBS}
A DSBS with disagreement probability $p \in \big(0, 1/2\big)$ is denoted by $(X_1, X_2) \sim {\rm DSBS}(p)$, where its PMF is given as $\mathbb{P}(X_1=X_2=0)=\mathbb{P}(X_1=X_2=1)=(1-p)/2$ and $\mathbb{P}(X_1=0,\ X_2=1)=\mathbb{P}(X_1=1,\ X_2=0)=p/2$, where both $X_1\sim {\rm Bern}(1/2)$ and $X_2\sim {\rm Bern}(1/2)$, and where $X_2$ is the output of a binary symmetric channel (BSC) with a crossover probability $p$, denoted as ${\rm BSC}(p)$, for a given input $X_1$~\cite{yu2023gray}.
\end{defi}

Our objective, as detailed next, is to determine the {\emph{region of achievable rates}} for the general matrix multiplication source network to achieve asymptotically\footnote{Achievability results for distributed matrix multiplication at practical blocklengths (e.g.,~\cite{irony2004communication}, \cite{shirani2014finite}) can be obtained using approximate or lossy ($\epsilon>0$) computation techniques leveraging Kolmogorov complexity~\cite[Ch. 14]{cover2012elements}.} lossless compression. We denote the achievable encoding rates by the pair $(R_1,R_2)$. Even when the receiver knows ${\bf A}$, $R_2 \geq H_q({\bf A}^{\intercal}{\bf B}\,\vert\, {\bf A})$ is still necessary. Similarly, we must have $R_1 \geq H_q({\bf A}^{\intercal}{\bf B}\,\vert\, {\bf B})$. Hence, the sum rate must satisfy $R_1+R_2\geq H_q({\bf A}^{\intercal}{\bf B}\,\vert\, {\bf A})+H_q({\bf A}^{\intercal}{\bf B}\,\vert\, {\bf B})$, whereas this lower bound may not be achievable with separate encoders in general. In \cite{SlepWolf1973}, Slepian and Wolf have provided the necessary and sufficient lossless coding rate for distributed compression of ${\bf A}^n$ and ${\bf B}^n$:
\begin{align}
\label{SW_condition}
R_1&\geq H_q({\bf A}\ \vert\ {\bf B}) \ ,\nonumber\\ R_2&\geq H_q({\bf B}\ \vert\ {\bf A}) \ ,\nonumber\\ R_1+R_2&\geq \RSWsum({\bf A}, {\bf B})\triangleq H_q({\bf A}, {\bf B}) \ .
\end{align} 
In their seminal work~\cite{han1987dichotomy}, Han and Kobayashi derived the necessary and sufficient conditions (restated in Lemma~\ref{lem:lemmas_1_2_Han_Kobayashi}) under which any achievable rate $(R_1,\ R_2)$ for the distributed computation of an arbitrary function $f({\bf A},{\bf B})$ of correlated variables ${\bf A}$ and ${\bf B}$, with entropy $\Rf\triangleq H_q(f({\bf A},{\bf B}))\leq \RSWsum$, coincides with the region in (\ref{SW_condition}). 
These conditions are not always satisfied in our setting, as we demonstrate below. Consequently, we identify regimes where the resulting sum rate $\RKMsum=R_1+R_2$ is strictly below the Slepian-Wolf bound $\RSWsum$ described by (\ref{SW_condition}), enabling the receiver to recover the matrix product $\mathbfcal{D}={\bf A}^{\intercal}{\bf B}$ without being able to decode $({\bf A}, {\bf B})$ in their entirety. To that end, we denote the gain of our scheme by $\eta\triangleq{\RSWsum}/{\RKMsum}$.

\section{Achievability}
\label{sec:achievability}
This section details achievable schemes for {\emph{structured distributed matrix multiplication}}. The first scheme in Proposition~\ref{prop:KW_sum_rate_for_symmetric_matrix_product} considers the symmetric case for $q>2$ and odd, and embeds the {\emph{dot-product computation}} problem. The second scheme, given in Theorem~\ref{theo:achievability_symmetric_matrix_products}, improves upon the sum rate of the first.  
For the general (non-symmetric) case, we propose two schemes. The first (Proposition~\ref{prop:KW_sum_rate_for_general_matrix_product}) targets the regime $q>2$ and odd, while the second (Theorem~\ref{theo:achievability_square_matrix_products}), extends the design to any $q\geq 2$ and achieves a lower sum rate. The achievability results hold for arbitrary source distributions, following, as we will see, the same reasoning as in~\cite[Theorem~1]{korner1979encode}.

\subsection{Structured Codes for Distributed Matrix Multiplication: the Symmetric Case}
\label{sec:achievability_results_symmetric_matrices}

Symmetric matrices (e.g., adjacency, Hessian, and covariance) are fundamental, particularly in machine learning and signal processing. Given distributed sources ${\bf A}\in\mathbb{F}_q^{m\times l}$ and ${\bf B}\in\mathbb{F}_q^{m\times l}$, with entries from $\mathbb{F}_q$ with $m$ even and $l\geq 1$ for $q>2$ and odd, we next consider distributed computing of ${\bf \mathbfcal{D}}={\bf A}^{\intercal}{\bf B}\in\mathbb{F}_q^{l\times l}$, which is a symmetric\footnote{Symmetry can be enforced through random symmetric transformations, such as constructing ${\bf B}$ as a linear transformation of ${\bf A}$ such that ${\bf B}={\bf A}{\bf M}$ where ${\bf M}\in\mathbb{F}_q^{l\times l}$ is symmetric, or constructing ${\bf A}$ and ${\bf B}$ as linear transformations ${\bf A}={\bf M}_1{\bf Q}$ and ${\bf B}={\bf M}_2{\bf Q}$ of ${\bf Q}\in\mathbb{F}_q^{k\times l}$ where the product ${\bf M}_1^{\intercal}{\bf M}_2$ is symmetric, by matrix averaging $\frac{1}{2}({\bf A}^{\intercal}{\bf B}\oplus_q {\bf B}^{\intercal}{\bf A})$ using the Toeplitz decomposition over a field where $2$ is invertible (i.e., $q>2$ is odd), or by structurally designing the matrices to enforce symmetry.} matrix, i.e., ${\bf \mathbfcal{D}}={\bf \mathbfcal{D}}^{\intercal}$.

\begin{prop}
\label{prop:KW_sum_rate_for_symmetric_matrix_product}
{\bf (Distributed computation of symmetric matrix products.)} 
For the matrix multiplication source network, in the symmetric case, for $q>2$ and odd, and for any $\epsilon\in(0,1)$, 
\begin{align}
\label{KW_sum_rate_for_symmetric_matrix_product}
\RKMsymsum({\bf A}, {\bf B}) = 2H_q({\bf U},\, {\bf V}, \, {\bf W}) 
\end{align}
denotes the achievable sum rate, where ${\bf A}$ and ${\bf B}$ have the following representations:
\begin{align}
\label{eq:A_B}
{\bf A}=[{\bf A}_1 ; {\bf A}_2]\in\mathbb{F}_q^{m\times l} \ ,\quad \mbox{and}\quad {\bf B}=[{\bf B}_1 ; {\bf B}_2]\in\mathbb{F}_q^{m\times l}\ 
\end{align}
of even $m$, with matrix partitions ${\bf A}_1,{\bf A}_2,{\bf B}_1,{\bf B}_2\in \mathbb{F}_q^{m/2\times l}$, and ${\bf U}\in \mathbb{F}_q^{m/2\times l}$, ${\bf V}\in \mathbb{F}_q^{m/2\times l}$, and ${\bf W}\in \mathbb{F}_q^{l\times l}$ are matrix variables, and they satisfy the following relations: 
\begin{align}
\label{UVW_symmetric_matrices}
{\bf U}&={\bf A}_2\oplus_q{\bf B}_1\in \mathbb{F}_q^{m/2\times l} \ ,\,\,\,\nonumber\\
{\bf V}&={\bf A}_1\oplus_q{\bf B}_2\in \mathbb{F}_q^{m/2\times l} \ , \,\,\,\nonumber\\
{\bf W}&={\bf A}_2^{\intercal} {\bf A}_1\oplus_q{\bf B}_1^{\intercal} {\bf B}_2\in \mathbb{F}_q^{l\times l}\ .
\end{align}
\end{prop}

\begin{proof}
{\bf Outline of the proof.} The proof of Proposition~\ref{prop:KW_sum_rate_for_symmetric_matrix_product} proceeds as follows. We first introduce the non-linear source transformations and explain how their additions in $\mathbb{F}_q$ can be recovered via structured linear encoding. The argument draws on Elias's lemma (Lemma~\ref{Elias_lemma}), the K\"orner-Marton theorem for modulo-two sum computation   (Theorem~\ref{theo:KM}), and its extension by Han and Kobayashi to standard field addition in $\mathbb{F}_q$ (Lemmas~\ref{Elias_lemma_generalization} and~\ref{Korner-Marton_generalization}), using an $(n,\epsilon,\delta)$-coding scheme (Definition~\ref{def:eps_delta_n_coding}). We also employ Lemmas~\ref{lem:increasing_k} and~\ref{Elias_lemma_vectors} to extend the setting to $q$-ary vector variables. We now begin the proof.

{\bf Encoding:}
Sources devise the respective non-linear mappings 
\begin{align}
\label{distributed_source_info_matrices_A_and_B}
{\bf X}'_1&=g_1({\bf A})=[{\bf A}_2 ; {\bf A}_1 ; {\bf A}_2^{\intercal} {\bf A}_1]
\ ,\nonumber\\
{\bf X}'_2&=g_2({\bf B})=[{\bf B}_1 ; {\bf B}_2 ; {\bf B}_1^{\intercal} {\bf B}_2]
\ .
\end{align}  
Vertically concatenating the columns of ${\bf X}'_1\in \mathbb{F}_q^{(m+l) \times l}$ and ${\bf X}'_2\in \mathbb{F}_q^{(m+l) \times l}$ in (\ref{distributed_source_info_matrices_A_and_B}) 
we obtain
\begin{align}
\label{concatenated_column_vectors_symmetric_matrix_product}
{\bf X}_1=[
{\bf X}'_1(:,1);
{\bf X}'_1(:,2);
\hdots;
{\bf X}'_1(:,l)
]
\ , \quad
{\bf X}_2=[
{\bf X}'_2(:,1);
{\bf X}'_2(:,2);
\hdots;
{\bf X}'_2(:,l)
]
\ .
\end{align}

Let ${\bf Z}={\bf X}_1\oplus_2{\bf X}_2\in \mathbb{F}_q^{(m+l)l}$. Exploiting the componentwise and non-linear mappings $g_1$ and $g_2$ in (\ref{distributed_source_info_matrices_A_and_B})  to the length-$n$ realizations ${\bf A}^n$ and ${\bf B}^n$, the sources compute ${{\bf X}'_1}^n=g_1({\bf A}^n)\in \mathbb{F}_2^{(m+l)l \times n}$ and ${{\bf X}'_2}^n=g_2({\bf B}^n) \in \mathbb{F}_2^{(m+l)l \times n}$. Using (\ref{concatenated_column_vectors_symmetric_matrix_product}), ${\bf X}_1^n\in \mathbb{F}_q^{(m+l)l \times n}$ and ${\bf X}_2^n\in \mathbb{F}_q^{(m+l)l \times n}$ denote the vertical concatenations of ${{\bf X}'_1}^n$ and ${{\bf X}'_2}^n$, respectively. The separate encoders devise mappings $f_1: \mathbfcal{X}_1^n\to \mathcal{R}_{f_1}$ and $f_2: \mathbfcal{X}_2^n\to \mathcal{R}_{f_2}$, respectively, which we will detail below. Let ${\bf Z}(j)\in \mathbb{F}_q$ be the $j$-th symbol of ${\bf Z}$, and ${\bf Z}^n(j)\in \mathbb{F}_q^{n}$, for $j\in[(m+l)l]$, be its length-$n$ i.i.d. realization.

{\bf Decoding:} 
We next detail below how to recover ${\bf Z}^n$, which is subsequently used to reconstruct ${\bf \mathbfcal{D}}^n$. To do so, we begin by establishing several sufficient conditions below (Lemmas~\ref{Elias_lemma}-\ref{lem:increasing_k}). Specifically, our encoding scheme requires a well-known fact of Elias~\cite{gallager1968information}, which is that the capacity of BSCs can be attained by linear codes, as stated next.

\begin{lem} 
\label{Elias_lemma}
{\bf (Elias's lemma~\cite{gallager1968information}.)} 
Let $\{Z_i\}_{i=1}^{\infty}$ be an i.i.d. binary sequence. For fixed $\epsilon>0$ and sufficiently large $n$, there exist a binary matrix\footnote{From~\cite[Appendix IV, Proof of Th. 10, p.~411]{ahlswede1983source}, a random linear mapping ${\bf \mathbfcal{C}}\in\mathbb{F}_2^{\kappa \times n}$, whose components are all chosen  independently and uniformly from $\mathbb{F}_2$, yields an $(n,\epsilon)$-coding scheme. The case where $q>2$ is considered in~\cite[Lemma~4]{han1987dichotomy}.} ${\bf \mathbfcal{C}}\in\mathbb{F}_2^{\kappa\times n}$ and a function $\psi:\mathbb{F}_2^{\kappa}\to\mathbb{F}_2^n$ such that
\begin{align}
\label{Lemma1_kappa}
\kappa<n(H(Z)+\epsilon) \ ,\quad
\mathbb{P}(\psi({\bf \mathbfcal{C}} Z^n)\neq Z^n)<\epsilon \ . 
\end{align}
\end{lem}

\begin{proof}[Proof of Lemma~\ref{Elias_lemma}]
This result, originally published in~\cite{elias1955coding}, has been detailed in~\cite[Section 6.2]{gallager1968information}. Its proof builds upon {\emph{coset codes}}. We restate the proof here to illustrate how linear codes yield tight error bounds. An $(n,\kappa)$ coset code is a code with $2^{\kappa}$ codewords of blocklength $n>\kappa$ in which the mapping from message $u^{\kappa}\in\mathbb{F}_2^{1\times \kappa}$ to codeword $z^n$ is given by $z^n=u^{\kappa}{\bf \mathbfcal{C}}\oplus_2 v^n\in\mathbb{F}_2^{1\times n}$, where ${\bf \mathbfcal{C}}\in\mathbb{F}_2^{\kappa\times n}$ is fixed but arbitrary binary encoding matrix and $v^n\in\mathbb{F}_2^{1\times n}$ is a fixed but arbitrary sequence. The codewords of a coset code are thus formed from the codewords of a corresponding {\emph{parity-check code}}, $\hat{z}^n=u^{\kappa}{\bf \mathbfcal{C}}$, by adding  $v^n$ to each codeword. For a BSC, with a transmitted sequence $z^n$ and a noise sequence $w^n$, the received word is $y^n=z^n\oplus_2 w^n$. After subtracting the fixed sequence $v^n$ from $y^n$ before decoding, we have ${\hat{y}}^n=y^n\oplus_2 v^n={\hat{z}}^n\oplus_2 w^n$. 

Since $w^n \independent z^n$ for a BSC, a {\emph{maximum likelihood decoder}} will correctly decode the same set of noise sequences as for the associated parity-check code~\cite[Section 6.2]{gallager1968information}. An $(n,\kappa)$  parity-check code is specified by a $(n-\kappa)\times n$ binary parity-check matrix ${\bf H}$, and it contains all vectors $\hat{z}^n$ whose syndrome $s^{n-\kappa}\triangleq {\bf H}\hat{z}^n$ is equal to zero, namely the set $\{\hat{z}^n\in\{0,1\}^{n}\ : \ {\bf H}\hat{z}^n=0\}$~\cite[Section~6.1]{gallager1968information}. Given some general syndrome $s^{n-\kappa}\in\{0,1\}^{n-\kappa}$, a coset is the set of all vectors $z^n$ satisfying ${\bf H}z^n=s^{n-\kappa}$. For any syndrome $s^{n-\kappa}$, {\emph{maximum likelihood decoding}} can be achieved by calculating $s^{n-k}={\bf H}y^n$, finding the minimum weight sequence $\Psi(s^{n-\kappa})$ that satisfies $s^{n-k}={\bf H}\Psi(s^{n-\kappa})$, and decoding to the codeword $\hat{z}^n=\Psi(s^{n-\kappa})\oplus_2 y^n$~\cite[Theorem~6.1.1]{gallager1968information}. 

To complete the proof of Elias's lemma, we need the {\emph{coding theorem for parity-check codes}} in~\cite[Theorem 6.2.1]{gallager1968information}. We present the main steps; the full proof is omitted here for brevity and is available in~\cite[pp.~206-207]{gallager1968information}. To this end, employ an ensemble of $(n,k)$ coset codes where each component of ${\bf \mathbfcal{C}}$ and $v^n$ is selected from $\mathbb{F}_2$, independently and with equal probability. Then, with {\emph{maximum likelihood decoding}}, the probability of error for each message for this ensemble of codes used on a BSC, where $Z^n\in\mathbb{F}_2^n$ is viewed as its input sequence, satisfies~\cite{gallager1968information}  
\begin{align}
\label{eq:error_prob_random_coding_exponent_bound}
\mathbb{P}(\psi({\bf \mathbfcal{C}} Z^n))\neq Z^n)\leq \exp(-n E_r(R)) \ ,
\end{align}
where  $E_r(R)=\ln(2)-2\ln(\sqrt{p_s}+\sqrt{1-p_s})-R$ is the {\emph{random coding exponent}}, as a function of the BSC crossover-probability $p_s=\mathbb{P}(s=1)$, and the transmission rate $R$. Because $\{Z_i\}_{i=1}^{\infty}$ is stationary, the minimum expected codeword length per symbol is bounded as~\cite[Theorem~5.4.2]{cover2012elements} 
\begin{align}
\label{eq:description_length}
R=(\kappa\ln(2))/n < H(Z)+1/n\ .
\end{align}
Note that $\exp(-n E_r(R))\overset{(a)}{=}2^{-n}\cdot(\sqrt{p_s}+\sqrt{1-p_s})^{2n}\cdot2^{\kappa}
\overset{(b)}{<}2^{\kappa-n}$, where $(a)$ uses the definition of $E_r(R)$ with $R$ from (\ref{eq:description_length}), and $(b)$ the triangle inequality. Setting $2^{\kappa-n} =\epsilon$ implies $\mathbb{P}(\psi({\bf \mathbfcal{C}} Z^n)\neq Z^n)<\epsilon$ via (\ref{eq:error_prob_random_coding_exponent_bound}). 
For large $n$, specifically $n>1/\epsilon$, we have $R<H(Z)+\epsilon$ by (\ref{eq:description_length}).   This concludes the proof of Lemma~\ref{Elias_lemma}.
\end{proof}

We continue with the proof of the proposition, and recall that K\"orner-Marton, in their seminal work~\cite{korner1979encode}, applied Lemma~\ref{Elias_lemma} to the modulo-two adder source network, given a symmetric source distribution with $(X_1, X_2) \sim {\rm DSBS}(p)$. We now restate their main theorem (the {\emph{direct part}}), which serves as a building block for our achievability schemes and will subsequently be used to prove the proposition.

\begin{theo}
\label{theo:KM}
{\bf (Distributed computation of the modulo-two sum of binary sources~\cite[Theorem~1]{korner1979encode}.)} For the modulo-two adder source network, for $q=2$, and for any $\epsilon\in(0,1)$,
\begin{align}
R_1\geq H(Z) \ , \quad R_2\geq H(Z)
\end{align}
denotes the set of achievable rates for computing $Z=X_1\oplus_2 X_2$.
\end{theo}

\begin{proof}[Proof of Theorem~\ref{theo:KM}]
We provide a sketch of the proof here to highlight the utility of algebraic codes. Let $\mathbfcal{C}(\cdot)$ denote the encoding function used for both $X_1^n$ and $X_2^n$, such that $f_1(X_1^n)\triangleq {\bf \mathbfcal{C}}(X_1^n)= \mathbfcal{C} X_1^n\in\mathbb{F}_2^{\kappa}$, and $f_2(X_2^n)\triangleq {\bf \mathbfcal{C}}(X_2^n)=  \mathbfcal{C} X_2^n\in\mathbb{F}_2^{\kappa}$. Next, define a function $\phi:\mathbb{F}_2^{\kappa}\times \mathbb{F}_2^{\kappa}\to \mathbb{F}_2^n$ as $\phi(a^{\kappa},b^{\kappa}) \triangleq \psi(a^{\kappa}\oplus_2 b^{\kappa})$, where $a^{\kappa}$ and $b^{\kappa}$ are binary sequences of length $\kappa$, and $a^{\kappa}\oplus_2 b^{\kappa}$ denotes their modulo-two sum. Since the encoding function $\mathbfcal{C}$ is linear, 
\begin{align}
\label{eq:phi_psi}
\mathbb{P}(\phi({\bf \mathbfcal{C}}X_1^n,{\bf \mathbfcal{C}}X_2^n)\neq Z^n)=
\mathbb{P}(\psi({\bf \mathbfcal{C}}Z^n)\neq Z^n)<\epsilon \ .
\end{align}
This demonstrates that $({\bf \mathbfcal{C}},{\bf \mathbfcal{C}})$ is an $(n,\epsilon)$-coding scheme, where rates $R_1<H_q(Z)+\epsilon$ and $R_2<H_q(Z)+\epsilon$ are achievable~\cite[Theorem~1]{korner1979encode}. This concludes the proof of Theorem~\ref{theo:KM}.
\end{proof}

Continuing with the proof of the proposition, we will make use of several generalizations of Elias's result~\cite{gallager1968information} and K\"orner-Marton's problem~\cite{korner1979encode}, which will subsequently allow us to extend Theorem~\ref{theo:KM} to additions of vector variables over $\mathbb{F}_q$. In addition, it is worth mentioning that the achievability result in~\cite[Theorem~1]{korner1979encode} holds for arbitrary source distributions, whereas the matching converse follows from employing the {\emph{strong converse to the source coding theorem}} with side information~\cite{ahlswede1976bounds} and the symmetric distribution of $(X_1,X_2)\sim {\rm DSBS}(p)$ (see Definition~\ref{def:DSBS}).

We now proceed with the proof of Proposition~\ref{prop:KW_sum_rate_for_symmetric_matrix_product}. Building on~\cite{han1987dichotomy}, we  extend the $(n,\epsilon)$-coding scheme in Definition~\ref{def:eps_n_coding} to an $(n,\epsilon,\delta)$-coding scheme. 

\begin{defi}[An $(n,\epsilon,\delta)$-coding scheme~\cite{han1987dichotomy}]
\label{def:eps_delta_n_coding}
The pair of source encoders $(f_1,f_2)$ is called an $(n,\epsilon,\delta)$-coding scheme if there exists a function $\phi: \mathcal{R}_{f_1}\times \mathcal{R}_{f_2} \to \mathcal{Z}^n$ that satisfies~(\ref{eq:phi_psi}), with the decoding function $\psi$ as given in  Lemma~\ref{Elias_lemma_generalization}, such that letting 
\begin{align}
\hat{Z}^n \triangleq \phi(f_1(X_1^n),f_2(X_2^n)) \ , 
\end{align}
we have $\mathbb{P}(\hat{Z}^n \neq Z^n)<\delta$, where the encoding rates satisfy $R_1<H_q(Z)+\epsilon$ and $R_2<H_q(Z)+\epsilon$.
\end{defi}

We next consider a generalization of Elias's result (Lemma~\ref{Elias_lemma}) in~\cite{gallager1968information} to $q$-ary variables. Its proof follows from a counting argument (cf.~Ahlswede-Han~\cite[p.~411]{ahlswede1983source}), and is omitted here.   

\begin{lem}
\label{Elias_lemma_generalization}
{\bf (Han-Kobayashi~\cite[Lemma~4]{han1987dichotomy}.)} 
Let $Z\in \mathbb{F}_q$ be any random variable.   
Set $Z^n =\{Z_i\}_{i=1}^n=[Z_1; Z_2; \dots;Z_n]\in\mathbb{F}_q^{n}$ to denote a length-$n$ i.i.d. realization of $Z$. Then for any $\epsilon > 0$, $\delta > 0$, and sufficiently large $n$, a $\kappa\times n$ matrix ${\bf \mathbfcal{C}}\in\mathbb{F}_q^{\kappa\times n}$ as a linear encoding function,  and a decoding function $\psi: \mathbb{F}_q^{\kappa} \to \mathbb{F}_q^n$ from Lemma~\ref{Elias_lemma} exist such that
\begin{align}
\label{Lemma2_kappa}
\kappa<n(H_q(Z)+\epsilon) \ ,\quad
\mathbb{P}(\psi({\bf \mathbfcal{C}} Z^n)\neq Z^n)<\delta \ .
\end{align}
\end{lem}

Having established Lemma~\ref{Elias_lemma_generalization}, we now proceed with the proof of Proposition~\ref{prop:KW_sum_rate_for_symmetric_matrix_product} by presenting a generalization of K\"orner-Marton's problem in~\cite{korner1979encode} to additions in $\mathbb{F}_q$. 

\begin{lem}
\label{Korner-Marton_generalization}
{\bf (Han-Kobayashi~\cite[Lemma~5]{han1987dichotomy}.)} 
Let $(X_1,\ X_2)$ be any correlated random variables over $\mathcal{X}_1\subseteq \mathbb{F}_q$ and $\mathcal{X}_2\subseteq \mathbb{F}_q$, respectively, encoded separately at different sources. Define $Z = f(X_1, X_2) = X_1 \oplus_q X_2$. Then, for the distributed encoding of the function $Z$, with linear encoding:  
\begin{align}
\label{eq:ach_rates_HK}
R_1 \geq H_q(Z)\ , \quad R_2 \geq H_q(Z) 
\end{align}
denotes the set of achievable rates for computing $Z=X_1\oplus_q X_2$.
\end{lem}

\begin{proof}[Proof of Lemma~\ref{Korner-Marton_generalization}]
The proof proceeds along the same lines as that of Theorem~\ref{theo:KM}, where the dimension $\kappa$ of the encoding matrix ${\bf \mathbfcal{C}}\in\mathbb{F}_q^{\kappa\times n}$ is chosen according to Lemma~\ref{Elias_lemma_generalization}. 
\end{proof}

With Lemma~\ref{Korner-Marton_generalization} in place, we continue with the proof of Proposition~\ref{prop:KW_sum_rate_for_symmetric_matrix_product}. To this end, we first introduce an intermediate result (Lemma~\ref{lem:increasing_k}), which guarantees the desired error probability in Lemma~\ref{Elias_lemma_generalization} by constructing a ${\kappa}'\times n$ encoding matrix with $\kappa'>\kappa$. This lemma 
serves as a stepping stone toward Lemma~\ref{Elias_lemma_vectors}, where Lemmas~\ref{Elias_lemma_generalization} and~\ref{Korner-Marton_generalization} are extended to $q$-ary vector variables, necessitating a larger encoding dimension.

\begin{lem}
\label{lem:increasing_k}
Consider the setting in Lemma~\ref{Elias_lemma_generalization}. Let ${\kappa}'=\kappa+\Delta$ for some $\Delta>0$, which enlarges the value of $\kappa$ guaranteed by Lemma~\ref{Elias_lemma_generalization}. Then, for any $\epsilon,\
\delta > 0$, and sufficiently large $n$, a ${\kappa}'\times n$ matrix ${\bf \mathbfcal{C}}'\in\mathbb{F}_q^{{\kappa}'\times n}$  and a decoding function $\psi': \mathbb{F}_q^{{\kappa}'} \to \mathbb{F}_q^n$ exist such that
\begin{align}
\label{Lemma4_kappa}
{\kappa}'<n(H_q(Z)+\epsilon) \ ,\quad
\mathbb{P}(\psi'({\bf \mathbfcal{C}}' Z^n)\neq Z^n)<\delta \ .
\end{align}
\end{lem}

\begin{proof}[Proof of Lemma~\ref{lem:increasing_k}]
Choose a random linear mapping ${\bf \mathbfcal{C}}$ independently and uniformly from $\mathbb{F}_q^{\kappa\times n}$, with $\kappa=n(H_q(Z)+\epsilon)$, as in Lemma~\ref{Elias_lemma_generalization}. Then, denoting by $T_{\varepsilon}(Z)$ the set of all $\varepsilon$-typical sequences for a random variable $Z$, to evaluate the probability of decoding error, we must consider the following events:
i) $E_1:\ Z^n\notin T_{\varepsilon}(Z)$, and 
ii) $E_2:\ f_1(z^n)=f_1(Z^n), \, \mbox{for some} \, z^n\neq Z^n \, \mbox{such that} \, z^n\in T_{\varepsilon}(Z)$, 
to evaluate the probability of decoding error: $P_e=\mathbb{P}(\psi({\bf \mathbfcal{C}} Z^n)\neq Z^n)=\mathbb{P}(E_1 \ , E_2)$. 
Because the pair $(X_1, X_2)$ uniquely determines the value of  $Z$, we have~$\mathbb{P}(E_1)=0$. Given ${\bf \mathbfcal{C}}\in\mathbb{F}_q^{\kappa\times n}$, by counting all the cases satisfying $f_1(z^n)=f_1(Z^n)$ it follows for any $z^n\neq Z^n$ that $\mathbb{P}(f_1(z^n)=f_1(Z^n)) = (q^{n-1}/q^n)^{\kappa}=q^{-\kappa}$. Hence, exploiting $\kappa=n(H_q(Z)+\epsilon)$, we have
\begin{align}
\label{error_prob_bound_k}
\mathbb{P}(E_2)&\leq  |T_{\varepsilon}(Z)|\cdot q^{-\kappa}\nonumber\\
&\leq \exp(n(H_q(Z)+\varepsilon)) \cdot q^{-\kappa}\nonumber\\
&=\exp(-n(\epsilon-\varepsilon))\leq  \delta \ ,
\end{align}
where $\delta$ can be made arbitrarily small by choosing $\varepsilon$ small and then $n$ large.

We now set ${\kappa}'=n(H_q(Z)+\epsilon)+\Delta>\kappa$ for some $\Delta>0$. Choosing ${\bf \mathbfcal{C}}'$ independently and uniformly from $\mathbb{F}_q^{{\kappa}'\times n}$, by counting all the cases satisfying $f_1(z^n)=f_1(Z^n)$, it follows for any $z^n\neq Z^n$ that $\mathbb{P}(f_1(z^n)=f_1(Z^n)) = (q^{n-1}/q^n)^{{\kappa}'}=q^{-{\kappa}'}$. Hence, exploiting ${\kappa}'>\kappa$, we have
\begin{align}
\label{error_prob_bound_kprime}
\mathbb{P}(E_2)&\leq  |T_{\varepsilon}(Z)|\cdot q^{-{\kappa}'}\nonumber\\
&\leq \exp(n(H_q(Z)+\varepsilon)) \cdot q^{-{\kappa}'}\nonumber\\
&=\exp(n({\kappa}'/n-\Delta/n-\epsilon+\varepsilon)-{\kappa}')\nonumber\\
&=\exp(-n(\epsilon-\varepsilon)-\Delta)< \exp(-n(\epsilon-\varepsilon))\leq \delta \ .
\end{align}
From (\ref{error_prob_bound_k}) and (\ref{error_prob_bound_kprime}) we infer that $\mathbb{P}(E_2)$ satisfies $P_e({\kappa}')< P_e(\kappa)< \delta$ whenever  ${\kappa}'>\kappa$. Hence, (\ref{Lemma4_kappa}) follows, which concludes the proof of Lemma~\ref{lem:increasing_k}.
\end{proof}

To proceed with the proof of  Proposition~\ref{prop:KW_sum_rate_for_symmetric_matrix_product}, we next extend Lemma~\ref{Korner-Marton_generalization} to {\em $q$-ary vector variables}. 

\begin{lem}
\label{Elias_lemma_vectors} 
Let $({\bf X}_1,\ {\bf X}_2)$ be any correlated random vectors over $\mathbb{F}_q^{m}$ each, encoded separately at different sources. Then, for the distributed encoding of ${\bf Z}= {\bf X}_1 \oplus_q {\bf X}_2$, with linear encoding:
\begin{align}
\label{eq:rate_vector_encoding}
R_1 \geq H_q({\bf Z})\ , \quad R_2 \geq H_q({\bf Z}) \ .
\end{align}
\end{lem}

\begin{proof}[Proof of Lemma~\ref{Elias_lemma_vectors}]
We prove this lemma using Lemma~\ref{lem:increasing_k}. Let ${\bf Z}={\bf X}_1 \oplus_q {\bf X}_2\in\mathbb{F}_q^{m}$ be any random vector. Then for any fixed $\epsilon>0$, $\delta>0$, $j\in[m]$, and sufficiently large $n$, from Lemmas~\ref{Elias_lemma_generalization} and~\ref{Korner-Marton_generalization}, a random linear encoding matrix ${\bf \mathbfcal{C}}_j\in\mathbb{F}_q^{\kappa_j\times n}$, and a decoding function $\psi_j: \mathbb{F}_q^{\kappa_j} \to \mathbb{F}_q^n$ exist such that 
\begin{align}
\label{eq:kappa_Elias_lemma_vectors}
\kappa_j<n(H({\bf Z}(j))+\epsilon) \ , \quad \mathbb{P}(\psi_j({\bf \mathbfcal{C}}_j {\bf Z}^n(j))\neq {\bf Z}^n(j))<\delta \ .
\end{align}
Thus, $({\bf \mathbfcal{C}}_j,{\bf \mathbfcal{C}}_j)$ is an $(n,\epsilon,\delta)$-coding scheme (cf. Definition~\ref{def:eps_delta_n_coding}). From Lemma~\ref{lem:increasing_k}, letting $\kappa'_j>\kappa_j$, with $\kappa_j$ given in  (\ref{eq:kappa_Elias_lemma_vectors}), a $\kappa'_j\times n$ matrix ${\bf \mathbfcal{C}}'_j$ exists such that $({\bf \mathbfcal{C}}'_j,{\bf \mathbfcal{C}}'_j)$ is also an $(n,\epsilon,\delta)$-coding scheme. A sequential decoding of $\{{\bf Z}(j)^n\}$ for ordered $j\in[m]$ is possible, with $\kappa_j<n(H_q({\bf Z}(j)\,\vert\, \{{\bf Z}(j')\}_{j'<j})+\epsilon)$ for a given $j$, which allows for reconstructing ${\bf Z}^n =\{{\bf Z}_i\}_{i=1}^{n}$, a length-$n$ i.i.d. realization of~${\bf Z}$. Thus, setting $\kappa\triangleq\sum\nolimits_{j\in[m]}\kappa_j<n(\sum\nolimits_{j\in[m]}H_q({\bf Z}(j)\,\vert\, \{{\bf Z}(j')\}_{j'<j})+\epsilon)=n(H_q({\bf Z})+m\epsilon)$, and using a linear encoding matrix ${\bf \mathbfcal{C}}$ drawn independently and uniformly from $\mathbb{F}_q^{\kappa\times n}$, leads to an  $(n,\epsilon)$-coding scheme with the achievable rate region given in (\ref{eq:rate_vector_encoding}). This concludes the proof of Lemma~\ref{Elias_lemma_vectors}.
\end{proof}

We are now ready to prove the statement of Proposition~\ref{prop:KW_sum_rate_for_symmetric_matrix_product}. To this end, employ Lemma~\ref{Elias_lemma_vectors} to the non-linear mappings ${\bf X}_1\in\mathbb{F}_q^{(m+l)l}$ and ${\bf X}_2\in\mathbb{F}_q^{(m+l)l}$ devised in (\ref{concatenated_column_vectors_symmetric_matrix_product}) for computing the symmetric product ${\bf \mathbfcal{D}}$, for $q>2$ and odd. Exploiting~\cite{korner1979encode} and~\cite[Lemma~5]{han1987dichotomy}, the sum rate
\begin{align}
\label{KW_sum_rate_for_symmetric_matrix_product_proof}
\RKMsymsum({\bf A}, {\bf B}) = 2H_q({\bf Z})= 2H_q({\bf U},\, {\bf V}, \, {\bf W}) 
\end{align}
is achievable for the receiver to recover ${\bf Z}^n={\bf X}_1^n\oplus_q{\bf X}_2^n\in \mathbb{F}_q^{(m+l)l \times n}$ with vanishing error. Using the decoded sequence ${\bf \hat{Z}}^n=\phi({\bf \mathbfcal{C}}{\bf X}_1^n,{\bf \mathbfcal{C}}{\bf X}_2^n)=\psi({\bf \mathbfcal{C}}{\bf X}_1^n\oplus_q{\bf \mathbfcal{C}}{\bf X}_2^n)$, the receiver computes 
\begin{align}
\label{sufficiency_symmetric_matrix_product}
\frac{1}{2}(({\bf U}^{\intercal}\cdot {\bf V}\ominus_q{\bf W})\oplus_q({\bf U}^{\intercal} \cdot{\bf V}\ominus_q{\bf W})^{\intercal})
\overset{(a)}{=}\frac{1}{2}({\bf \mathbfcal{D}}\oplus_q{\bf \mathbfcal{D}}^{\intercal})
\overset{(b)}{=}{\bf \mathbfcal{D}}\ ,
\end{align}
where $(a)$ follows from employing ${\bf U}^{\intercal}\cdot {\bf V}\ominus_q{\bf W}={\bf A}_2^{\intercal}{\bf B}_2\oplus_q {\bf B}_1^{\intercal}{\bf A}_1$, a reordering of the terms, and ${\bf \mathbfcal{D}}={\bf A}_1^{\intercal}{\bf B}_1\oplus_q{\bf A}_2^{\intercal}{\bf B}_2$, $(b)$ from employing the Toeplitz decomposition to uniquely write any symmetric matrix ${\bf \mathbfcal{D}}={\bf A}^{\intercal}{\bf B}\in\mathbb{F}_q^{l\times l}$ over a field where $2$ is invertible (i.e., $q$ is odd) as ${\bf \mathbfcal{D}}=({\bf \mathbfcal{D}}\oplus_q{\bf \mathbfcal{D}}^{\intercal})/2$. Thus, (\ref{KW_sum_rate_for_symmetric_matrix_product_proof}) is achievable for computing the symmetric matrix product ${\bf \mathbfcal{D}}$. 
\end{proof}

In Proposition~\ref{prop:KW_sum_rate_for_symmetric_matrix_product}, based on (\ref{sufficiency_symmetric_matrix_product}), the receiver computes the matrix product ${\bf U}^{\intercal}\cdot {\bf V}$, whose inner dimension is half that of ${\bf A}^{\intercal}{\bf B}$, {\emph{indicating a computational split between the sources and receiver.}} Notably, lossless reconstruction of ${\bf Z}^n$ does not imply full recovery of ${\bf X}_1^n$ and ${\bf X}_2^n$, allowing computation of ${\bf \mathbfcal{D}}={\bf A}^{\intercal}{\bf B}$ without revealing the sources in their entirety --- a feature with security implications. The extension to odd $m$ is straightforward and omitted for brevity.

We derive a corollary for the {\emph{dot-product source network}}, where given even-length vectors ${\bf A}=\left(a_{i}\right)_{i\in [m]}\in\mathbb{F}_q^{m}$ and ${\bf B}=\left(b_{i}\right)_{i\in [m]}\in\mathbb{F}_q^{m}$, the receiver computes $d=\langle \,{\bf A} ,{\bf B} \,\rangle=\sum\nolimits_{i=1}^{m} a_i b_i\in\mathbb{F}_q$ for any $q\geq 2$. This generalizes the result from  Proposition~\ref{prop:KW_sum_rate_for_symmetric_matrix_product} originally established for $q>2$ and odd. This is a special instance where $l=1$, and thus its proof is omitted.

\begin{cor}
\label{cor:KW_sum_rate_for_inner_product}
{\bf (Distributed computation of dot products.)} 
For the matrix multiplication source network, in the symmetric case where $l=1$, for any $q\geq 2$ and for any $\epsilon\in(0,1)$, the sum rate 
\begin{align}
\label{KW_sum_rate_for_inner_product}
\RKMsymsum({\bf A}, {\bf B}) = 2H_q({\bf U},\, {\bf V}, \, W) 
\end{align}
is achievable, and the vector variables ${\bf U}\in \mathbb{F}_q^{m/2}$ and ${\bf V}\in \mathbb{F}_q^{m/2}$, and $W\in \mathbb{F}_q$ satisfy
\begin{align}
\label{UVW}
{\bf U}={\bf A}_2\oplus_q{\bf B}_1 \ ,\quad
{\bf V}={\bf A}_1\oplus_q{\bf B}_2 \ , \quad
W={\bf A}_2^{\intercal} {\bf A}_1\oplus_q{\bf B}_1^{\intercal} {\bf B}_2 \ ,
\end{align}
and $\langle \,{\bf A} ,{\bf B} \,\rangle={\bf U}^{\intercal}{\bf V}\ominus_q W$ for any $q\geq 2$, given the representations of ${\bf A}$ and ${\bf B}$ as in (\ref{eq:A_B}). 
\end{cor}

Denoting by $\eta_{\rm sym}$ the gain $\eta$ in the symmetric case (including dot products and symmetric matrix products), from Corollary~\ref{cor:KW_sum_rate_for_inner_product}, the receiver can compute $\langle \,{\bf A} ,{\bf B} \,\rangle$ without recovering $({\bf A},{\bf B})$ when $\eta_{\rm sym}>1$.  
We next provide an example under a specific PMF for binary-valued distributed source vectors $({\bf A} ,{\bf B})$ to show that the achievability result in Corollary~\ref{cor:KW_sum_rate_for_inner_product} does not coincide with (\ref{SW_condition}). In particular, $\RKMsymsum$ can be substantially smaller than $\RSWsum$, yielding $\eta_{\rm sym}>1$.

\begin{ex}
\label{ex:innerproduct_length_m_binary}
{\bf (Distributed dot product computation over structured source vectors.)} 
For the matrix multiplication source network, in the symmetric case where $l=1$, for $q=2$, consider ${\bf A}\in\mathbb{F}_2^{m}$ and ${\bf B}\in\mathbb{F}_2^{m}$ with the following asymmetric DSBS model:
\begin{align}
\label{conditions_cross_elements}
(a_{\frac{m}{2}+i},\ b_i)\sim {\rm DSBS}(p) \ ,\quad (a_i,\ b_{\frac{m}{2}+i})\sim {\rm DSBS}(p) \  \,\,  \mbox{are i.i.d. across}\,\, i\in[m/2]\ .
\end{align}

Employing the definitions of ${\bf X}_1$ and ${\bf X}_2$ in (\ref{concatenated_column_vectors_symmetric_matrix_product}) and letting ${\bf Z}={\bf X}_1\oplus_2{\bf X}_2\in \mathbb{F}_q^{m+1}$, from Corollary~\ref{cor:KW_sum_rate_for_inner_product}, the achievable sum rate for the receiver to recover ${\bf Z}=\begin{bmatrix}{\bf U};\, {\bf V};\, W\end{bmatrix}$ is given as
\begin{align}
\label{KW_sum_rate_inner_product}
\RKMsymsum({\bf A}, {\bf B}) &= 2H({\bf U},\, {\bf V}, \, W) 
=2H({\bf U})+2H({\bf V})+2H({\bf A}_2^{\intercal} {\bf A}_1\oplus_2{\bf B}_1^{\intercal} {\bf B}_2 \, \vert \, {\bf U}, {\bf V})\nonumber\\
&\overset{(a)}{=}mh(p)+mh(p)+2H({\bf A}_2^{\intercal} {\bf A}_1\oplus_2{\bf B}_1^{\intercal} {\bf B}_2 \, \vert \, {\bf U}, {\bf V})\nonumber\\
&\overset{(b)}{=}2mh(p)+2H({\bf U}^{\intercal}{\bf A}_1\oplus_2 {\bf A}_2^{\intercal}{\bf V}\, \vert \, {\bf U}, {\bf V})
\overset{(c)}{=}2mh(p)+2H({\bf Q}^{\intercal}{\bf A} \, \vert  \, {\bf Q}) \nonumber\\
&\overset{(d)}{=}2mh(p)+2\sum\limits_{j\in[m]} \binom{m}{j} p^j(1-p)^{m-j} H\Big(\sum\limits_{i\in\{i_1,i_2,\dots,i_j\}} a_i\Big)\nonumber\\
&\overset{(e)}{=}2mh(p)+2(1-(1-p)^m) \ ,
\end{align}
where $(a)$ uses ${\bf U}=\left(u_{i}\right)_{i\in [m/2]}$ and ${\bf V}=\left(v_{i}\right)_{i\in [m/2]}$, defined in (\ref{UVW}) with components $u_i, v_i \sim {\rm Bern}(p)$, i.i.d. across $i\in[m/2]$; $(b)$ uses ${\bf U}={\bf A}_2\oplus_q{\bf B}_1$ and ${\bf V}={\bf A}_1\oplus_q{\bf B}_2$ in (\ref{UVW}) to rewrite ${\bf A}_2^{\intercal} {\bf A}_1\oplus_2{\bf B}_1^{\intercal} {\bf B}_2$; $(c)$ follows from ${\bf Q}=\begin{bmatrix}{\bf U} ; {\bf V}\end{bmatrix}\in\mathbb{F}_2^{m}$ and observing ${\bf A}_2^{\intercal}{\bf V}={\bf V}^{\intercal}{\bf A}_2\in\mathbb{F}_2$; and $(d)$ from $H({\bf Q}^{\intercal}{\bf A} \,\vert\, {\bf Q})= H({\bf Q}^{\intercal}{\bf A} \,\vert\, {\bf Q}^{\intercal}{\bm 1}_m)$, given the model in (\ref{conditions_cross_elements}), where the nonzero components of ${\bf Q}$ are indexed by $\{i_1,i_2,\dots,i_j\}$ when ${\bf Q}^{\intercal}{\bm 1}_m=j$. Hence, 
\begin{align}
H({\bf Q}^{\intercal}{\bf A} \,\vert\, {\bf Q}^{\intercal}{\bm 1}_m=j)=
\begin{cases}
0 \ ,\quad j=0 \ ,\\  
H({\bf Q}^{\intercal}{\bf A} \,\vert\, {\bf Q}^{\intercal}{\bm 1}_m=j)=H\Big(\sum\limits_{i\in\{i_1,i_2,\dots,i_j\}} a_i\Big) \ ,\quad j\geq 1 \ .
\end{cases}
\end{align}
Finally, step $(e)$ holds because, given ${\bf Q}\neq {\bm 0}_m$, which occurs with probability $1-(1-p)^m$, the DSBS model implies that the entries of ${\bf A}$ are independent and uniformly distributed, i.e., $a_i\overset{i.i.d.}{\sim} {\rm Bern}(1/2)$. Hence, incorporating $H\Big(\sum\nolimits_{i\in\{i_1,i_2,\dots,i_j\}} a_i\Big)=1$, $j\geq 1$, we obtain  (\ref{KW_sum_rate_inner_product}).

The encoding rate for the asymptotically lossless compression of ${\bf A}$ and ${\bf B}$ is given by 
\begin{align}
\label{SW_sum_rate_inner_product}
\RSWsum({\bf A}, {\bf B})
\overset{(a)}{=}2H({\bf A}_1,{\bf B}_2)
=2\cdot\frac{m}{2} (1+h(p))
=m(1+h(p)) \ ,
\end{align}
using the Slepian-Wolf theorem~\cite{SlepWolf1973}, where $(a)$ is because $H({\bf A})=H({\bf A}_1,\ {\bf A}_2)$ and $H({\bf B})=H({\bf B}_1, \ {\bf B}_2)$, noting that ${\bf A}_1 \independent {\bf A}_2$ and ${\bf B}_1 \independent {\bf B}_2$, for ${\bf A}$ and ${\bf B}$ with i.i.d.-valued entries.

For this asymmetric DSBS setting,  the ratio of $\RSWsum$ in (\ref{SW_sum_rate_inner_product}) over $\RKMsymsum$ in (\ref{KW_sum_rate_inner_product}) is
\begin{align}
\label{gain_DSBS_channel_KM}
\eta_{\rm sym}=\frac{\RSWsum({\bf A}, {\bf B})}{\RKMsymsum({\bf A}, {\bf B})}=\frac{m(1+h(p))}{2mh(p)+2(1-(1-p)^m)} \ .
\end{align}

It is necessary from~\cite[Lemmas~1-2]{han1987dichotomy} (see Conditions~\ref{cond:3.1} and~\ref{cond:3.11} from  Lemma~\ref{lem:lemmas_1_2_Han_Kobayashi} in Section~\ref{sec:converses}) that $R_1,\ R_2\geq mh(p)$, for the joint PMF in (\ref{conditions_cross_elements}). Our structured scheme for computing ${\bf A}^{\intercal}{\bf B}$ incurs $1-(1-p)^m$ additional bits per source versus this lower bound, approaching $1$ as $m\to\infty$. Furthermore,  $\lim\limits_{p\to 0}\eta_{\rm sym} =\infty$, $\lim\limits_{p\to 1}\eta_{\rm sym} =m/2$, and $\lim\limits_{m\to \infty}\eta_{\rm sym}= \frac{1+h(p)}{2h(p)}\geq 1$ matches the gain in~\cite{korner1979encode}, which tends to infinity as $p\to 0$ or $p\to 1$. 

In Figure~\ref{fig:distributedgeneralmatrixproduct_symmetric}, we showcase $\RKMsymsum$ and $\RSWsum$ versus $p$ (in $\log$ scale), using the asymmetric DSBS model in (\ref{conditions_cross_elements}) for each pair ${\bf A}(:,j)$ and ${\bf B}(:,j)$, for $j\in[l]$, with $q=2$, for distributed computing of symmetric matrices ${\bf \mathbfcal{D}}={\bf A}^{\intercal}{\bf B}$ for different $m$ with $l=m$, under two further assumptions: (i) ${\bf A}_1^{\intercal}{\bf B}_1={\bf B}_1^{\intercal}{\bf A}_1$, i.e., ${\bf W}={\bm 0}_{l\times l}$, and (ii) ${\bf A}_2^{\intercal}{\bf A}_1={\bf B}_1^{\intercal}{\bf B}_2$. Then, ${\bf U}^{\intercal}{\bf V}=({\bf A}_2\oplus_2{\bf B}_1)^{\intercal}\cdot ({\bf A}_1\oplus_2{\bf B}_2)
\overset{(a)}{=}{\bf A}_2^{\intercal}{\bf A}_1\oplus_2 {\bf A}_2^{\intercal}{\bf B}_2\oplus_2{\bf A}_1^{\intercal}{\bf B}_1\oplus_2{\bf B}_1^{\intercal}{\bf B}_2
\overset{(b)}{=}{\bf A}_1^{\intercal}{\bf B}_1\oplus_2 {\bf A}_2^{\intercal}{\bf B}_2={\bf \mathbfcal{D}}$, where $(a)$ and $(b)$ follow from (i) and (ii), respectively. (i)-(ii) ensure $\eta_{\rm sym}$ to grow exponentially fast, as $p\to 0$ or $p\to 1$. For $q>2$ and odd, without (i)-(ii), ${\bf A}^{\intercal}{\bf B}$ can still be recovered  (Proposition~\ref{prop:KW_sum_rate_for_symmetric_matrix_product}).
\end{ex}

\begin{figure}[t!]
\centering
\includegraphics[width=0.7\textwidth]{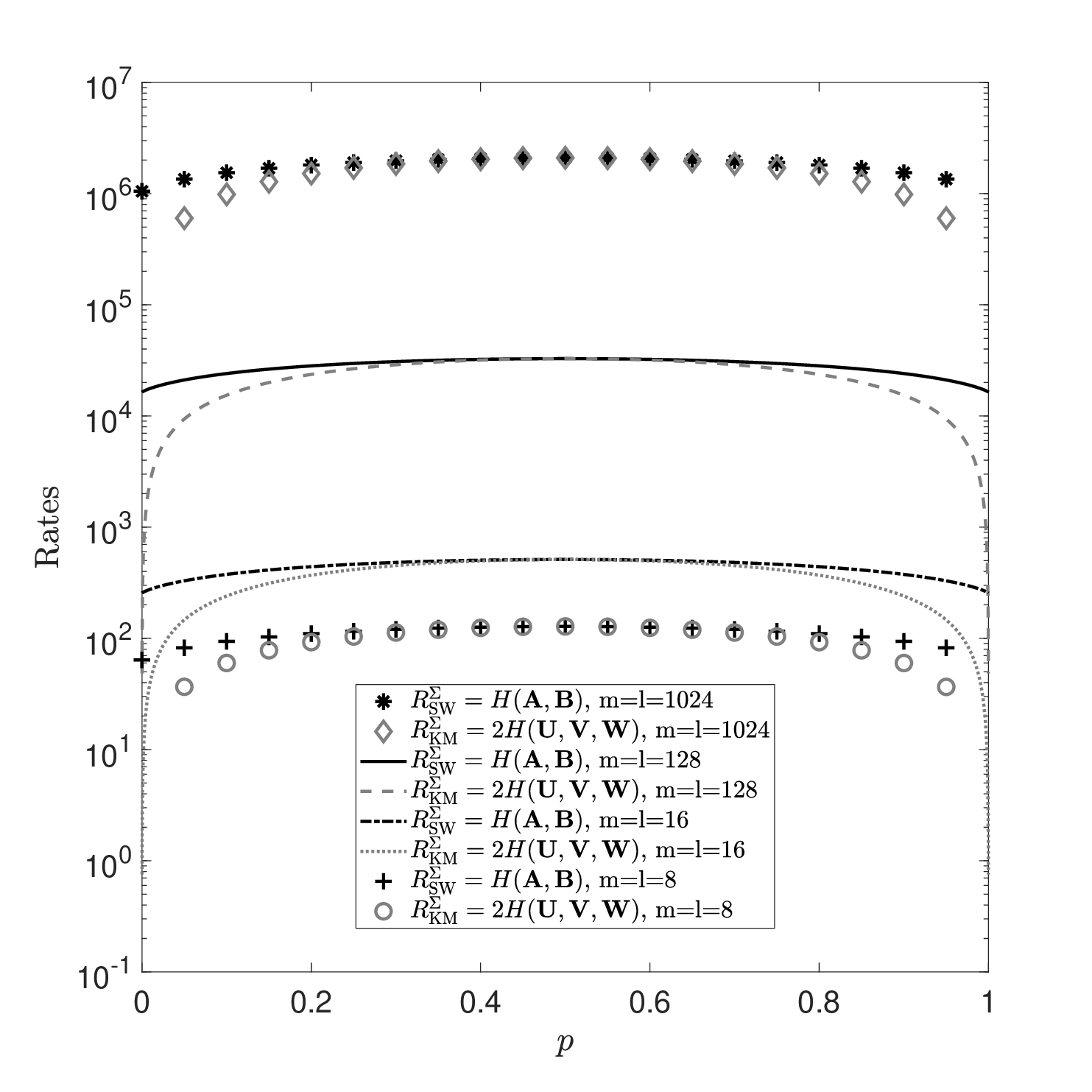}
\caption{Rate (in log scale) versus $p$ for the symmetric matrix multiplication network, to compute ${\bf \mathbfcal{D}}={\bf A}^{\intercal}{\bf B}={\bf B}^{\intercal}{\bf A}$, given ${\bf A}, {\bf B}\in \mathbb{F}_2^{m\times l}$, for different $m$ with $l=m$, where the joint source PMF is given in Example~\ref{ex:innerproduct_length_m_binary} (cf.~Corollary~\ref{cor:KW_sum_rate_for_inner_product}).}
\label{fig:distributedgeneralmatrixproduct_symmetric}
\end{figure}

Example~\ref{ex:innerproduct_length_m_binary} only captures a restricted class of source vectors, whereas Corollary~\ref{cor:KW_sum_rate_for_inner_product} holds for any possible joint distribution, i.e., any correlation structure between ${\bf A}, {\bf B}\in\mathbb{F}_q^{m}$, for $q\geq 2$, where the sum rate required for the partially secure or information-theoretically secure distributed computation of $\langle \,{\bf A} ,{\bf B} \,\rangle$ may approach or even exceed $\RSWsum$. To that end, we now turn to a potentially more realistic scenario (Example~\ref{ex:innerproduct_length_m_binary_elementwise}), with i.i.d.-valued distributed ${\bf A} ,{\bf B}\in\mathbb{F}_2^{m}$ with elementwise correlation, and show that the sum rate $\RKMsymsum$ may still fall below $\RSWsum$ in (\ref{SW_condition}).

\begin{ex}
\label{ex:innerproduct_length_m_binary_elementwise}
{\bf (Distributed dot product computation over elementwise-correlated vectors.)} 
For the matrix multiplication source network, in the symmetric case where $l=1$, for $q=2$, consider ${\bf A}\in\mathbb{F}_2^{m}$ and ${\bf B}\in\mathbb{F}_2^{m}$, with the following elementwise DSBS model:
\begin{align}
\label{eq:elementwise_DSBS}
(a_{i},\ b_{i})\sim {\rm DSBS}(p)\  \,\,  \mbox{are i.i.d. across}\,\, i\in\big[m\big] \ .
\end{align}

For a given $p\in \big[0,\ {1}/{2}\big]$, we note that (i) $(a_i,\ b_{i})\sim {\rm DSBS}(p)$, (ii) $(a_i,\  b_j)\sim {\rm DSBS}(1/2)$ for $i\neq j$, 
and (iii) $(a_i\oplus_2 b_i,\ a_{m/2+i}\oplus_2 b_{m/2+i})\sim {\rm DSBS}(2p(1-p))$ from (i). From Corollary~\ref{cor:KW_sum_rate_for_inner_product},  
\begin{align}
\label{KM_rate_elementwise_correlation_inner_product}
\RKMsymsum({\bf A}, {\bf B}) 
&\overset{(a)}{=}2H({\bf U},\, {\bf U}\oplus_2{\bf V}, \, W)\nonumber\\
&\overset{(b)}{\leq }2\cdot \Big(\frac{m}{2}+\frac{m}{2}h(2p(1-p))+1\Big)
=m(1+h(2p(1-p)))+2 \ ,
\end{align}
where $(a)$ is because joint entropy is invariant under bijective transformations, and $(b)$ follows from the relations (ii)-(iii) above, which imply $H({\bf U})=m/2$ and $H({\bf U}\oplus_2{\bf V})=(m/2)\cdot h(2p(1-p))$, respectively, and also from (\ref{UVW}), which yields $H(W)\leq 1$ for $W\in\mathbb{F}_2$. 

Employing (\ref{eq:elementwise_DSBS}), the scheme of Slepian-Wolf in~\cite{SlepWolf1973} yields the sum rate
\begin{align}
\label{SW_rate_elementwise_correlation_inner_product}
\RSWsum({\bf A}, {\bf B})=m(1+h(p)) \ .
\end{align}

For this elementwise DSBS setting, the ratio of $\RSWsum$ in (\ref{SW_rate_elementwise_correlation_inner_product})  over $\RKMsymsum$ in (\ref{KM_rate_elementwise_correlation_inner_product}) is
\begin{align}
\label{gain_DSBS_channel_KM_elementwise}
\eta_{\rm sym}=\frac{\RSWsum({\bf A}, {\bf B})}{\RKMsymsum({\bf A}, {\bf B})}\geq \frac{m(1+h(p))}{m(1+h(2p(1-p)))+2} \ .
\end{align}

It is necessary from~\cite[Lemmas~1-2]{han1987dichotomy} (see Conditions~\ref{cond:3.1} and~\ref{cond:3.11} from  Lemma~\ref{lem:lemmas_1_2_Han_Kobayashi} in Section~\ref{sec:converses}) that $R_1,\ R_2\geq m h(p)$, for the joint PMF in (\ref{eq:elementwise_DSBS}), for which the sum rate is upper-bounded by (\ref{SW_rate_elementwise_correlation_inner_product}). Our structured scheme for computing ${\bf A}^{\intercal}{\bf B}$ incurs fewer than $(m/2)(1+h(2p(1-p))-2h(p))+1$ additional bits per source versus this lower bound, approaching $m/2+1$ as $p\to 0$ or $p\to 1$, and $1$ as $p\to 1/2$. Furthermore, $\lim\limits_{m\to \infty}\eta_{\rm sym} \geq \frac{1+h(p)}{1+h(2p(1-p))}$, and $\lim\limits_{m\to \infty}\eta_{\rm sym} \geq 1$ in the limit as $p\to 1/2$. When $p=1/2$ in (\ref{eq:elementwise_DSBS}), resulting in ${\bf A}$ and ${\bf B}$ being independent, any achievable $(R_1, R_2)$ for ${\bf A}^{\intercal}{\bf B}$ has to satisfy $R_1+R_2\geq 2m=\RSWsum$ from the necessary conditions. 
\end{ex}

In Example~\ref{ex:innerproduct_length_m_binary_elementwise}, while the elementwise DSBS model of (\ref{eq:elementwise_DSBS}) may require the encoders to operate at $\RKMsymsum>\RSWsum$, the structured encoding scheme ensures that ${\bf A}$ and ${\bf B}$ are not fully disclosed.

We next derive a necessary condition for the receiver to compute the symmetric matrix product ${\bf \mathbfcal{D}}={\bf A}^{\intercal}{\bf B}\in\mathbb{F}_q^{l\times l}$ without fully recovering $({\bf A},\ {\bf B})$, thus ensuring $\eta_{\rm sym}>1$. 

\begin{prop}
\label{prop:KW_sum_rate_for_inner_product+vs_SW_sum_rate}
{\bf (Necessary condition for achieving $\eta_{\rm sym}>1$ in distributed computation of symmetric matrix products.)} 
For the matrix multiplication source network, in the symmetric case, for $q>2$ and odd, and for any $\epsilon\in(0,1)$,
the condition 
\begin{align}
\label{eq:condition_KM_less_SW}
H_q({\bf \mathbfcal{D}})+H_q({\bf U}\ ,{\bf V}\,\vert\,{\bf \mathbfcal{D}})< H_q({\bf A}\,\vert\,{\bf U}\ ,{\bf V}\ ,{\bf \mathbfcal{D}}) \ ,
\end{align}
with $m$ even and $l\geq 1$, where ${\bf U}, {\bf V}\in \mathbb{F}_q^{m/2\times l}$ are defined in (\ref{UVW_symmetric_matrices}), Proposition~\ref{prop:KW_sum_rate_for_symmetric_matrix_product} (and Corollary~\ref{cor:KW_sum_rate_for_inner_product} for $l=1$, and for any $q\geq 2$) ensures that the sum rate $\RKMsymsum$ is strictly less than $\RSWsum$ in (\ref{SW_condition}).
\end{prop}

\begin{proof} 
We first consider the dot product $d=\langle \,{\bf A} ,{\bf B} \,\rangle$ case from Corollary~\ref{cor:KW_sum_rate_for_inner_product}, representing the symmetric setting where $l=1$, for any $q\geq 2$. We then have from (\ref{KW_sum_rate_for_inner_product})
\begin{align}
\label{KM_expansion_corollary}
\RKMsymsum({\bf A}, {\bf B})&=2H_q({\bf U},\, {\bf V}, \,  W)\nonumber\\
&=2H_q({\bf U},\, {\bf V}, \, d)
=2H_q(d)+2H_q({\bf Q}\,\vert\,d)\ , \\
\label{SW_expansion_corollary}
\RSWsum({\bf A}, {\bf B})&=H_q({\bf A},{\bf B})\overset{(a)}{=}H_q({\bf A},{\bf B},{\bf Q}, \, d)\nonumber\\
&\overset{(b)}{=}H_q({\bf Q}, \, d)+H_q({\bf A},{\bf B}\,\vert\,{\bf U},\, {\bf V},d)\nonumber\\
&\overset{(c)}{=}H_q(d)+H_q({\bf Q} \,\vert\, d)+H_q({\bf A}\,\vert\,{\bf Q},d)\ ,
\end{align}
where $(a)$ holds because both ${\bf Q}$ and $d$ are deterministic functions of $({\bf A},{\bf B})$; $(b)$ by employing ${\bf Q}=\begin{bmatrix}{\bf U} ; {\bf V}\end{bmatrix}$; and $(c)$ from the condition $H_q({\bf B}\,\vert\,{\bf A},{\bf Q})=0$. By contrasting (\ref{KM_expansion_corollary}) with (\ref{SW_expansion_corollary}), the condition (\ref{eq:condition_KM_less_SW}) guaranties that $\RKMsymsum<\RSWsum$.

We then turn to the  general symmetric case ${\bf \mathbfcal{D}}={\bf A}^{\intercal}{\bf B}\in\mathbb{F}_q^{l\times l}$ from Proposition~\ref{prop:KW_sum_rate_for_symmetric_matrix_product} (see (\ref{KW_sum_rate_for_symmetric_matrix_product})), where $l>1$, and $q>2$ and odd. Following the same reasoning as in (\ref{KM_expansion_corollary}) and (\ref{SW_expansion_corollary}), we obtain:
\begin{align}
\label{KM_expansion}
\RKMsymsum({\bf A}, {\bf B})&=2H_q({\bf \mathbfcal{D}})+2H_q({\bf Q}\,\vert\,{\bf \mathbfcal{D}})\ , \\
\label{SW_expansion}
\RSWsum({\bf A}, {\bf B})&=H_q({\bf \mathbfcal{D}})+H_q({\bf Q} \,\vert\, {\bf \mathbfcal{D}})+H_q({\bf A}\,\vert\,{\bf Q},{\bf \mathbfcal{D}})\ .
\end{align}
Thus, by contrasting (\ref{KM_expansion}) with (\ref{SW_expansion}), the condition (\ref{eq:condition_KM_less_SW}) guarantees that $\RKMsymsum<\RSWsum$.
\end{proof}

From Proposition~\ref{prop:KW_sum_rate_for_inner_product+vs_SW_sum_rate}, $\RKMsymsum=2H_q({\bf \mathbfcal{D}})$ is achievable when $H_q({\bf Q}\,\vert\,{\bf \mathbfcal{D}})=0$ in~\eqref{eq:condition_KM_less_SW}. In this case, $H_q({\bf \mathbfcal{D}})< H_q({\bf A}\,\vert\,{\bf Q},{\bf \mathbfcal{D}})$, so ${\bf \mathbfcal{D}}$ can be computed at a sum-rate below that required for lossless reconstruction of both ${\bf A}$ and ${\bf B}$, although partial information about them may still be revealed. 
For general, potentially non-structured source PMFs, the condition in~\eqref{eq:condition_KM_less_SW} may fail. In such cases, Proposition~\ref{prop:KW_sum_rate_for_symmetric_matrix_product} may require $\RKMsymsum>\RSWsum$, while still not necessarily revealing ${\bf A}$ and ${\bf B}$ completely.

Drawing on Lemmas~\ref{Elias_lemma}-\ref{lem:increasing_k} and building on Corollary~\ref{cor:KW_sum_rate_for_inner_product}, we next present another achievable region, where given distributed sources ${\bf A}\in\mathbb{F}_q^{m\times l}$ and ${\bf B}\in\mathbb{F}_q^{m\times l}$, the receiver aims to compute the symmetric matrix ${\bf \mathbfcal{D}}={\bf A}^{\intercal}{\bf B}\in \mathbb{F}_q^{l\times l}$ for $q>2$ and odd, achieving an improved $\RKMsymsum$.

\begin{theo}
\label{theo:achievability_symmetric_matrix_products}
{\bf (Distributed computation of symmetric matrix products.)} 
For the matrix multiplication source network, in the symmetric case, for $q>2$ and odd, and for any $\epsilon\in(0,1)$, the sum rate 
\begin{align}
\label{KW_sum_rate_for_symmetric_matrix_product_generalized}
\RKMsymrfdsum({\bf A},\ {\bf B}) = 2H_q({\bf Z}) 
\end{align}
is achievable, where ${\bf Z}={\bf X}_1\oplus_q{\bf X}_2\in \mathbb{F}_q^{(m+l)l}$, with  
vectors ${\bf X}_1,\ {\bf X}_2\in \mathbb{F}_q^{(m+l)l}$ that satisfy
\begin{align}
\label{concatenated_column_vectors}
{\bf X}_1=\begin{bmatrix}
{\bf X}'_1(:,1);
{\bf X}'_1(:,2);
\hdots;
{\bf X}'_1(:,l)
\end{bmatrix}
\ ,\quad 
{\bf X}_2=\begin{bmatrix}
{\bf X}'_2(:,1);
{\bf X}'_2(:,2);
\hdots;
{\bf X}'_2(:,l)
\end{bmatrix}
\ ,
\end{align}
with matrices ${\bf X}'_1\in \mathbb{F}_q^{(m+l) \times l}$ and ${\bf X}'_2\in \mathbb{F}_q^{(m+l) \times l}$ derived from the following encoder mappings:
\begin{align}
\label{distributed_source_info_matrices_matrix_form_AB}
{\bf X}'_1=g_1({\bf A})\!=\!\begin{bmatrix}{\bf A}_2 ; {\bf A}_1 ; {\bf A}_2^{\intercal} {\bf A}_1\oplus_q {\bf A}_1^{\intercal} {\bf A}_2\end{bmatrix}
\ ,\quad
{\bf X}'_2=g_2({\bf B})\!=\!\begin{bmatrix}{\bf B}_1 ; {\bf B}_2 ; {\bf B}_1^{\intercal} {\bf B}_2\oplus_q {\bf B}_2^{\intercal} {\bf B}_1\end{bmatrix} \ .
\end{align}
Then for any $\epsilon > 0$, $\delta > 0$, and sufficiently large $n$, a $\kappa\times n$ matrix ${\bf \mathbfcal{C}}\in\mathbb{F}_q^{\kappa\times n}$  and decoding functions $\psi_j: \mathbb{F}_q^{\kappa_j} \to \mathbb{F}_q^n$, $j\in[(m+l)l]$ exist such that
\begin{align}
\label{k_value}
&\kappa=\max\Big\{\sum\limits_{j\in[ml]} \kappa_j \ , \sum\limits_{j\in[ml+1,\ (m+l)l]} \kappa_j \Big\}\ , \\
\label{individual_k_independence_across_j}
&\kappa_j<n(H_q({\bf Z}(j))+\epsilon) \ ,\quad j\in[(m+l)l] \ ,\\
\label{error_probability_decoding_j}
&\mathbb{P}(\{\psi_j({\bf \mathbfcal{C}}{\bf Z}^n(j))\neq {\bf Z}^n(j)\}_{j\in[(m+l)l]})<\delta \ ,
\end{align}
where ${\bf Z}(j)$, $j\in[(m+l)l]$, is the $j$-th component of ${\bf Z}$, and ${\bf Z}^n(j) = ({\bf Z}_1(j), \dots,{\bf Z}_n(j))\in\mathbb{F}_q^{n}$. 
\end{theo}

\begin{proof}
{\bf Outline of the proof.} The proof of Theorem~\ref{theo:achievability_symmetric_matrix_products} proceeds as follows. We first introduce the non-linear source transformations and explain how their additions in $\mathbb{F}_q$ can be recovered via structured linear encoding, drawing on Lemmas~\ref{Elias_lemma}-\ref{Elias_lemma_vectors}. We now begin the proof.

{\bf Encoding:} 
Following Proposition~\ref{prop:KW_sum_rate_for_symmetric_matrix_product}, let ${\bf Z}={\bf X}_1\oplus_q{\bf X}_2\in \mathbb{F}_q^{(m+l)l}$. For fixed $\epsilon, \ \delta>0$ and sufficiently large $n$, the encoders operate componentwise for each $j\in[(m+l)l]$, using the same matrix~${\bf \mathbfcal{C}}_j$, drawn independently and uniformly from $\mathbb{F}_q^{\kappa_j\times n}$ (cf.~Ahlswede-Han~\cite{ahlswede1983source}). They compute ${\bf \mathbfcal{C}}_j {\bf X}^n_1(j)\in\mathbb{F}_q^{\kappa_j}$ and ${\bf \mathbfcal{C}}_j {\bf X}^n_2(j)\in\mathbb{F}_q^{\kappa_j}$, where, by Lemmas~\ref{Elias_lemma_generalization} and~\ref{Korner-Marton_generalization}, $\kappa_j<n (H_q({\bf Z}(j)\,\vert\, \{{\bf Z}(j')\}_{j'<j})+\epsilon)$. From Lemma~\ref{Elias_lemma_vectors}, we then set $\kappa=\sum\nolimits_{j\in[m]}\kappa_j<n(H_q({\bf Z})+m\epsilon)$. Using a linear encoding matrix~${\bf \mathbfcal{C}}$ drawn independently and uniformly from $\mathbb{F}_q^{\kappa\times n}$ achieves $R_1\geq H_q({\bf Z})$ and $R_2\geq H_q({\bf Z})$.

{\bf Decoding:} 
From~\cite{ahlswede1983source}, there exists a decoding function $\psi_j:\mathbb{F}_q^{\kappa_j}\to \mathbb{F}_q^{n}$ that satisfies:
\begin{align}
{\bf \hat{Z}}^n(j)\triangleq\phi({\bf \mathbfcal{C}}_j {\bf X}^n_1(j),{\bf \mathbfcal{C}}_j {\bf X}^n_2(j)) \triangleq \psi_j({\bf \mathbfcal{C}}_j ({\bf X}^n_1(j)\oplus_q {\bf X}^n_2(j)))
\end{align}
such that 
i) $\kappa_j<n(H({\bf Z}(j)\,\vert\, \{{\bf Z}(j')\}_{j'<j})+\epsilon)$, and 
ii) $\mathbb{P}(\psi_j({\bf \mathbfcal{C}}_j{\bf Z}^n(j))\neq {\bf Z}^n(j))<\delta$.

Using the achievability result in~\cite{korner1979encode}, and  Lemmas~\ref{Elias_lemma}-\ref{Elias_lemma_generalization}, the achievable sum rate for the receiver to recover the matrix sequence ${\bf Z}^n={\bf X}_1^n\oplus_q{\bf X}_2^n\in\mathbb{F}_q^{(m+l)l \times n}$ with vanishing error is:
\begin{align}
\label{KW_sum_rate_for_symmetric_matrix_product_achieve_proof}
\RKMsymrfdsum({\bf A},\ {\bf B}) = 2H_q({\bf U},\, {\bf V}, \, {\bf W}_S) \ ,
\end{align}
where the matrix variables ${\bf U},\, {\bf V}, \, {\bf W}_S$ are given as follows:
\begin{align}
\label{UVW_symmetric_matrices_refined}
{\bf U}&={\bf A}_2\oplus_q{\bf B}_1\in \mathbb{F}_q^{m/2\times l} \ ,\quad
{\bf V}={\bf A}_1\oplus_q{\bf B}_2\in \mathbb{F}_q^{m/2\times l} \ , \nonumber\\
&{\bf W}_S={\bf A}_2^{\intercal} {\bf A}_1\oplus_q{\bf A}_1^{\intercal} {\bf A}_2\oplus_q{\bf B}_1^{\intercal} {\bf B}_2\oplus_q{\bf B}_2^{\intercal} {\bf B}_1\in \mathbb{F}_q^{l\times l}\ .
\end{align}
Using ${\bf \hat{Z}}^n$ and exploiting  (\ref{UVW_symmetric_matrices_refined}), the receiver computes
\begin{align}
\label{sufficiency_symmetric_matrix_product_achieve}
\frac{1}{2}({\bf U}^{\intercal}\cdot {\bf V} \oplus_q{\bf V}^{\intercal} \cdot{\bf U} \ominus_q {\bf W}_S)
\overset{(a)}{=}\frac{1}{2}(({\bf A}_1^{\intercal}{\bf B}_1\oplus_q{\bf A}_2^{\intercal}{\bf B}_2)\oplus_q({\bf A}_1^{\intercal}{\bf B}_1\oplus_q{\bf A}_2^{\intercal}{\bf B}_2)^{\intercal})
\overset{(b)}{=}{\bf \mathbfcal{D}}
\ ,
\end{align}
where $(a)$ follows from reordering the terms and using (\ref{UVW_symmetric_matrices_refined}), and $(b)$ from  
the symmetry of ${\bf \mathbfcal{D}}={\bf A}^{\intercal}{\bf B}\in\mathbb{F}_q^{l\times l}$, rendering the sum rate in (\ref{KW_sum_rate_for_symmetric_matrix_product_achieve_proof}) achievable for $q>2$ and odd.

Concatenating the columns of the matrices ${\bf U}$, ${\bf V}$, and ${\bf W}_S$ each, defined in (\ref{UVW_symmetric_matrices_refined}), denote by 
\begin{align}
\label{concatenated_U}
&\begin{bmatrix}
{\bf U}(:,1);
{\bf U}(:,2);
\hdots; 
{\bf U}(:,l)
\end{bmatrix}=\begin{bmatrix}
{\bf Z}(1);
{\bf Z}(2);
\hdots; 
{\bf Z}(ml/2)
\end{bmatrix}\in \mathbb{F}_q^{ml/2} \ ,\\
\label{concatenated_V}
&\begin{bmatrix}
{\bf V}(:,1);
{\bf V}(:,2);
\hdots; 
{\bf V}(:,l)
\end{bmatrix}=\begin{bmatrix}
{\bf Z}(ml/2+1);
{\bf Z}(ml/2+2);
\hdots;
{\bf Z}(ml))
\end{bmatrix}\in \mathbb{F}_q^{ml/2}\ ,\\
\label{concatenated_W}
&\begin{bmatrix}
{\bf W}_S(:,1);
{\bf W}_S(:,2);
\hdots;
{\bf W}_S(:,l)
\end{bmatrix}=\begin{bmatrix}
{\bf Z}(ml+1);
{\bf Z}(ml+2);
\hdots;
{\bf Z}((m+l)l)
\end{bmatrix}\in \mathbb{F}_q^{l^2}\ , 
\end{align}
where ${\bf Z}(j)={\bf A}_2(:,j)\oplus_q
{\bf B}_1(:,j)$, for $j\in[ml/2]$,  ${\bf Z}(j)={\bf A}_1(:,j)\oplus_q {\bf B}_2(:,j)$, for $j\in[ml/2+1,ml]$, and ${\bf Z}(ml+j)={\bf W}_S(j-l\cdot\big(\lceil\frac{j}{l}\rceil-1\big),\lceil\frac{j}{l}\rceil)\in\mathbb{F}_q$, for $j\in[l^2]$.  
We note the following. (\ref{concatenated_U}) and (\ref{concatenated_V}) can be combined into ${\bf Z}(j)= g_{1,j}({\bf A})\oplus_q g_{2,j}({\bf B}), \,\, j\in [ml]$, for some $g_{1,j}$ and $g_{2,j}$, where $\{{\bf Z}(j)\}_{j\in [ml]}$ are mutually independent. Similarly, ${\bf Z}(j)= g'_{1,j}({\bf A})\oplus_q g'_{2,j}({\bf B}), \,\, j\in [ml+1,\ (m+l)l]$, where $g'_{1,j}\neq g_{1,j}$ and $g'_{2,j}\neq g_{2,j}$, by directly comparing (\ref{concatenated_U}), (\ref{concatenated_V}), and (\ref{concatenated_W}).

From (\ref{UVW_symmetric_matrices_refined}), the computation of the linear parts in (\ref{concatenated_U}) and (\ref{concatenated_V}) requires
\begin{align}
\label{eq:relation_Z_linear}
H_q({\bf Z}(j),\,j\in[ml])=H_q({\bf U}\ , {\bf V})\leq lm \ ,
\end{align}
and we also note that the computation of the non-linear part in (\ref{concatenated_W}) requires
\begin{align}
\label{eq:relation_Z_nonlinear}
H_q({\bf Z}(j),&\,j\in[ml+1,\ (m+l)l] \,\vert\, {\bf Z}(j),\,j\in[ml])
\overset{(a)}{=}H_q({\bf W}_S \,\vert\, {\bf U}\ , {\bf V})\nonumber\\
&\overset{(b)}{=}H_q({\bf U}^{\intercal}\cdot {\bf V} \oplus_q{\bf V}^{\intercal} \cdot{\bf U} \ominus_q {\bf W}_S\,\vert\, {\bf U}\ , {\bf V})
\overset{(c)}{=}H_q({\bf A}^{\intercal}{\bf B}\,\vert\, {\bf U}\ , {\bf V})
\leq l^2 \ ,
\end{align}
where $(a)$ is due to (\ref{UVW_symmetric_matrices_refined}), 
and from using definitions in (\ref{concatenated_U}), (\ref{concatenated_V}), and (\ref{concatenated_W}), $(b)$ follows by conditioning, and $(c)$ from employing (\ref{sufficiency_symmetric_matrix_product_achieve}). 
Hence, for ${\bf X}_1\in\mathbb{F}_q^{(m+l)l}$ and ${\bf X}_2\in\mathbb{F}_q^{(m+l)l}$ given in (\ref{concatenated_column_vectors}), employing  Lemmas~\ref{lem:increasing_k} and~\ref{Elias_lemma_vectors}, and from (\ref{eq:relation_Z_linear}) and (\ref{eq:relation_Z_nonlinear}), the following rate per source can be achieved for computing the symmetric matrix ${\bf \mathbfcal{D}}={\bf A}^{\intercal}{\bf B}\in\mathbb{F}_q^{l\times l}$, for $q>2$ and odd:
\begin{align}
\label{eq:general_achievability_symmetric}
\frac{\kappa}{n}&<\max\{H_q({\bf Z}(j),\,j\in[ml])\ ,\, 
H_q({\bf Z}(j),\,j\in[ml+1,\ (m+l)l] \,\vert\, {\bf Z}(j),\,j\in[ml]) \}+\epsilon  \nonumber\\
&=\max\{H_q({\bf U}\ , {\bf V}\,\vert\, {\bf \mathbfcal{D}}={\bf \mathbfcal{D}}^{\intercal})\ ,\, H_q({\bf A}^{\intercal}{\bf B}\,\vert\, {\bf U}\ , {\bf V},\ {\bf \mathbfcal{D}}={\bf \mathbfcal{D}}^{\intercal}) \}+\epsilon \ , 
\end{align}
for a $\kappa\times n$ linear encoding matrix ${\bf \mathbfcal{C}}\in\mathbb{F}_q^{\kappa\times n}$ with $\kappa=\max\big\{\sum\nolimits_{j\in[ml]} \kappa_j\ , \ \sum\nolimits_{j\in[ml+1,\ (m+l)l]} \kappa_j\big\}$.

From Lemma~\ref{lem:increasing_k}, and~\cite{korner1979encode},~\cite{han1987dichotomy}, and~\cite{ahlswede1983source}, if we choose ${\bf \mathbfcal{C}}$ independently and uniformly from $\mathbb{F}_q^{\kappa\times n}$, and $\kappa$ as in (\ref{k_value}), $({\bf \mathbfcal{C}},{\bf \mathbfcal{C}})$ forms an $(n,\epsilon,\delta)$-coding scheme for decoding ${\bf Z}^n$.
\end{proof}

\subsection{Structured Codes for Distributed Matrix Multiplication: the General non-Symmetric Case} 
\label{sec:achievability_results_general_matrices} 
Focusing on general ${\bf A},\ {\bf B}\in\mathbb{F}_q^{m\times l}$, for $q>2$ and odd, and $m, \ l>1$, we next devise distributed encoding schemes for computing ${\bf \mathbfcal{D}}={\bf A}^{\intercal}{\bf B}\in\mathbb{F}_q^{l\times l}$. Unlike Proposition~\ref{prop:KW_sum_rate_for_symmetric_matrix_product} and Theorem~\ref{theo:achievability_symmetric_matrix_products}, this setting necessitates new techniques due to the non-symmetric form of ${\bf \mathbfcal{D}}$, as we detail below.

\begin{prop}
\label{prop:KW_sum_rate_for_general_matrix_product}
{\bf (Distributed computation of square matrix products.)} 
For the general matrix multiplication source network, for $q>2$ and odd, and for any $\epsilon\in(0,1)$, the sum rate
\begin{align}
\label{KM_sum_rate_for_general_matrix_product}
\RKMalternativesum({\bf A}, {\bf B}) = 2H_q(\{{\bf A}\oplus_q{\bf \tilde{B}}_j\}_{j=1}^l\, , \, \{{\bf A}^{\intercal}{\bf A}\oplus_q{\bf \tilde{B}}_j^{\intercal}{\bf \tilde{B}}_j\}_{j=1}^l) 
\end{align}
is achievable, where ${\bf \tilde{B}}_j={\bf B}_j{\bm 1}_{1\times l}\in\mathbb{F}_q^{m\times l}$, with ${\bf B}_j\in\mathbb{F}_q^{m}$ denoting the $j$-th column of ${\bf B}$.
\end{prop}

\begin{proof}
We set vector ${\bf \mathbfcal{D}}_j$ to be ${\bf \mathbfcal{D}}_j=(d_{ij})_{i\in [l]}={\bf A}^{\intercal}{\bf B}_j\in\mathbb{F}_q^{l}$ for $j\in[l]$. 
Following the steps of Lemma~\ref{Elias_lemma} and the Proofs of Corollary \ref{cor:KW_sum_rate_for_inner_product} and  Proposition~\ref{prop:KW_sum_rate_for_symmetric_matrix_product}, the receiver can recover $\{{\bf A}\oplus_q{\bf \tilde{B}}_j\}_{j=1}^l$, $\{{\bf A}^{\intercal}{\bf A}\oplus_q{\bf \tilde{B}}_j^{\intercal}{\bf \tilde{B}}_j\}_{j=1}^l$, and then compute the following $l\times l$ matrix:
\begin{align}
({\bf A}\oplus_q{\bf \tilde{B}}_j)^{\intercal} ({\bf A}\oplus_q{\bf \tilde{B}}_j)\ominus_q({\bf A}^{\intercal}{\bf A}\oplus_q&{\bf \tilde{B}}_j^{\intercal}{\bf \tilde{B}}_j)\nonumber\\
&={\bm 1}_{l\times 1}{\bf B}_j^{\intercal}{\bf A}\oplus_q \begin{bmatrix}{\bf A}_1^{\intercal}; {\bf A}_2^{\intercal};\hdots, {\bf A}_l^{\intercal}\end{bmatrix} {\bf B}_j{\bm 1}_{1\times l}\nonumber\\
&={\bm 1}_{l\times 1}\big((d_{ij})_{i\in [l]}\big)^{\intercal}\oplus_q (d_{ij})_{i\in[l]}{\bm 1}_{1\times l}\nonumber\\
&=(d_{ij}\oplus_q d_{i'j})_{i,\ i'\in[l]}\in\mathbb{F}_q^{l\times l} 
\end{align}
which is a symmetric matrix with $l$ unknowns and $l(l-1)/2\geq l$ linearly independent equations for $l\geq 2$ and $q> 2$. Hence,  ${\bf \mathbfcal{D}}_j$, for each $j\in[l]$, as well as ${\bf \mathbfcal{D}}$ can be recovered.
\end{proof}

To demonstrate the performance of Proposition~\ref{prop:KW_sum_rate_for_general_matrix_product}, we next consider an example.  
\begin{ex}
\label{ex:general_matrix_q3}
{\bf (Computing a non-symmetric matrix product of structured sources.)} 
For the general matrix multiplication source network, where $l=2$, for $q=3$, consider ${\bf A},\ {\bf B}\in\mathbb{F}_3^{m\times 2}$, with entries that satisfy $a_{ij}\sim \big({1}/{2}-\zeta ,\ 2\zeta ,\ {1}/{2}-\zeta \big)$, for $i\in[m]$ and $j\in\{1,2\}$, i.i.d. across $i$ for some $\zeta\in[0,{1}/{2}]$, and the joint PMF of $(a_{i1},\ b_{i1})$, i.i.d. across $i\in[m]$, is given as
\begin{align}
\label{eq:example_q3_joint_PMF}
\small{P_{a_{i1},b_{i1}}=
\begin{bmatrix}
(\frac{1}{2}-\zeta)(1-p)& (\frac{1}{2}-\zeta)p & 0 \\
2\zeta p & 0 & 2\zeta(1-p)\\
0 & (\frac{1}{2}-\zeta)(1-p)& (\frac{1}{2}-\zeta)p
\end{bmatrix}}\ .
\end{align}
We further assume that $b_{i1}=b_{i2}=-a_{i2}$. Thus, $H_3({\bf A},{\bf B})=mH_3(a_{i1},a_{i2},b_{i1},b_{i2})=mH_3(a_{i1},b_{i1})$.

Therefore, the sum rate for distributed encoding of $({\bf A},{\bf B})$ is given as~\cite{SlepWolf1973}
\begin{align}
\label{SW_sum_rate_for_general_matrix_product_q3}
\RSWsum({\bf A}, {\bf B})&=m(h(a_{i1})+h(b_{i1}\,\vert\,a_{i1}))\nonumber\\
&=m(h(2\zeta)+(1-2\zeta)+h(p)) \ .
\end{align}

Exploiting Proposition~\ref{prop:KW_sum_rate_for_general_matrix_product} to compute ${\bf \mathbfcal{D}}={\bf A}^{\intercal}{\bf B}\in\mathbb{F}_3^{2\times 2}$, we can achieve a sum rate of 
\begin{align}
\label{KM_sum_rate_for_general_matrix_product_q3}
\RKMalternativesum({\bf A}, {\bf B})\leq  2m h\Big(2\big(\frac{1}{2}-\zeta\big)(1-p)+2\zeta(1-p),\ 
2\big(\frac{1}{2}-\zeta\big)p+2\zeta p\Big)+2\log_2(3) \ .
\end{align}
For details on the evaluation of $\RKMalternativesum$, we refer the reader to~\cite[Appendix~A-J]{malak2024structured}. These details consist of standard algebraic manipulations and are omitted here.   

In Figure~\ref{fig:distributedgeneralmatrixproduct_square}, using (\ref{eq:example_q3_joint_PMF}), with $q=3$ and $\zeta=0.2$, we display the sum rate $\RKMalternativesum$ in (\ref{KM_sum_rate_for_general_matrix_product}) and $\RSWsum$ versus $p$ (in $\log$ scale). The gain $\eta$ grows exponentially as $p$ approaches $0$ or $1$. 
\begin{figure}[t!]
\centering
\includegraphics[width=0.7\textwidth]{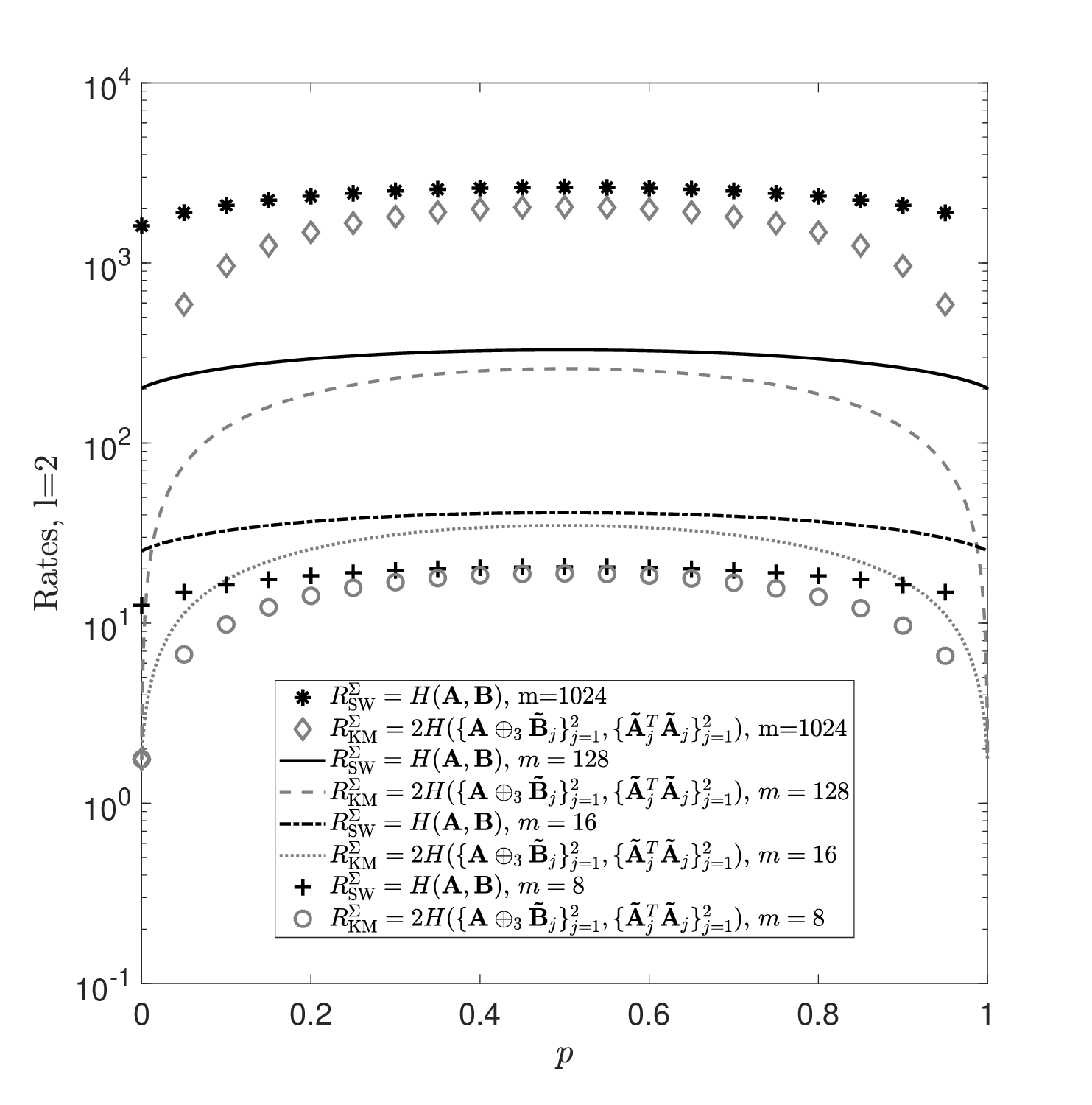}
\caption{Rate (in log scale) versus $p$ for the matrix multiplication source network, to compute ${\bf \mathbfcal{D}}={\bf A}^{\intercal}{\bf B}\in\mathbb{F}_3^{2\times 2}$, given ${\bf A}, {\bf B}\in \mathbb{F}_3^{m\times l}$, for different $m$ and $l=2$, where the joint source PMF is given in  Example~\ref{ex:general_matrix_q3} (cf.~Proposition~\ref{prop:KW_sum_rate_for_general_matrix_product}).}
\label{fig:distributedgeneralmatrixproduct_square}
\end{figure}

\end{ex}

Inspired by the Ahlswede-Han scheme~\cite{ahlswede1983source}, which blends unstructured coding~\cite{SlepWolf1973} with structured coding~\cite{korner1979encode}, we design a new encoding scheme for the distributed computation of ${\bf \mathbfcal{D}}={\bf A}^{\intercal}{\bf B}\in\mathbb{F}_q^{l\times l}$, for $q\geq 2 $ and $m, \ l>1$, generalizing  Proposition~\ref{prop:KW_sum_rate_for_general_matrix_product}, originally stated for odd $q>2$.

\begin{theo}
\label{theo:achievability_square_matrix_products}
{\bf (Distributed computation of square matrix products.)} 
For the general matrix multiplication source network, for any $q\geq 2$, and for any $\epsilon\in(0,1)$, the following sum rate is achievable:
\begin{align}
\label{eq:RAHsum_square_matrices}
\RAHsum({\bf A},\ {\bf B})&=H_q({\bf A}_1, {\bf B}_2)+2\max\{H_q({\bf A}_{2}\oplus_q{\bf B}_{1}\,\vert\, {\bf A}_1, {\bf B}_2) , \ \nonumber\\
&\hspace{4.34cm} H_q({\bf A}_{1}^{\intercal}{\bf A}_{2}\oplus_q{\bf B}_{1}^{\intercal}{\bf B}_{2}\,\vert\, {\bf A}_1, {\bf B}_2, {\bf A}_{2}\oplus_q{\bf B}_{1}) \} \ .
\end{align}
\end{theo}

\begin{proof}
We next exploit the achievable rate region of Ahlswede-Han in~\cite{ahlswede1983source}. Let ${\bf S}_1$, ${\bf S}_2$ be finite-valued variables such that ${\bf S}_1-{\bf A}-{\bf B}-{\bf S}_2$ forms a Markov chain. For $q=2$, from~\cite{ahlswede1983source},  
\begin{align}
\label{AH_rate_region}
R_1&\geq I( {\bf S}_1; {\bf A} \ \vert\ {\bf S}_2) + H( {\bf A}\oplus_2{\bf B} \ \vert\ {\bf S}_1,\ {\bf S}_2)\ ,\nonumber\\
R_2&\geq I( {\bf S}_2; {\bf B} \ \vert\ {\bf S}_2) + H( {\bf A}\oplus_2{\bf B} \ \vert\ {\bf S}_1,\ {\bf S}_2)\ , \nonumber\\
R_1+R_2 &\geq \RAHsum({\bf A},\ {\bf B})=I({\bf S}_1,\ {\bf S}_2; {\bf A},\ {\bf B})+2H({\bf A}\oplus_2{\bf B}\,  \vert \, {\bf S}_1,\ {\bf S}_2) \ ,
\end{align}
which reduces to the rate region of~\cite{SlepWolf1973} for ${\bf S}_1={\bf A}$, ${\bf S}_2={\bf B}$, and to~\cite{korner1979encode} for ${\bf S}_1={\bf S}_2={\bm 0}_{m\times l}$.

In the general case $q>2$, the square matrix product ${\bf A}^{\intercal}{\bf B}\in\mathbb{F}_q^{l\times l}$ can be expressed as
\begin{align}
{\bf A}^{\intercal}{\bf B}
\label{eq:sq_matrix_product_representation_2}
={\bf A}_{1}^{\intercal}({\bf A}_{2}\oplus_q{\bf B}_{1})\oplus_q({\bf A}_{2}\oplus_q{\bf B}_{1})^{\intercal}{\bf B}_{2}\ominus_q({\bf A}_{1}^{\intercal}{\bf A}_{2}\oplus_q{\bf B}_{1}^{\intercal}{\bf B}_{2}) \ ,
\end{align}
where this decomposition --- given in (\ref{eq:sq_matrix_product_representation_2}) --- does not rely on the symmetry of the matrix product. Exploiting (\ref{eq:sq_matrix_product_representation_2}) together with the sum-rate expression $\RAHsum$ in (\ref{AH_rate_region}), we consider the setting ${\bf S}_1={\bf A}_1$, ${\bf S}_2={\bf B}_2$. In this case, the receiver can recover the quantities ${\bf A}_1, {\bf B}_2$, ${\bf A}_2\oplus_q {\bf B}_1$, and ${\bf A}_{1}^{\intercal}{\bf A}_{2}\oplus_q{\bf B}_{1}^{\intercal}{\bf B}_{2}$ as specified in (\ref{eq:sq_matrix_product_representation_2}), thus enabling the recovery of ${\bf A}^{\intercal}{\bf B}$ at the rate given by (\ref{eq:RAHsum_square_matrices}). The use of $\max\{\cdot, \cdot\}$ in (\ref{eq:RAHsum_square_matrices}) is justified by the same arguments as in (\ref{eq:general_achievability_symmetric}).
\end{proof}

\subsection{Distributed Computation of Matrix Products using Hybrid Coding}
\label{sec:achievability_results_dot_products_hybrid}

We next introduce a hybrid encoding scheme based on {\emph{K\"orner's characteristic graphs}}~\cite{Kor73}, tailored for scenarios with {\emph{linearly separable side information}}. The framework in~\cite{Kor73} has been used to characterize the communication rate by identifying which source values can be grouped under the same codeword without ambiguity at the receiver, enabling asymptotically lossless function computation, as shown in prior work~\cite{Kor73, AO96, OR01, feizi2014network,malak2022fractional,malak2023Structured, salehi2023achievable, malak2024multi, malak2024multiAllerton, deylam2025graph}. To motivate our approach, we begin by defining {\emph{characteristic graphs}} in a point-to-point setting with receiver side information.

\begin{defi}[Characteristic graph~\cite{OR01}]
\label{def-char-graph}
Consider a point-to-point model where the source observes $X_1\in \mathbb{F}_q$ and the receiver has side information $X_2\in \mathbb{F}_q$, aiming to compute $f(X_1, X_2)$. The characteristic graph $G_{X_1}=G(\mathcal{V}, \mathcal{E})$, constructed by the source using $X_1$ with respect to $X_2$, $P_{X_1,X_2}$, and $f$, is defined as follows:
 \begin{enumerate}
 \item $\mathcal{V}$ denotes the vertex set such that $\mathcal{V}=\{x_1^{(k)}\in \mathbb{F}_q\}$, and  
 \item $\mathcal{E}$ denotes the edge set such that $\mathcal{E}=\{(x_1^{(1)},x_1^{(2)})\ : x_1^{(1)},x_1^{(2)} \in \mathcal{V}, \ x_1^{(1)}\neq x_1^{(2)} , \ \exists x_2\in \mathbb{F}_q\,\, : \,\,P_{X_1,X_2}(x_1^{(1)},x_2^{(1)})\,\cdot\,P_{X_1,X_2}(x_1^{(2)}, x_2^{(2)})>0 \,\,\mbox{and}\,\, f(x_1^{(1)},x_2^{(1)})\neq f(x_1^{(2)},x_2^{(2)})\}$~\cite{feizi2014network}.
 \end{enumerate}  

\end{defi}

\begin{defi}[Independent set~\cite{beigel1999finding}]
\label{def-In-MIS independent set}
An independent set (IS) of a graph $G(\mathcal{V}, \mathcal{E})$ is a subset of $\mathcal{V}$ with no adjacent pairs. A maximal independent set (MIS) is an IS that cannot be extended by including any additional vertex from $\mathcal{V}$~\cite{beigel1999finding}, and $\Gamma(G)$ denotes the set of all MISs of $G$.  
\end{defi}

As established by Orlitsky and Roche~\cite{OR01}, the minimum compression rate for computing $f(X_1, X_2)$ given side information $X_2$ with vanishing error is given by 
\begin{align}
\label{eq:Rate_ach_conditional_graph_entropy}
H_{G_{X_1}}(X_1\,\vert\,X_2)\triangleq    \min \{I(X_1;\mathcal{W}\mid X_2) \mid \mathcal{W}-X_1-X_2 \ , \ X_1\in \mathcal{W}\in \Gamma(G_{X_1})\}\ ,
\end{align}
which is the {\emph{conditional characteristic graph entropy}}, where $\mathcal{W}-X_1-X_2$ indicates a Markov chain,  $I(X_1; \mathcal{W} \mid X_2)=H_{q}(\mathcal{W}\mid X_2)-H_{q}(\mathcal{W}\mid X_1)$, and  $X_1\in \mathcal{W}\in \Gamma(G_{X_1})$ means that the minimization is over all $P_{\mathcal{W}, X_1}(\omega,x_1)>0$ such that $\mathcal{W}$ is an IS of $G_{X_1}$.


\begin{prop}
\label{prop:KM-OR_sum_rate_for_matrix_vector_product}
{\bf (Hybrid encoding for distributed computation of matrix products.)} For the matrix multiplication source network, for any $q\geq 2$, and for any $\epsilon\in(0,1)$, the following sum rate is achievable:
\begin{align}
\label{eq:sum_rate_KM_OR}
\RKMORsum({\bf Y})=2H_q({\bf Y})+H_{G_{\bf A}}({\bf A}\,\vert\, {\bf Y})  \ ,
\end{align}
where ${\bf Y}=\theta_1({\bf A})\oplus_q\theta_2({\bf B})$ for some functions $\theta_1$ and $\theta_2$, reflecting a linearly separable structure.
\end{prop}

\begin{proof}
Let ${\bf Y}$ be the side information at the receiver, which is expressed as ${\bf Y}=\theta_1({\bf A})\oplus_q\theta_2({\bf B})$. If ${\bf Y}={\bf A}\oplus_q {\bf B}$, then ${\bf A}^{\intercal}{\bf B}={\bf A}^{\intercal}({\bf Y}\ominus_q{\bf A})$. Exploiting (\ref{eq:Rate_ach_conditional_graph_entropy}), the minimum compression rate for computing $g({\bf A},{\bf Y})$ given side information ${\bf Y}$ is equal to $H_{G_{\bf A}}({\bf A}\,\vert\, {\bf Y})$. To achieve (\ref{eq:sum_rate_KM_OR}), we employ a hybrid coding scheme: first, the structured coding scheme in~\cite{korner1979encode} is used to compute ${\bf Y}$; this is followed by the unstructured coding approach of~\cite{OR01} to compute $g({\bf A},{\bf Y})$. 
\end{proof}

\begin{figure*}[t!]
\centering
\includegraphics[width=0.45\textwidth]{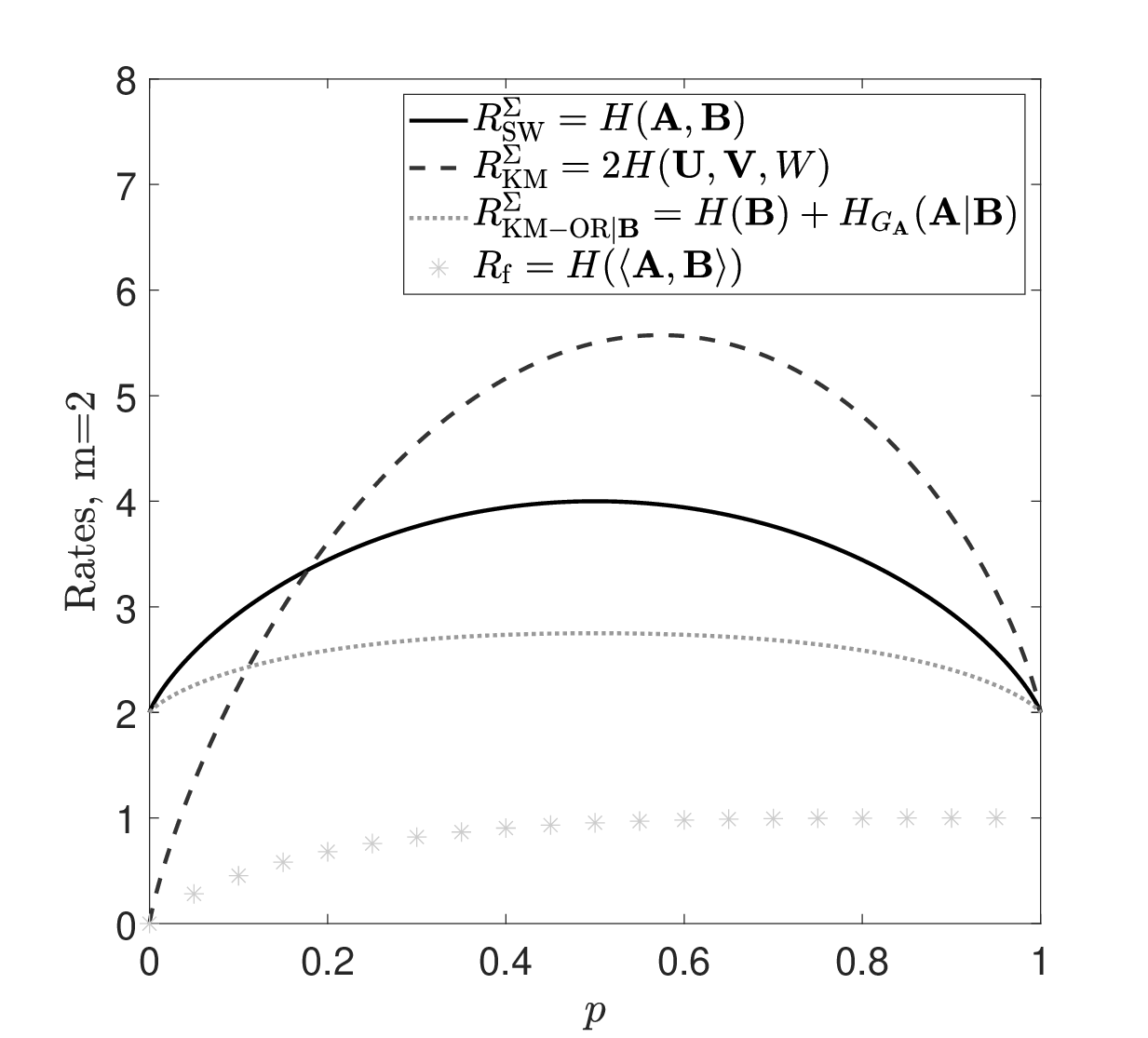}
\includegraphics[width=0.45\textwidth]{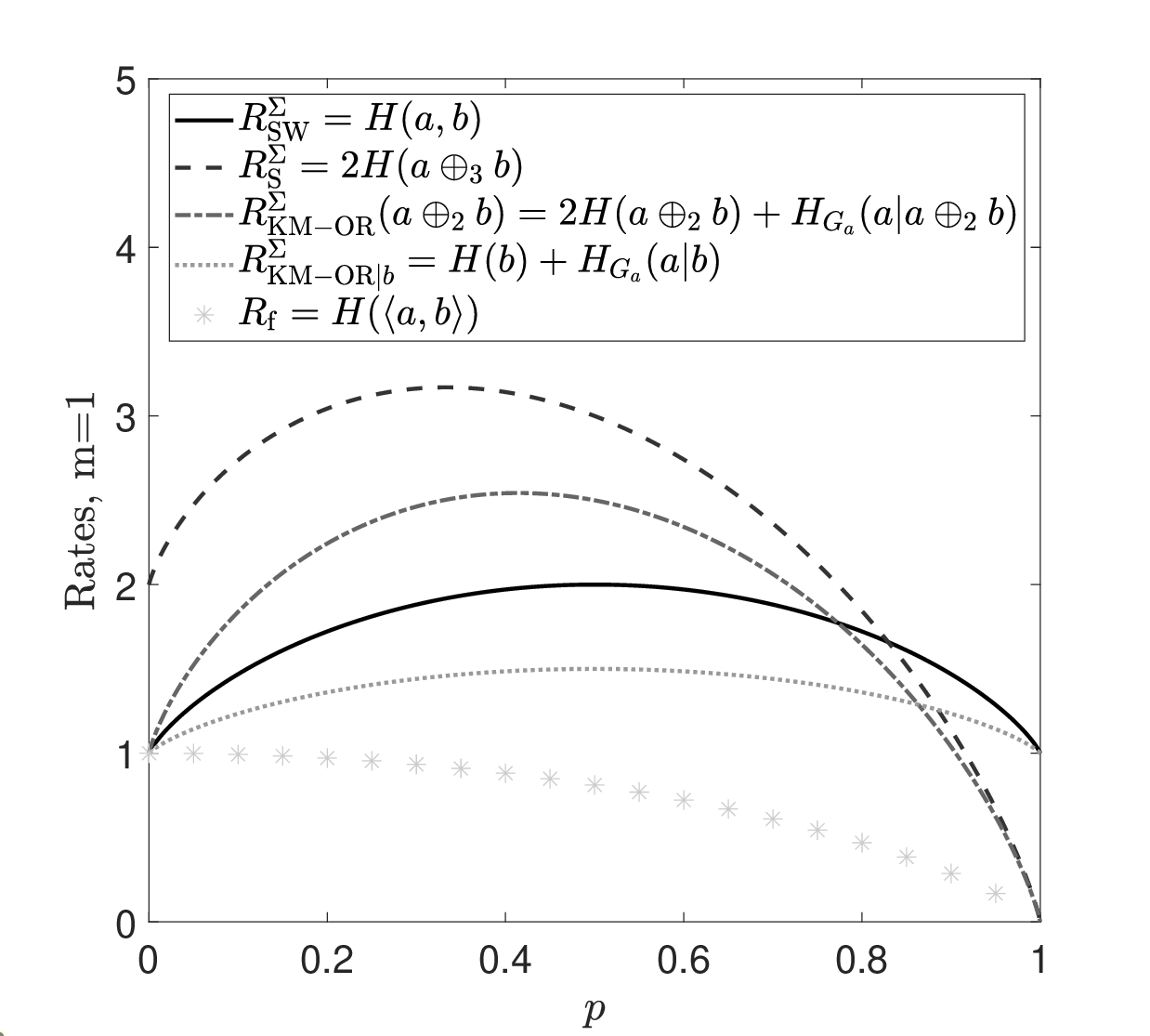}
\includegraphics[width=0.45\textwidth]{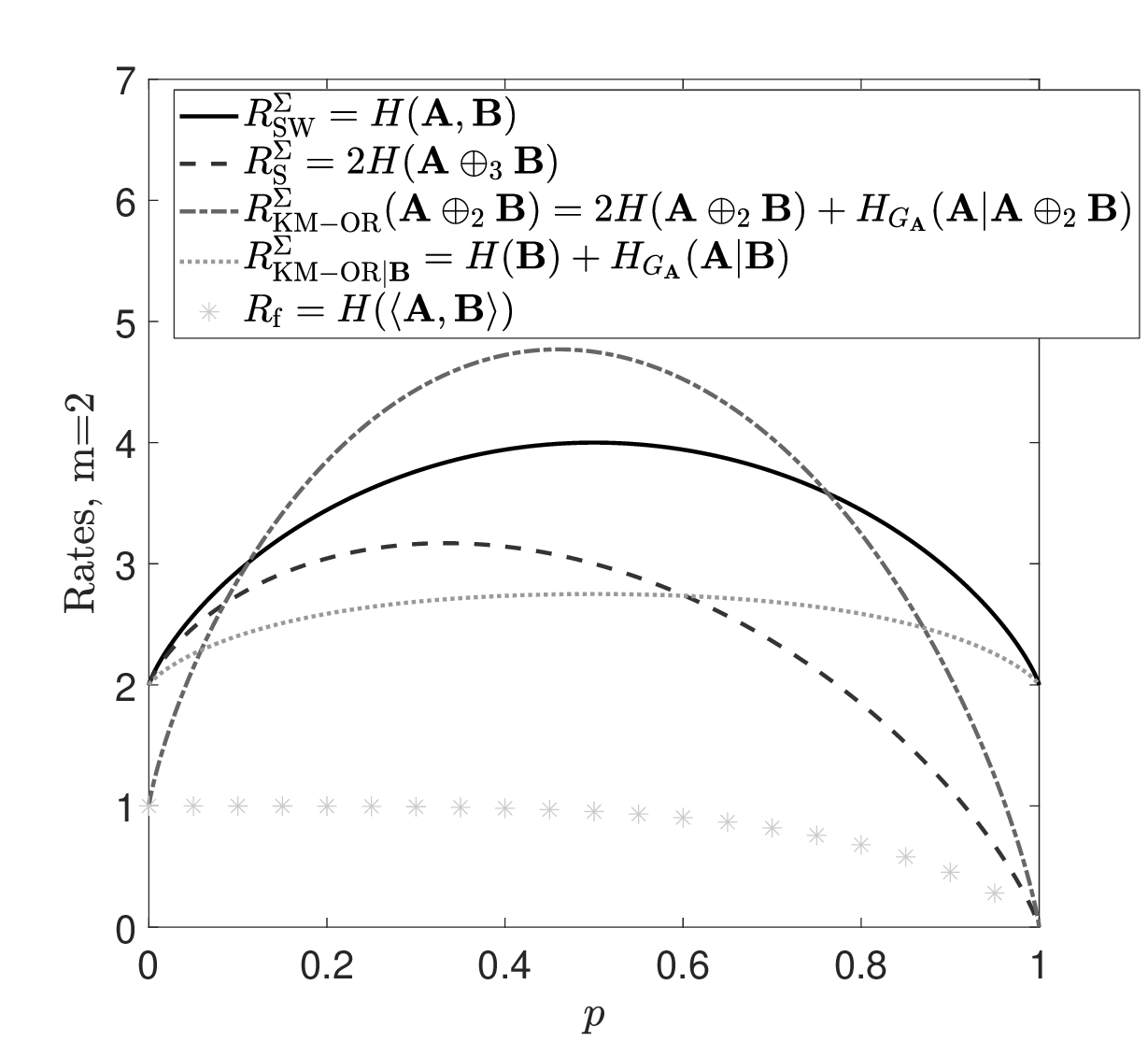} 
\caption{Rate comparisons for various $P_{{\bf A},{\bf B}}$. 
(Top-Left) The asymmetric DSBS model in Example~\ref{ex:innerproduct_length_m_binary} with $m=2$. (Top-Right) $m=1$, where $(a, b)\sim {\rm DSBS}(p)$. (Bottom) The elementwise DSBS model in Example~\ref{ex:innerproduct_length_m_binary_elementwise} with $m=2$.}
\label{fig:innerproducts_general_comparison}
\end{figure*}

For the matrix multiplication source network, in the symmetric case where $l=1$ and ${\bf A}\in\mathbb{F}_2^{m}$ and ${\bf B}\in\mathbb{F}_2^{m}$, in Figure~\ref{fig:innerproducts_general_comparison}, we contrast the sum-rate performances of various special cases captured by Corollary~\ref{cor:KW_sum_rate_for_inner_product} with a corresponding sum rate $\RKMsymsum$, and $\RSsum=2H({\bf A}\oplus_3 {\bf B})$, which models the sum rate required to compute ${\bf A}^{\intercal}{\bf B}$ by embedding the source variables in $\mathbb{F}_3$. This comparison includes the scheme of Slepian-Wolf in~\cite{SlepWolf1973} with sum rate $\RSWsum=H({\bf A}, {\bf B})$, and the characteristic graph-based approach with a sum rate $\RKMORsum=2H_q({\bf Y})+H_{G_{\bf A}}({\bf A}\,\vert\, {\bf Y})$ where ${\bf Y}={\bf A}\oplus_2{\bf B}$, and with $\RKMORsumB=H({\bf B})+H_{G_{\bf A}}({\bf A}\,\vert\,{\bf B})\leq \RSWsum$, which follows from letting ${\bf Y}={\bf B}$ in Proposition~\ref{prop:KM-OR_sum_rate_for_matrix_vector_product}. We also provide the lower bound given by $\Rf=H({\bf A}^{\intercal}{\bf B})$. In Figure~\ref{fig:innerproducts_general_comparison}-(Top-Left), we depict Example~\ref{ex:innerproduct_length_m_binary} with $m=2$, where $(a_1,b_2)\sim {\rm DSBS}(p)$, and $(a_2,b_1)\sim {\rm DSBS}(p)$. We do not indicate $\RSsum$ and $\RKMORsum$, which perform poorly. $\RKMsymsum$ performs well at low $p$, and converges to $\Rf$ as $p\to 0$, and to $\RSWsum$ as $p\to 1$. In Figure~\ref{fig:innerproducts_general_comparison}-(Top-Right), we use $m=1$, where $(a, b)\sim {\rm DSBS}(p)$. We indicate $\RSsum
$ and $\RKMORsum$. At low $p$ values, $\RKMORsum$ and $\RSWsum$ converge to $\Rf$, whereas $\RSsum$ performs poorly. For large $p$, structured coding yields low $\RSsum$ and $\RKMORsum$. In Figure~\ref{fig:innerproducts_general_comparison}-(Bottom), we depict Example~\ref{ex:innerproduct_length_m_binary_elementwise} with $m=2$, where $(a_i,b_i)\sim {\rm DSBS}(p)$, for each $i\in\{1,2\}$. We also indicate $\RKMORsum$. The rate $\RKMsymsum$ exceeds $\RSWsum$ and is omitted. For any $p$, $\RSsum<\RSWsum$, and $\RKMORsum$ approaches $\Rf$ for small and large $p$.

\subsection{Distributed Computation of General Matrix Products via Recursive Dot Products}
\label{sec:achievability_results_general_matrices_q_2}

We next recursively apply the {\emph{distributed dot-product computation technique}} from Corollary~\ref{cor:KW_sum_rate_for_inner_product} to compute ${\bf \mathbfcal{D}}={\bf A}^{\intercal}{\bf B}$, given the general matrix multiplication source network for $q\geq 2$. We then derive the corresponding sum rate, extending beyond the results of Propositions~\ref{prop:KW_sum_rate_for_symmetric_matrix_product} and~\ref{prop:KW_sum_rate_for_general_matrix_product}. 

\begin{prop}
\label{prop:recursive_inner_product_KM}
{\bf{(Distributed computation of square matrix products via recursive application of dot products.)}} 
For the general matrix multiplication source network, for any $q\geq 2$, with even $l$, and for any $\epsilon\in(0,1)$, the sum rate
\begin{align}
\label{KW_sum_rate_inner_product_based_matrix_computation}
\RKMsumrecursiveinnerproduct({\bf A}, {\bf B}) = 2H_q(\{{\bf U}_{ij},\, {\bf V}_{ij}, \, W_{ij}\}_{\,i\leq j\in[l]}) 
\end{align}
is achievable, where ${\bf A}$ and ${\bf B}$ are defined as in (\ref{eq:A_B}) and
\begin{align}
\label{UVW_set}
{\bf U}_{ij}={\bf A}_{2i}\oplus_q{\bf B}_{1j}&\in \mathbb{F}_q^{m/2} \ ,\quad 
{\bf V}_{ij}={\bf A}_{1i}\oplus_q{\bf B}_{2j}\in \mathbb{F}_q^{m/2} \ , \nonumber\\
W_{ij}&={\bf A}_{2i}^{\intercal} {\bf A}_{1i}\oplus_q{\bf B}_{1j}^{\intercal} {\bf B}_{2j}\in \mathbb{F}_q \ .
\end{align}
\end{prop}

\begin{proof}
Given ${\bf A},\ {\bf B}\in \mathbb{F}_q^{m\times l}$, and ${\bf \mathbfcal{D}}=(d_{ij})_{i,\ j\in[l]}={\bf A}^{\intercal}{\bf B}\in\mathbb{F}_q^{l\times l}$, it holds that $d_{ij}={\bf A}_i^{\intercal}{\bf B}_j$, for $i,\ j \in [l]$, where ${\bf A}_i=\begin{bmatrix}{\bf A}_{1i} ; {\bf A}_{2i} \end{bmatrix}\in\mathbb{F}_q^{m}$ and ${\bf B}_j=\begin{bmatrix} {\bf B}_{1j} ; {\bf B}_{2j}\end{bmatrix}\in\mathbb{F}_q^{m}$ represent the $i$-th and the $j$-th columns of ${\bf A}$ and ${\bf B}$, respectively, where ${\bf A}_{1i}=\left(a_{ji}\right)_{j\in [m/2]}\in\mathbb{F}_q^{\frac{m}{2}}$ and ${\bf A}_{2i}=\left(a_{ji}\right)_{j\in [m/2+1,m]}\in\mathbb{F}_q^{\frac{m}{2}}$, similarly for ${\bf B}_{1j}$ and ${\bf B}_{2j}$. Exploiting Corollary~\ref{cor:KW_sum_rate_for_inner_product} and (\ref{UVW_set}), we observe  
\begin{align}
\label{d_ij_general}
d_{ij}&={\bf U}_{ij}^{\intercal}\cdot {\bf V}_{ij}\ominus_q W_{ij}\in\mathbb{F}_q \ , 
\end{align}
where each component $d_{i'j'}$, for $i'\neq i$ and $j'\neq j$, can be recursively derived using the tuples $\{{\bf U}_{ij},\, {\bf V}_{ij}, \, W_{ij}\}$, $\{{\bf U}_{i'j},\, {\bf V}_{i'j}, \, W_{i'j}\}$, and $\{{\bf U}_{ij'},\, {\bf V}_{ij'}, \, W_{ij'}\}$, which follows from capturing 
\begin{align}
\label{UVW_connections}
{\bf U}_{i'j'}=({\bf U}_{i'j}\oplus_q{\bf U}_{ij'})\ominus_q{\bf U}_{ij}&\in \mathbb{F}_q^{m/2} \ , \quad
{\bf V}_{i'j'}=({\bf V}_{i'j}\oplus_q{\bf V}_{ij'})\ominus_q {\bf V}_{ij}\in \mathbb{F}_q^{m/2} \ ,\nonumber\\
W_{i'j'}&=(W_{i'j}\oplus_q W_{ij'})\ominus_q W_{ij} \in \mathbb{F}_q\ .
\end{align}
Hence, recursively applying Corollary~\ref{cor:KW_sum_rate_for_inner_product} that exploits the structured coding mechanism of K\"orner-Marton~\cite{korner1979encode}, the achievable sum rate for the receiver to recover ${\bf A}^{\intercal}{\bf B}$ can be determined as $\RKMsumrecursiveinnerproduct = 2H_q(\{{\bf U}_{ij},\, {\bf V}_{ij}, \, W_{ij}\}_{\, i\leq j\in[l]})$, which gives the achievability result we seek.
\end{proof}

From Proposition~\ref{prop:recursive_inner_product_KM}, the complete computation of ${\bf \mathbfcal{D}}$ relies on the set $\{{\bf U}_{ij},\, {\bf V}_{ij}, \, W_{ij}\}_{\, i\leq j\in[l]}$. For this setting, we next upper bound the achievable sum rate and the computational complexity of deriving ${\bf \mathbfcal{D}}$,  measured in terms of the number of multiplications.

\begin{cor}
\label{cor:recursive_inner_product_KM}
The achievable sum rate by the encoding scheme outlined in Proposition~\ref{prop:recursive_inner_product_KM} for the asymptotically lossless computation of ${\bf \mathbfcal{D}}={\bf A}^{\intercal}{\bf B}\in\mathbb{F}_q^{l\times l}$ is upper bounded by
\begin{align}
\label{rates_KW_sum_rate_inner_product_based_matrix_computation}
\RKMsumrecursiveinnerproduct({\bf A}, {\bf B}) \leq (m+1)l(l+1) \ . 
\end{align}

The number of multiplications that the receiver needs to perform to derive ${\bf \mathbfcal{D}}$ is
\begin{align}
\label{complexity_KW_sum_rate_inner_product_based_matrix_computation}
\frac{1}{4}ml(l+1) \ .
\end{align}
\end{cor}

\begin{proof}
{\bf Rate.} We employ $\RKMsumrecursiveinnerproduct$ from (\ref{KW_sum_rate_inner_product_based_matrix_computation}), where each tuple $\{{\bf U}_{ij},\, {\bf V}_{ij}, \, W_{ij}\}$, for $i\leq j\in[l]$, in (\ref{UVW_set}) has a total dimension of $2\cdot m/2+1=m+1$. Since there are $l(l+1)/2$ such tuples, the total number of bits per source is upper bounded by $(m+1)l(1+l)$. Thus,  (\ref{rates_KW_sum_rate_inner_product_based_matrix_computation}) follows.

{\bf Receiver complexity.} 
(\ref{complexity_KW_sum_rate_inner_product_based_matrix_computation}) follows by applying the definitions of $\{{\bf U}_{ij} \ ,  {\bf V}_{ij}\ , W_{ij}\}$ from (\ref{UVW_set}) and the relation $d_{ij}={\bf U}_{ij}^{\intercal}\cdot {\bf V}_{ij}\ominus_q W_{ij}$ in (\ref{d_ij_general}) from Proposition~\ref{prop:recursive_inner_product_KM} for each of the $l(l+1)/2$ tuples, along with the dot product computation cost $m/2$ for ${\bf U}_{ij}\in\mathbb{F}_q^{m/2}$ and ${\bf V}_{ij}\in\mathbb{F}_q^{m/2}$. 
\end{proof}

We next propose a further {\emph{recursive application of the dot-product computation technique}} in Corollary~\ref{cor:KW_sum_rate_for_inner_product} for distributed computation of a symmetric matrix ${\bf \mathbfcal{D}}={\bf A}^{\intercal}{\bf B}$, with $q>2$. The diagonal entries $\{d_{ii}\}_{i\in[l]}$ are calculated first similarly as in Proposition~\ref{prop:recursive_inner_product_KM}. The additional rate required for computing the off-diagonal entries $\{d_{ij}\}_{i< j\in[l]}$ is decided using the symmetry in ${\bf \mathbfcal{D}}$.

\begin{prop}
\label{prop:achievability_results_symmetric_matrices_q_2}
{\bf{(Distributed computation of symmetric matrix products via recursive application of dot products.)}} 
For the matrix multiplication source network, in the symmetric case, for $q>2$ and odd, and for any $\epsilon\in(0,1)$, the following sum rate is achievable:
\begin{align}
\label{KW_sum_rate_inner_product_based_symmetric_matrix_computation}
\RKMsumrecursiveinnerproductsymmetric({\bf A},\ {\bf B}) = 2H_q(\{{\bf U}_{ii},\, {\bf V}_{ii}, \, W_{ii}\}_{i\in[l]}\ , \{{\bf U}_{ij},\, {\bf V}_{ij}\}_{\,  i< j\in[l]}) \ .
\end{align}
\end{prop}

\begin{proof}
Take two indices $i,\ j\in[l]$ such that $i<j$. When ${\bf \mathbfcal{D}}={\bf A}^{\intercal}{\bf B}$ is symmetric, it holds that
\begin{align}
\label{d_ij_symmetric}
d_{ij}={\bf U}_{ij}^{\intercal}\cdot {\bf V}_{ij}\ominus_q W_{ij}
={\bf U}_{ji}^{\intercal}\cdot {\bf V}_{ji}\ominus_q W_{ji}=d_{ji}\in\mathbb{F}_q \ .
\end{align} 
Exploiting the symmetry in ${\bf \mathbfcal{D}}$, and using (\ref{UVW_connections}), it holds that
\begin{align}
\label{d_ji_for_d_ij_symmetric}
d_{ji}=(({\bf U}_{ii}\oplus_q {\bf U}_{jj})\ominus_q {\bf U}_{ij})^{\intercal}\cdot (({\bf V}_{ii}\oplus_q {\bf V}_{jj})\ominus_q {\bf V}_{ij})\ominus_q ((W_{ii}\oplus_q W_{jj})\ominus_q W_{ij})\in\mathbb{F}_q \ .
\end{align}

The receiver, using the relations (\ref{d_ij_symmetric}) and (\ref{d_ji_for_d_ij_symmetric}), can compute directly the term  
\begin{align}
\label{2dij_expression}
d_{ij}=\frac{1}{2}(d_{ij}\oplus_q d_{ji}) \ ,
\end{align}
provided that $q>2$ and odd. Given $\{{\bf U}_{ii},\, {\bf V}_{ii}, \, W_{ii}\}$, the receiver can decode $d_{ij}$ if in addition $\{{\bf U}_{ij},\, {\bf V}_{ij}\}$ are also given. Exploiting the symmetry in ${\bf \mathbfcal{D}}$ and  (\ref{2dij_expression}), it suffices to determine the diagonal and the upper triangular entries, namely $\{d_{ij}\}_{ i\leq j\in[l]}$, to achieve the sum rate in  (\ref{KW_sum_rate_inner_product_based_symmetric_matrix_computation}).
\end{proof}

From Proposition~\ref{prop:achievability_results_symmetric_matrices_q_2}, the computation of ${\bf \mathbfcal{D}}$ relies on the diagonal tuples $\{{\bf U}_{ii},\, {\bf V}_{ii}, \, W_{ii}\}_{i\in[l]}$ to determine $\{d_{ii}\}_{i\in[l]}$, and on the off-diagonal tuples $\{{\bf U}_{ij},\, {\bf V}_{ij}\}_{ i<j\in[l]}$, noting that ${\bf \mathbfcal{D}}$ is symmetric. 

\begin{cor}
\label{cor:achievability_results_symmetric_matrices_q_2}
The achievable sum rate by the encoding scheme outlined in  Proposition~\ref{prop:achievability_results_symmetric_matrices_q_2} for the asymptotically lossless computation of ${\bf \mathbfcal{D}}={\bf A}^{\intercal}{\bf B}\in\mathbb{F}_q^{l\times l}$ is upper bounded by 
\begin{align}
\label{rates_KW_sum_rate_inner_product_based_symmetric_matrix_computation}
\RKMsumrecursiveinnerproductsymmetric({\bf A},\ {\bf B}) \leq ml(l+1)+2l \ .
\end{align}

The number of multiplications that the receiver needs to perform to derive ${\bf \mathbfcal{D}}$ is 
\begin{align}
\label{complexity_KW_sum_rate_inner_product_based_symmetric_matrix_computation}
\frac{1}{2}ml^2 \ . 
\end{align}
\end{cor}

\begin{proof}
{\bf Rate.} We employ  $\RKMsumrecursiveinnerproductsymmetric$ from (\ref{KW_sum_rate_inner_product_based_symmetric_matrix_computation}), where each tuple $\{{\bf U}_{ii},\, {\bf V}_{ii}, \, W_{ii}\}$, for $i\in[l]$, has a total dimension of $m+1$, and there are $l$  such tuples, and each tuple $\{{\bf U}_{ij},\, {\bf V}_{ij}\}$, for $ i<j\in[l]$, has a total dimension of $m$, and there are $(l^2-l)/2$  such tuples, the total number of bits per source is upper bounded by $ml(l+1)/2+l$. This leads directly to (\ref{rates_KW_sum_rate_inner_product_based_symmetric_matrix_computation}).

{\bf Receiver complexity.} Equation (\ref{complexity_KW_sum_rate_inner_product_based_symmetric_matrix_computation}) follows by noting that each diagonal term $d_{ii}$, $i\in[l]$, requires a dot product of size $m/2$. Due to the symmetry of ${\bf \mathbfcal{D}}$, the remaining $(l^2-l)/2$ off-diagonal terms can be computed using the relation $d_{ij}={\bf U}_{ij}^{\intercal}\cdot {\bf V}_{ij}\ominus_q W_{ij}=d_{ji}$ from (\ref{d_ij_symmetric}), as well as the alternative expression in (\ref{2dij_expression}), each involving two dot products of size $m/2$.
\end{proof}

We next tighten the result in Proposition~\ref{prop:achievability_results_symmetric_matrices_q_2} via a {\emph{nested application of the dot-product computation technique}} in Corollary~\ref{cor:KW_sum_rate_for_inner_product}. In this approach, $\{d_{ii}\}_{i\in[l]}$ are calculated first using the same technique as in Proposition~\ref{prop:recursive_inner_product_KM}, and then the additional rate required for computing $\{d_{ij}\}_{i< j\in[l]}$ is determined as a function of $\{{\bf U}_{ij},\ {\bf V}_{ij},\ W_{ij}\}_{i,\ j\in[l]}$, as described next.

\begin{prop}
\label{prop:achievability_results_symmetric_matrices_nested_KM_q_2}
{\bf{(Distributed computation of symmetric matrix products via nested application of dot products.)}} 
For the matrix multiplication source network, in the symmetric case, for $q>2$ and odd, and for any $\epsilon\in(0,1)$, the following sum rate is achievable:
\begin{align}
\label{nested_KW_sum_rate_inner_product_based_symmetric_matrix_computation}
\RKMsumnestedinnerproductsymmetric({\bf A}, {\bf B}) = 2H_q\Big(&\{{\bf U}_{ii},\, {\bf V}_{ii}, \, W_{ii}\}_{i\in[l]}\ , \nonumber\\
&\Big\{{\bf U}_{ij}(\frac{m}{4}+1:\frac{m}{2})\oplus_q {\bf V}_{ij}(1:\frac{m}{4}) \ ,{\bf U}_{ij}(1:\frac{m}{4})\oplus_q {\bf V}_{ij}(\frac{m}{4}+1:\frac{m}{2}) \ ,\nonumber\\
&{\bf U}_{ij}(\frac{m}{4}+1:\frac{m}{2})^{\intercal}\cdot {\bf U}_{ij}(1:\frac{m}{4})\oplus_q
{\bf V}_{ij}(1:\frac{m}{4})^{\intercal}\cdot {\bf V}_{ij}(\frac{m}{4}+1:\frac{m}{2})\nonumber\\
&\ominus_q ({\bm \alpha}^{\intercal}({\bf U}_{ii}, {\bf U}_{jj}){\bf U}_{ij}+{\bm \beta}^{\intercal}({\bf V}_{ii}, {\bf V}_{jj}){\bf V}_{ij})\Big\}_{\, i<j\in[l]}\Big) \ ,
\end{align}
where $\{{\bf U}_{ij},\ {\bf V}_{ij},\ W_{ij}\}_{i,\ j\in[l]}$ are defined similarly to (\ref{UVW_set}) in Proposition~\ref{prop:recursive_inner_product_KM}, and ${\bm \alpha}({\bf U}_{ii}, {\bf U}_{jj})\in \mathbb{F}_q^{m/2}$ and ${\bm \beta}({\bf V}_{ii}, {\bf V}_{jj})\in \mathbb{F}_q^{m/2}$ represent coefficient matrices, which are given as follows:
\begin{align}
\label{alpha_matrix}
{\bm \alpha}({\bf U}_{ii}, {\bf U}_{jj})&=\frac{1}{2}\begin{bmatrix}{\bf U}_{ii}(\frac{m}{4}+1:\frac{m}{2})\oplus_q {\bf U}_{jj}(\frac{m}{4}+1:\frac{m}{2}) ; {\bf U}_{ii}(1:\frac{m}{4})\oplus_q {\bf U}_{jj}(1:\frac{m}{4})\end{bmatrix}
\ ,\\
\label{beta_matrix}
{\bm \beta}({\bf V}_{ii}, {\bf V}_{jj})&=\frac{1}{2}\begin{bmatrix} {\bf V}_{ii}(\frac{m}{4}+1:\frac{m}{2})\oplus_q {\bf V}_{jj}(\frac{m}{4}+1:\frac{m}{2}) ; {\bf V}_{ii}(1:\frac{m}{4})\oplus_q {\bf V}_{jj}(1:\frac{m}{4})\end{bmatrix}
\ .
\end{align} 
\end{prop}

\begin{proof}
Given diagonal tuples $\{{\bf U}_{ii},\, {\bf V}_{ii}, \, W_{ii}\}_{i\in[l]}$, and substituting the expansion from (\ref{2dij_expression}), the additional rate needed for the receiver to reconstruct $d_{ij}\in\mathbb{F}_q$, for $i< j\in[l]$, is:
\begin{align}
\label{conditional_entropy_d_ij}
H_q(d_{ij} \,\vert\, &\{{\bf U}_{ii},\, {\bf V}_{ii}, \, W_{ii}\}_{i\in[l]})\nonumber\\
&=H_q(({\bf U}_{ii}\oplus_q {\bf U}_{jj})^{\intercal}\cdot ({\bf V}_{ii}\oplus_q {\bf V}_{jj})\ominus_q {\bf U}_{ij}^{\intercal}\cdot ({\bf V}_{ii}\oplus_q {\bf V}_{jj})\ominus_q ({\bf U}_{ii}\oplus_q {\bf U}_{jj})^{\intercal}\cdot{\bf V}_{ij}\nonumber\\
&\ominus_q (W_{ii}\oplus_q W_{jj}) \,\oplus_q\, 2{\bf U}_{ij}^{\intercal}\cdot {\bf V}_{ij} \,\vert\, \{{\bf U}_{ii},\, {\bf V}_{ii}, \, W_{ii}\}_{i\in[l]})\nonumber\\
&=H_q({\bf U}_{ij}^{\intercal}\cdot {\bf V}_{ij}\ominus_q \frac{1}{2}{\bf U}_{ij}^{\intercal}\cdot ({\bf V}_{ii}\oplus_q {\bf V}_{jj})\ominus_q \frac{1}{2}({\bf U}_{ii}\oplus_q {\bf U}_{jj})^{\intercal}\cdot{\bf V}_{ij} \,\vert\, \{{\bf U}_{ii},\, {\bf V}_{ii}, \, W_{ii}\}_{i\in[l]})\nonumber\\
&=H_q\big(\bar{{\bf U}}(ij)^{\intercal} \cdot \bar{{\bf V}}(ij)\,\big\vert\, \{{\bf U}_{ii},\, {\bf V}_{ii}, \, W_{ii}\}_{i\in[l]}\big) \ ,
\end{align}
where the last step follows from the following substitutions: 
\begin{align}
\label{Ubar_and_Vbar}
\bar{{\bf U}}(ij)={\bf U}_{ij}\ominus_q \frac{1}{2}({\bf U}_{ii}\oplus_q {\bf U}_{jj})\in \mathbb{F}_q^{m/2} \ ,\quad
\bar{{\bf V}}(ij)={\bf V}_{ij}\ominus_q \frac{1}{2}({\bf V}_{ii}\oplus_q {\bf V}_{jj})\in \mathbb{F}_q^{m/2} \ . 
\end{align}
Note that $\bar{{\bf U}}(ij)^{\intercal}\bar{{\bf V}}(ij)=\langle \,\bar{{\bf U}}(ij) ,\bar{{\bf V}}(ij) \,\rangle$. Leveraging Corollary~\ref{cor:KW_sum_rate_for_inner_product} on dot-product computation, and letting $\bar{{\bf U}}(ij)^{\intercal}=\begin{bmatrix}
\bar{{\bf U}}_1(ij)^{\intercal}&
\bar{{\bf U}}_2(ij)^{\intercal}\end{bmatrix}$ and $\bar{{\bf V}}(ij)^{\intercal}=\begin{bmatrix}
\bar{{\bf V}}_1(ij)^{\intercal}&
\bar{{\bf V}}_2(ij)^{\intercal}\end{bmatrix}$ where $\bar{{\bf U}}_1(ij),\bar{{\bf U}}_2(ij),\bar{{\bf V}}_1(ij),\bar{{\bf V}}_2(ij)\in\mathbb{F}_q^{m/4}$, we see from (\ref{conditional_entropy_d_ij}) and (\ref{Ubar_and_Vbar}) that the receiver can reconstruct $d_{ij}$ using $\bar{{\bf U}}(ij)^{\intercal}\bar{{\bf V}}(ij)$, which can be derived from the following two linear terms:
\begin{align}
\label{d_ij_inner_product_based_reconstruction_no_SI_1}
\bar{{\bf U}}_2(ij)\oplus_q \bar{{\bf V}}_1(ij)&={\bf U}_{ij}(\frac{m}{4}+1:\frac{m}{2})\ominus_q \frac{1}{2}\big({\bf U}_{ii}(\frac{m}{4}+1:\frac{m}{2})\oplus_q {\bf U}_{jj}(\frac{m}{4}+1:\frac{m}{2})\big)\nonumber\\
&\oplus_q {\bf V}_{ij}(1:\frac{m}{4})\ominus_q \frac{1}{2}\big({\bf V}_{ii}(1:\frac{m}{4})\oplus_q {\bf V}_{jj}(1:\frac{m}{4})\big)\in \mathbb{F}_q^{m/4} \ , \\
\label{d_ij_inner_product_based_reconstruction_no_SI_2}
\bar{{\bf U}}_1(ij)\oplus_q \bar{{\bf V}}_2(ij) &={\bf U}_{ij}(1:\frac{m}{4})\ominus_q \frac{1}{2}\big({\bf U}_{ii}(1:\frac{m}{4})\oplus_q {\bf U}_{jj}(1:\frac{m}{4})\big)\nonumber\\
&\oplus_q {\bf V}_{ij}(\frac{m}{4}+1:\frac{m}{2})\ominus_q \frac{1}{2}\big({\bf V}_{ii}(\frac{m}{4}+1:\frac{m}{2})\oplus_q {\bf V}_{jj}(\frac{m}{4}+1:\frac{m}{2})\big)\in \mathbb{F}_q^{m/4}\ ,
\end{align}
and the following non-linear term
\begin{align}
\bar{{\bf U}}_2(ij)^{\intercal}&\bar{{\bf U}}_1(ij)\oplus_q \bar{{\bf V}}_1(ij)^{\intercal}\bar{{\bf V}}_2(ij)
={\bf U}_{ij}(\frac{m}{4}+1:\frac{m}{2})^{\intercal}\cdot {\bf U}_{ij}(1:\frac{m}{4})\nonumber\\
&\ominus_q{\bf U}_{ij}(\frac{m}{4}+1:\frac{m}{2})^{\intercal}\cdot \frac{1}{2}\big({\bf U}_{ii}(1:\frac{m}{4})\oplus_q {\bf U}_{jj}(1:\frac{m}{4})\big)\nonumber\\
&\ominus_q\frac{1}{2}\big({\bf U}_{ii}(\frac{m}{4}+1:\frac{m}{2})\oplus_q {\bf U}_{jj}(\frac{m}{4}+1:\frac{m}{2})\big)^{\intercal}\cdot{\bf U}_{ij}(1:\frac{m}{4})\nonumber\\
&\oplus_q \frac{1}{2}\big({\bf U}_{ii}(\frac{m}{4}+1:\frac{m}{2})\oplus_q {\bf U}_{jj}(\frac{m}{4}+1:\frac{m}{2})\big)^{\intercal}\cdot\frac{1}{2}\big({\bf U}_{ii}(1:\frac{m}{4})\oplus_q {\bf U}_{jj}(1:\frac{m}{4})\big)\nonumber\\
&\oplus_q
{\bf V}_{ij}(1:\frac{m}{4})^{\intercal}\cdot {\bf V}_{ij}(\frac{m}{4}+1:\frac{m}{2})\nonumber\\
&\ominus_q{\bf V}_{ij}(1:\frac{m}{4})^{\intercal}\cdot\frac{1}{2}\big({\bf V}_{ii}(\frac{m}{4}+1:\frac{m}{2})\oplus_q {\bf V}_{jj}(\frac{m}{4}+1:\frac{m}{2})\big)\nonumber\\
&\ominus_q\frac{1}{2}\big({\bf V}_{ii}(1:\frac{m}{4})\oplus_q {\bf V}_{jj}(1:\frac{m}{4})\big)^{\intercal}\cdot{\bf V}_{ij}(\frac{m}{4}+1:\frac{m}{2})\nonumber\\
\label{d_ij_inner_product_based_reconstruction_no_SI_nonlinear_part}
&\oplus_q\frac{1}{2}\big({\bf V}_{ii}(1:\frac{m}{4})\oplus_q {\bf V}_{jj}(1:\frac{m}{4})\big)^{\intercal}\cdot\frac{1}{2}\big({\bf V}_{ii}(\frac{m}{4}+1:\frac{m}{2})\oplus_q {\bf V}_{jj}(\frac{m}{4}+1:\frac{m}{2})\big)\in \mathbb{F}_q \ .
\end{align}
Exploiting the side information $\{{\bf U}_{ii},\, {\bf V}_{ii}, \, W_{ii}\}_{i=1}^l$, we can significantly simplify  (\ref{d_ij_inner_product_based_reconstruction_no_SI_1}), (\ref{d_ij_inner_product_based_reconstruction_no_SI_2}), and (\ref{d_ij_inner_product_based_reconstruction_no_SI_nonlinear_part}), and deduce that the receiver can reconstruct $d_{ij}$ from 
\begin{subequations}
\label{d_ij_inner_product_based_reconstruction_with_SI}
\begin{align}
&{\bf U}_{ij}(\frac{m}{4}+1:\frac{m}{2})\oplus_q {\bf V}_{ij}(1:\frac{m}{4})\in \mathbb{F}_q^{m/4} \ ,  \label{eq:i}\\
&{\bf U}_{ij}(1:\frac{m}{4})\oplus_q {\bf V}_{ij}(\frac{m}{4}+1:\frac{m}{2})\in \mathbb{F}_q^{m/4} \ , \label{eq:ii}\\
&{\bf U}_{ij}(\frac{m}{4}+1:\frac{m}{2})^{\intercal}\cdot {\bf U}_{ij}(1:\frac{m}{4})\oplus_q
{\bf V}_{ij}(1:\frac{m}{4})^{\intercal}\cdot {\bf V}_{ij}(\frac{m}{4}+1:\frac{m}{2})\nonumber\\
&\ominus_q({\bm \alpha}^{\intercal}({\bf U}_{ii}, {\bf U}_{jj})\cdot{\bf U}_{ij}+{\bm \beta}^{\intercal}({\bf V}_{ii}, {\bf V}_{jj})\cdot{\bf V}_{ij})\in \mathbb{F}_q\ , \label{eq:iii}
\end{align}
\end{subequations}
where ${\bm \alpha}({\bf U}_{ii}, {\bf U}_{jj})\in \mathbb{F}_q^{m/2}$ given in (\ref{alpha_matrix}) and ${\bm \beta}({\bf V}_{ii}, {\bf V}_{jj})\in \mathbb{F}_q^{m/2}$ given in (\ref{beta_matrix}) represent coefficient matrices that are determined as functions of ${\bf U}_{ii}, {\bf U}_{jj}$ and ${\bf V}_{ii}, {\bf V}_{jj}$, respectively.

From (\ref{conditional_entropy_d_ij})-(\ref{d_ij_inner_product_based_reconstruction_with_SI}), which detail the reconstruction of $\{d_{ij}\}_{i<j\in[l]}$, we get the final result.
\end{proof}

Using the sum rate $\RKMsumnestedinnerproductsymmetric$ given in (\ref{nested_KW_sum_rate_inner_product_based_symmetric_matrix_computation}) of Proposition~\ref{prop:achievability_results_symmetric_matrices_nested_KM_q_2}, we next derive the rate and complexity of distributed lossless computation of ${\bf \mathbfcal{D}}={\bf A}^{\intercal}{\bf B}\in\mathbb{F}_q^{l\times l}$.

\begin{cor}
\label{cor:achievability_results_symmetric_matrices_nested_KM_q_2}
The achievable sum rate by the encoding scheme outlined in  Proposition~\ref{prop:achievability_results_symmetric_matrices_nested_KM_q_2} for the asymptotically lossless computation of ${\bf \mathbfcal{D}}={\bf A}^{\intercal}{\bf B}\in\mathbb{F}_q^{l\times l}$ is upper bounded by
\begin{align}
\label{rates_nested_KW_sum_rate_inner_product_based_symmetric_matrix_computation}
\RKMsumnestedinnerproductsymmetric({\bf A}, {\bf B})\leq \frac{1}{2}ml(l+3)+l(l+1) \ .
\end{align}

The number of multiplications that the receiver needs to perform to derive ${\bf \mathbfcal{D}}$ is 
\begin{align}
\label{complexity_nested_KW_sum_rate_inner_product_based_symmetric_matrix_computation}
\frac{1}{8}ml(7l-3) \ .
\end{align}
\end{cor}

\begin{proof}
{\bf Rate.} We employ $\RKMsumnestedinnerproductsymmetric$ from (\ref{nested_KW_sum_rate_inner_product_based_symmetric_matrix_computation}), where each $\{{\bf U}_{ii},\, {\bf V}_{ii}, \, W_{ii}\}$, for $i\in [l]$, has dimension $m+1$, contributing a total of $(m+1)l$ bits. For each off-diagonal tuple with $i<j$, the dimension is $m/2+1$, as seen from (\ref{d_ij_inner_product_based_reconstruction_with_SI}) in  Proposition~\ref{prop:achievability_results_symmetric_matrices_nested_KM_q_2}, and there are $l(l-1)/2$ such tuples. Hence, the rate per source is at most $(m+1)l+(m/2+1)l(l-1)/2$, which yields (\ref{rates_nested_KW_sum_rate_inner_product_based_symmetric_matrix_computation}).

{\bf Receiver complexity.} Employing the definitions of $\{{\bf U}_{ij} \ ,  {\bf V}_{ij}\ , W_{ij}\}$ from  (\ref{UVW_set}) and $d_{ij}$ in (\ref{d_ij_general}) (Proposition~\ref{prop:recursive_inner_product_KM}), the total complexity of computing $\{d_{ii}\}_{i\in[l]}$ from $\{{\bf U}_{ii} \ ,  {\bf V}_{ii}\ , W_{ii}\}$ is 
\begin{align}
\label{complexity_diagonal}
\frac{1}{2}ml \ .
\end{align}
Given $\{d_{ii}\}_{i\in[l]}$, to reconstruct $\{d_{ij}\}_{i< j\in[l]}$, we use (\ref{conditional_entropy_d_ij}) along with the linear terms in (\ref{d_ij_inner_product_based_reconstruction_no_SI_1})-(\ref{d_ij_inner_product_based_reconstruction_no_SI_2}), and the non-linear term in  (\ref{d_ij_inner_product_based_reconstruction_no_SI_nonlinear_part}), as detailed in  Proposition~\ref{prop:achievability_results_symmetric_matrices_nested_KM_q_2}. The linear terms incur no multiplicative cost. The non-linear term in  (\ref{d_ij_inner_product_based_reconstruction_no_SI_nonlinear_part}) involves two dot products --- $\bar{{\bf U}}_2(ij)^{\intercal}\cdot\bar{{\bf U}}_1(ij)$ and $\bar{{\bf V}}_1(ij)^{\intercal}\cdot\bar{{\bf V}}_2(ij)$ --- each of complexity $m/4$. Additionally, reconstructing $d_{ij}$ via (\ref{d_ij_inner_product_based_reconstruction_with_SI}) requires one dot product from (\ref{eq:i})-(\ref{eq:ii}), with complexity $m/4$, and four dot products in (\ref{eq:iii}), each with complexity $m/4$, yielding a total per-pair cost of $2\cdot m/4+ m/4+ 4\cdot m/4=7m/4$. Because there are $(l^2-l)/2$  such tuples, the total complexity of reconstructing $\{d_{ij}\}_{i< j\in[l]}$ is 
\begin{align}
\label{complexity_lower triangular}
\frac{7}{8}ml(l-1) \ .
\end{align}
Combining this with the diagonal complexity from (\ref{complexity_diagonal}) yields the expression in (\ref{complexity_nested_KW_sum_rate_inner_product_based_symmetric_matrix_computation}).
\end{proof}

\begin{table*}[t!]\small
\setlength{\extrarowheight}{4pt}
\begin{center}
\begin{tabular}{| l | l | l |}
\hline
 & Sum-rate upper bound & Complexity
\\
\hline
Recursive dot prods. for general matrix prods. & $(m+1)l(l+1)$ 
(Eq.~(\ref{rates_KW_sum_rate_inner_product_based_matrix_computation})) & $ml(l+1)/4$ 
(Eq.~(\ref{complexity_KW_sum_rate_inner_product_based_matrix_computation}))
\\
\hline
Recursive dot prods. for symmetric matrix prods. & $ml(l+1)+2l$ (Eq.~(\ref{rates_KW_sum_rate_inner_product_based_symmetric_matrix_computation})) & $ml^2/2$ (Eq.~(\ref{complexity_KW_sum_rate_inner_product_based_symmetric_matrix_computation}))
\\
\hline
Nested dot prods. for symmetric matrix prods. & $ml(l+3)/2+l(l+1)$ (Eq.~(\ref{rates_nested_KW_sum_rate_inner_product_based_symmetric_matrix_computation})) & $ml(7l-3)/8$ (Eq.~(\ref{complexity_nested_KW_sum_rate_inner_product_based_symmetric_matrix_computation}))\\
\hline
\end{tabular}
\end{center}
\caption{Distributed computation of general matrix products via recursive dot products: comparison of sum rates and receiver-side computational complexity for computing ${\bf \mathbfcal{D}}$.}
\label{table:complexity_recursive_inner_product}
\end{table*}

Our structured schemes in Propositions~\ref{prop:recursive_inner_product_KM}-\ref{prop:achievability_results_symmetric_matrices_nested_KM_q_2} reduce the computational complexity of distributed matrix multiplication (see Table~\ref{table:complexity_recursive_inner_product}). 
While Strassen-like algorithms~\cite{strassen1969gaussian} and learning-based techniques~\cite{fawzi2022discovering} fall outside our scope, they can be incorporated to further reduce the complexity.

\section{Converses} 
\label{sec:converses}

In this section, for the proposed {\emph{matrix multiplication source network}}, drawing on Lemmas~1-3 of~\cite{han1987dichotomy}, we first establish necessary conditions on square matrix products, as $q\to\infty$ (Theorem~\ref{theo:strong_converse_general_matrix_product}) and products of full-rank matrices, with $q\geq 2$ (Theorem~\ref{theo:converse_general_matrix_product_Han_Kobayashi}) on the rate pair $(R_1,\ R_2)$, 
calibrating the strong converse bound of Ahlswede-G{\'a}cs-K\"orner~\cite{ahlswede1976bounds} and the Han-Kobayashi approach~\cite{han1987dichotomy}, respectively. To investigate the optimality gaps of our achievability schemes in Section~\ref{sec:achievability},  
we subsequently specialize these conditions to the products of independently and uniformly drawn source matrices from $\mathbb{F}_q^{m\times l}$, as $q \to \infty$ (Corollary~\ref{cor:achievable_rate_lem:entropy_general_matrix_product}), then to symmetric matrix products  (Proposition~\ref{prop:multiplicative_gain_for_binary_symmetric_matrix_products}), and square matrix products (Proposition~\ref{prop:multiplicative_gain_for_binary_square_matrix_products}), each with $q=2$, respectively.

We next provide Lemmas~1-3 of~\cite{han1987dichotomy}, which establish converses on $(R_1,R_2)$ for distributed computing of an arbitrary function $Z=f(X, Y)$ taking values in $\mathcal{Z}\subseteq \mathbb{F}_q$, given separately encoded correlated 
$(X,\ Y)$ with values in $\mathcal{X}\subseteq \mathbb{F}_q$ and $\mathcal{Y}\subseteq \mathbb{F}_q$, respectively.  
Let the set $\mathcal{P}$ 
be defined by $(X, Y)\in \mathcal{P}$ if and only if $P_{X,Y}(k,\ell)=\mathbb{P}(X=k,\ Y=\ell)>0$, for all $k \in \mathcal{X},\ \ell \in \mathcal{Y}$. 
For this setting, the {\emph{function matrix}} denoted by $(f(k, \ell))_{k\in \mathcal{X},\ \ell\in\mathcal{Y}}$ is an $|\mathcal{X}| \times |\mathcal{Y}|$ matrix with the $(k, \ell)$-th component being $f(k, \ell)$, where $k$ and $\ell$ indicate the rows and columns, respectively.

\begin{lem}
\label{lem:lemmas_1_2_Han_Kobayashi}
{\bf (Han-Kobayashi~\cite[Lemmas~1-3]{han1987dichotomy}.)} 
Let $(X, Y)$ be any element of $\mathcal{P}$.

\begin{enumerate}
[label=(Condition \arabic*),
                  ref=\arabic*,
                  align=left,
                  leftmargin=*,
                  labelsep=0.5em]
\item \label{cond:3.1}

{\bf \cite[Lemma~1]{han1987dichotomy}.}
If any two distinct rows of the function matrix $f$ are different, then any achievable $(R_1, R_2)$ for $f$ has to satisfy $R_1\geq H_q( X\,\vert\, Y)$.

\item \label{cond:3.11}

{\bf \cite[Lemma~2]{han1987dichotomy}.}
If any two distinct columns of the function matrix $f$ are different, then any achievable $(R_1, R_2)$ for $f$ has to satisfy $R_2\geq H_q(Y\,\vert\, X)$.

\item \label{cond:3.13}

{\bf \cite[Lemma~3]{han1987dichotomy}.}
If, in addition to Conditions~\ref{cond:3.1} and~\ref{cond:3.11}, the condition $f(k_1,\ell_1)\neq f(k_2,\ell_2)$ for any $(k_1,\ell_1)$, $(k_2,\ell_2)$ with $k_1\neq k_2$, $\ell_1\neq \ell_2$ holds, then any achievable $(R_1, R_2)$ for $f$ has to satisfy
$R_1+R_2\geq H_q(X,Y)$.
\end{enumerate}
\end{lem}

In their seminal work~\cite{han1987dichotomy}, Han and Kobayashi have identified necessary and sufficient conditions --- namely, that $f(k,\ell)$, for $k\in\mathcal{X}$, $\ell\in\mathcal{Y}$, satisfies the three properties in Lemma~\ref{lem:lemmas_1_2_Han_Kobayashi} --- for the achievable rate region for distributed computing of $Z=f(X,Y)$ to coincide with the Slepian-Wolf region~\cite{SlepWolf1973} whenever $P_{X,Y}(k,\ell)>0$ for all $k\in\mathcal{X}$, $\ell\in\mathcal{Y}$~\cite[Theorem~1]{han1987dichotomy}.

To compute the dot product $d=\langle \,{\bf A} ,{\bf B} \,\rangle\in\mathbb{F}_q$ in the proposed matrix multiplication source network, Conditions~\ref{cond:3.1}-\ref{cond:3.11} hold, while Condition~\ref{cond:3.13} does not. Thus, by~\cite[Theorem~1]{han1987dichotomy}, the sum rate $R_1+R_2$ can fall below $\RSWsum({\bf A},\ {\bf B})$. Likewise, for computing ${\bf \mathbfcal{D}}={\bf A}^{\intercal}{\bf B}\in\mathbb{F}_q^{l\times l}$, each entry of ${\bf A}^{\intercal}{\bf B}$ being a dot product ensures that the same conditions in~\cite{han1987dichotomy} apply.

We next present a {\emph{strong converse}} to the  {\emph{general matrix multiplication source network}}, for ${\bf A}$ and ${\bf B}$, chosen independently and uniformly from $\mathbb{F}_q^{m\times l}$, as $q\to\infty$. We will subsequently derive an additional converse (Theorem~\ref{theo:converse_general_matrix_product_Han_Kobayashi}) for products of full-rank matrices, for any $q\geq 2$.

\begin{theo}
\label{theo:strong_converse_general_matrix_product}
{\bf (Strong converse for a square matrix product.)} 
For the general matrix multiplication source network, for random matrices ${\bf A}$ and ${\bf B}$, chosen independently and uniformly from $\mathbb{F}_q^{m\times l}$, as $q\to\infty$, the set of achievable rates must satisfy
\begin{align}
\label{eq:strong_converse_general}
R_1,\ R_2\geq l\cdot\min\{l,m\} \ .
\end{align}
\end{theo}

\begin{proof}
By the strong converse to the source coding theorem with side information~\cite{ahlswede1976bounds}, if one variable is available as side information, then we must have
\begin{align}
\label{eq:strong_converse_bounds}
R_1\geq H_q({\bf A}^{\intercal}{\bf B}\,\vert\, {\bf B}) \ , \quad
R_2\geq H_q({\bf A}^{\intercal}{\bf B}\,\vert\, {\bf A}) \ .
\end{align}

We next restate~\cite[Lemma~2]{jia2021capacity}, which we exploit to subsequently present the {\emph{strong converse}}.

\begin{lem}
\label{lem:entropy_general_matrix_product}
{\bf (Entropy of square matrix products~\cite[Lemma~2]{jia2021capacity}.)} 
Let ${\bf A}$ and ${\bf B}$ be random matrices drawn independently and uniformly from $\mathbb{F}_q^{m\times l}$. Then, 
\begin{align}
\label{matrix_product_entropy_q_infty}
\lim\limits_{q\to\infty}\ H_q({\bf A}^{\intercal}{\bf B})&=
2l\cdot\min\{m,\ l\}-\min\{m,\ l\}^2 \ ,\\
\label{cond_matrix_product_entropy_q_infty}
\lim\limits_{q\to\infty}\ H_q({\bf A}^{\intercal}{\bf B}\,\vert\,{\bf A})&=\lim\limits_{q\to\infty}\ H_q({\bf A}^{\intercal}{\bf B}\,\vert\,{\bf B})=l\cdot\min\{l,m\} \ .
\end{align}
\end{lem}

\begin{proof}[Proof of Lemma~\ref{lem:entropy_general_matrix_product}]
We provide a sketch of the proof by employing Lemmas~8-10 from~\cite{jia2021capacity}. As $q\to\infty$, when ${\bf A}$ and ${\bf B}$ are independent, and when ${\bf B}$ is drawn independently and uniformly from $\mathbb{F}_q^{m\times m}$,~\cite[Lemma~8]{jia2021capacity} yields $R_1\geq \lim\limits_{q\to\infty}\ H_q({\bf A}^{\intercal}{\bf B}\,\vert\, {\bf B})=H_q({\bf A})$. If both ${\bf A}$ and ${\bf B}$ are drawn independently and uniformly over $\mathbb{F}_q^{m\times l}$, if $m\geq l$, then~\cite[Lemma~9]{jia2021capacity} yields $R_1\geq \lim\limits_{q\to\infty}\ H_q({\bf A}^{\intercal}{\bf B}\,\vert\, {\bf B})=l^2$, and if $m<l$, then~\cite[Lemma~10]{jia2021capacity} yields $R_1\geq \lim\limits_{q\to\infty}\ H_q({\bf A}^{\intercal}{\bf B}\,\vert\, {\bf B})=H_q({\bf A})=lm$. This concludes the proof of Lemma~\ref{lem:entropy_general_matrix_product}.
\end{proof}

We now proceed with the proof of Theorem~\ref{theo:strong_converse_general_matrix_product}. Adapting the strong converse in Lemma~\ref{lem:entropy_general_matrix_product} (cf.~(\ref{cond_matrix_product_entropy_q_infty})) to the case where both ${\bf A}$ and ${\bf B}$ are independently and uniformly distributed over $\mathbb{F}_q^{m\times l}$, in the limit as $q\to\infty$, we obtain (\ref{eq:strong_converse_general}).
\end{proof}

In Theorem~\ref{theo:strong_converse_general_matrix_product}, ${\bf A}$ and ${\bf B}$ drawn independently and uniformly from $\mathbb{F}_q^{m\times l}$, as $q\to\infty$. Hence, in this setup, $\RSWsum=\lim\limits_{q\to\infty}\ H_q({\bf A},{\bf B})=2lm$, and no further compression is possible in direct transmission when the goal is to jointly recover $({\bf A}^n,{\bf B}^n)$. However, Condition~\ref{cond:3.13} of Lemma~\ref{lem:lemmas_1_2_Han_Kobayashi} indicates that compression can yield gains in the distributed computation of ${\bf A}^{\intercal}{\bf B}$, as opposed to directly transmitting the sources. From Theorem~\ref{theo:strong_converse_general_matrix_product},  structured source coding for computing ${\bf A}^{\intercal}{\bf B}$, rather than directly compressing $({\bf A},{\bf B})$, may yield savings only when $l<m$, where (\ref{eq:strong_converse_general}) yields $R_1+R_2\geq 2l^2$. We next tighten   Theorem~\ref{theo:strong_converse_general_matrix_product}, under which ${\bf A}$ and ${\bf B}$ are full rank with probability $1$ as $q\to\infty$~\cite{waterhouse1987often}, for full-rank matrix-product operations with any $q\geq 2$.

\begin{theo}
\label{theo:converse_general_matrix_product_Han_Kobayashi}
{\bf (A tight converse for the product of full-rank matrices.)} 
For the general matrix multiplication source network, for any $q\geq 2$, with $m, \ l>1$, the achievable rates must satisfy
\begin{align}
\label{eq:converse_general_matrix_product_Han_Kobayashi}
R_1\geq H_q({\bf A}\ \vert\ {\bf B})\ , \quad 
R_2\geq H_q({\bf B}\ \vert\ {\bf A}) \ ,
\end{align}
establishing the following lower bound on the achievable sum rate for computing ${\bf A}^{\intercal}{\bf B}$: 
\begin{align}
\label{HK}
\RHKsum({\bf A},\ {\bf B})\geq H_q({\bf A}\ \vert\ {\bf B})+H_q({\bf B}\ \vert\ {\bf A}) \ .
\end{align}
\end{theo}

\begin{proof}
For distributed computing of ${\bf \mathbfcal{D}}={\bf A}^{\intercal}{\bf B}$, pick $k_1$ and $k_2$ be two distinct rows of $f(k,\ell)$ corresponding to source matrices ${\bf A}^{(k_1)}$ and ${\bf A}^{(k_2)}$. Hence, $f(k_1,\ell)-f(k_2,\ell)=({\bf A}^{(k_1)}-{\bf A}^{(k_2)}){\bf B}^{(\ell)}$ varies as a function of ${\bf B}^{(\ell)}$, rendering Condition~\ref{cond:3.1} true. Similarly, picking $\ell_1$ and $\ell_2$ be two distinct columns of $f(k,\ell)$ corresponding to ${\bf B}^{(\ell_1)}$ and ${\bf B}^{(\ell_2)}$, it can be shown that Condition~\ref{cond:3.11} holds true. Hence, from Lemma~\ref{lem:lemmas_1_2_Han_Kobayashi}, any achievable rate must satisfy (\ref{eq:converse_general_matrix_product_Han_Kobayashi}). 
\end{proof}

Exploiting the relation $H_q({\bf A}, {\bf B})\geq H_q({\bf A}^{\intercal}{\bf B}, {\bf B})$ and subtracting $H_q({\bf B})$ from both sides, we infer that $H_q({\bf A}\, \vert \, {\bf B})\geq H_q({\bf A}^{\intercal}{\bf B}\, \vert\, {\bf B})$. Similarly, $H_q({\bf B}\,\vert\, {\bf A})\geq H_q({\bf A}^{\intercal}{\bf B}\,\vert\, {\bf A})$. Thus, given full-rank ${\bf A}, {\bf B}\in\mathbb{F}_q^{m\times l}$, as $q\to\infty$, the rate bounds in~Theorem~\ref{theo:converse_general_matrix_product_Han_Kobayashi} are tighter than those in~Theorem~\ref{theo:strong_converse_general_matrix_product}.

We next demonstrate the tightness of Theorem~\ref{theo:achievability_square_matrix_products} under the setting described in Lemma~\ref{lem:entropy_general_matrix_product}.

\begin{cor}
\label{cor:achievable_rate_lem:entropy_general_matrix_product}
{\bf (Optimality gap for square matrix products in the limit as $q\to\infty$.)} 
Consider the general matrix multiplication source network, as $q\to \infty$, and for any $\epsilon\in(0,1)$, for random matrices ${\bf A}$ and ${\bf B}$ drawn independently and uniformly from $\mathbb{F}_q^{m\times l}$. Then, the optimality gap of the scheme in Theorem~\ref{theo:achievability_square_matrix_products} from the strong converse in Theorem~\ref{theo:strong_converse_general_matrix_product} is 
\begin{align}
\label{eq:RAHsumratiotoconverse}
\frac{\lim\limits_{q\to\infty}\ \RAHsum({\bf A},\ {\bf B})}{\lim\limits_{q\to\infty}\ (H_q({\bf A}^{\intercal}{\bf B}\,\vert\, {\bf B})+H_q({\bf A}^{\intercal}{\bf B}\,\vert\, {\bf A}))}=1 \ .
\end{align}
\end{cor}

\begin{proof}

To prove the optimality gap of (\ref{eq:RAHsum_square_matrices}) in Theorem~\ref{theo:achievability_square_matrix_products}, under the setting described in Lemma~\ref{lem:entropy_general_matrix_product}, 
from (\ref{eq:strong_converse_general}) in Theorem~\ref{theo:strong_converse_general_matrix_product}, we consider two cases, namely $m\geq l$, and $m<l$, respectively.

\underline{{\bf Case 1 ($m\geq l$).}} 
For the setting of Lemma~\ref{lem:entropy_general_matrix_product}, we evaluate ${\bf A}^{\intercal}{\bf B}$ using an inner product-based characterization exploiting the following row-block representation:
\begin{align}
{\bf A}=\begin{bmatrix}
{\bf A}_F ;
{\bf A}_{\Delta}
\end{bmatrix}\in\mathbb{F}_q^{m\times l}\ ,\quad 
{\bf B}=\begin{bmatrix}
{\bf B}_F ;
{\bf B}_{\Delta}
\end{bmatrix}\in\mathbb{F}_q^{m\times l} \ ,
\end{align}
where the probability of a matrix drawn uniformly from $\mathbb{F}_q^{l\times l}$ being singular is equal to $1-\prod\nolimits_{i=1}^l (1-q^{-i})$, which goes to $0$ as $q\to\infty$~\cite{waterhouse1987often}. Hence, the matrices ${\bf A}_F\in \mathbb{F}_q^{l\times l}$ and ${\bf B}_F\in \mathbb{F}_q^{l\times l}$ are full-rank, while the matrices ${\bf A}_{\Delta}\in \mathbb{F}_q^{(m-l)\times l}$ and ${\bf B}_{\Delta}\in \mathbb{F}_q^{(m-l)\times l}$ can be directly derived from ${\bf A}_F$ and ${\bf B}_F$, respectively. More specifically, ${\bf A}_{\Delta}\in \mathbb{F}_q^{(m-l)\times l}$ and ${\bf B}_{\Delta}\in \mathbb{F}_q^{(m-l)\times l}$ can be given as
\begin{align}
\label{linearly_dependent_matrices}
{\bf A}_{\Delta}={\bf G}_1 {\bf A}_F\in \mathbb{F}_q^{(m-l)\times l}\ ,\quad
{\bf B}_{\Delta}={\bf G}_2 {\bf B}_F\in \mathbb{F}_q^{(m-l)\times l} \ ,
\end{align}
deterministic mappings ${\bf G}_1\in\mathbb{F}_q^{(m-l)\times l}$ and ${\bf G}_2\in\mathbb{F}_q^{(m-l)\times l}$  known to source one and source two, respectively. Exploiting (\ref{linearly_dependent_matrices}), we can rewrite the desired matrix product ${\bf A}^{\intercal}{\bf B}$ as
\begin{align}
\label{product_large_m}
{\bf A}^{\intercal}{\bf B}={\bf A}_F^{\intercal}{\bf B}_F\oplus_q {\bf A}_{\Delta}^{\intercal}{\bf B}_{\Delta}
={\bf A}_F^{\intercal}{\bf B}_F\oplus_q {\bf A}_F^{\intercal}{\bf G}_1^{\intercal}{\bf G}_2{\bf B}_F \ .
\end{align}

To devise the optimality gap under the setting of Lemma~\ref{lem:entropy_general_matrix_product}, we evaluate $\RAHsum$ in 
(\ref{eq:RAHsum_square_matrices}) for computing the term ${\bf A}_F^{\intercal}{\bf B}_F$ in (\ref{product_large_m}), substituting ${\bf A}$ and ${\bf B}$ by ${\bf A}_F=\begin{bmatrix}{\bf A}_{F1} ; {\bf A}_{F2}\end{bmatrix}$ and ${\bf B}_F=\begin{bmatrix}{\bf B}_{F1} ; {\bf B}_{F2}\end{bmatrix}$, respectively, with ${\bf A}_F$ and ${\bf B}_F$ drawn independently and uniformly from $\mathbb{F}_q^{l\times l}$:
\begin{align}
\label{eq:RAHsum_square_matrices_large_m}
\RAHsum({\bf A}_F,\ {\bf B}_F)&=H_q({\bf A}_{F1}, {\bf B}_{F2})+2\max\{H_q({\bf A}_{F2}\oplus_q{\bf B}_{F1}\,\vert\, {\bf A}_{F1}, {\bf B}_{F2}) , \ \nonumber\\
&\hspace{4.65cm}H_q({\bf A}_{F1}^{\intercal}{\bf A}_{F2}\oplus_q{\bf B}_{F1}^{\intercal}{\bf B}_{F2}\,\vert\, {\bf A}_{F1}, {\bf B}_{F2}, {\bf A}_{F2}\oplus_q{\bf B}_{F1}) \} \nonumber\\
&\leq 2\cdot \frac{l^2}{2}+2\max\Big\{\frac{l^2}{2},\ H_q({\bf A}_{F2})\Big\}\nonumber\\
&= 2\cdot \frac{l^2}{2}+2\max\Big\{\frac{l^2}{2},\ \frac{l^2}{2}\Big\}
= 2l^2 \ ,
\end{align}
where the calculation steps follow the same reasoning as in (\ref{eq:RAHsum_square_matrices_small_m}), and the notation $\RAHsum({\bf A}_F,\ {\bf B}_F)$ emphasizes that the structured coding is performed for the pair $({\bf A}_F,\ {\bf B}_F)$ versus $({\bf A}, {\bf B})$.

To establish the optimality gap, it remains to show the achievability of ${\bf A}^{\intercal}{\bf B}$, employing the relation in (\ref{product_large_m}). Exploiting the structured encoding scheme of  (\ref{eq:RAHsum_square_matrices_large_m}) to recover ${\bf A}_F^{\intercal}{\bf B}_F$, we denote the side information available to the receiver by ${\rm SI}_F\triangleq\{{\bf A}_{F1},\ {\bf B}_{F2},\ {\bf U}_F={\bf A}_{F2}\oplus_q{\bf B}_{F1},\ {\bf A}_{F1}^{\intercal}{\bf A}_{F2}\oplus_q{\bf B}_{F1}^{\intercal}{\bf B}_{F2}\}$. As a result, the necessary and sufficient rate for the receiver to recover the matrices ${\bf A}_{\Delta}, \ {\bf B}_{\Delta}$, denoted by $\RSWsum({\bf A}_{\Delta},\ {\bf B}_{\Delta}\,\vert\, {\rm SI}_F)$, is given as
\begin{align}
\label{eq:RSWsum_square_matrices_Delta_part_large_m}
&\RSWsum({\bf A}_{\Delta},\ {\bf B}_{\Delta}\,\vert\, {\rm SI}_F)\triangleq H_q({\bf A}_{\Delta}, \ {\bf B}_{\Delta}\,\vert\, {\rm SI}_F)\nonumber\\
&=H_q({\bf G}_1 {\bf A}_F,\ {\bf G}_2 {\bf B}_F\,\vert\, {\bf A}_{F1},\ {\bf B}_{F2},\ {\bf A}_{F2}\oplus_q{\bf B}_{F1},\ {\bf A}_{F1}^{\intercal}{\bf A}_{F2}\oplus_q{\bf B}_{F1}^{\intercal}{\bf B}_{F2})\nonumber\\
&\overset{(a)}{=}H_q({\bf G}_{11} {\bf A}_{F1}\oplus_q {\bf G}_{12} {\bf A}_{F2},\ {\bf G}_{21} {\bf B}_{F1}\oplus_q {\bf G}_{22} {\bf B}_{F2}\,\vert\, {\bf A}_{F1},\ {\bf B}_{F2},\ {\bf U}_F,\ {\bf A}_{F1}^{\intercal}{\bf A}_{F2}\oplus_q{\bf B}_{F1}^{\intercal}{\bf B}_{F2})\nonumber\\
&=H_q({\bf G}_{12} {\bf A}_{F2},\ {\bf G}_{21} {\bf B}_{F1}\,\vert\, {\bf A}_{F1},\ {\bf B}_{F2},\ {\bf U}_F,\ {\bf A}_{F1}^{\intercal}{\bf A}_{F2}\oplus_q{\bf B}_{F1}^{\intercal}{\bf B}_{F2})\nonumber\\
&=H_q({\bf G}_{12} {\bf A}_{F2},\ {\bf G}_{21} ({\bf U}_F\ominus_q{\bf A}_{F2})\,\vert\, {\bf A}_{F1},\ {\bf B}_{F2},\ {\bf U}_F,\ 
{\bf A}_{F1}^{\intercal}{\bf A}_{F2}\oplus_q({\bf U}_F\ominus_q{\bf A}_{F2})^{\intercal}{\bf B}_{F2})\nonumber\\
&\overset{(b)}{\leq} H_q({\bf G}_{12} {\bf A}_{F2},\ {\bf G}_{21} {\bf A}_{F2}\,\vert\, {\bf A}_{F1},\ {\bf B}_{F2},\ {\bf A}_{F1}^{\intercal}{\bf A}_{F2}\ominus_q{\bf A}_{F2}^{\intercal}{\bf B}_{F2})\nonumber\\
&\overset{(c)}{=}0 \ ,
\end{align}
where $(a)$ follows from letting ${\bf G}_1=\begin{bmatrix} {\bf G}_{11} & {\bf G}_{12} \end{bmatrix}$, and ${\bf G}_2=\begin{bmatrix} {\bf G}_{21} & {\bf G}_{22} \end{bmatrix}$, where the submatrices satisfy ${\bf G}_{11}, {\bf G}_{12}, {\bf G}_{21}, {\bf G}_{22}\in\mathbb{F}_q^{(m-l)\times \frac{l}{2}}$. 
Step $(b)$ follows from eliminating ${\bf U}_F$ from the set of conditional random variables, and it is clear that $\RSWsum({\bf A}_{\Delta},\ {\bf B}_{\Delta}\,\vert\, {\rm SI}_F\backslash\{{\bf U}_F\})\leq H_q({\bf A}_{F2})=l^2/2$. 
In step $(c)$, we note that given the set of matrices ${\bf A}_{F1}=(a^{F1}_{ij})\in\mathbb{F}_q^{\frac{l}{2}\times l}$, ${\bf B}_{F2}=(b^{F2}_{ij})\in\mathbb{F}_q^{\frac{l}{2}\times l}$, and ${\bf A}_{F1}^{\intercal}{\bf A}_{F2}\ominus_q{\bf A}_{F2}^{\intercal}{\bf B}_{F2}=\Big(\sum\limits_{k\in[l/2]}a^{F1}_{ki}a^{F2}_{kj}\ominus_q\sum\limits_{k\in[l/2]}a^{F2}_{ki}b^{F2}_{kj}\Big)\in\mathbb{F}_q^{l\times l}$ available as side information, in the limit as $q$ tends to infinity, with probability $1$, the receiver has $l^2$ linearly independent equations in ${\bf A}_{F2}=(a^{F2}_{ij})\in\mathbb{F}_q^{\frac{l}{2}\times l}$ with $l^2/2$ unknowns. Thus, as $q\to\infty$, given ${\bf A}_{F1}$ and ${\bf B}_{F2}$, and ${\bf A}_{F1}$, drawn independently and uniformly from $\mathbb{F}_q^{\frac{l}{2}\times l}$, the receiver can solve for ${\bf A}_{F2}$.

From (\ref{eq:RAHsum_square_matrices_large_m}) and (\ref{eq:RSWsum_square_matrices_Delta_part_large_m}), we infer that  
${\bf A}^{\intercal}{\bf B}$ in the case $m\geq l$ can be recovered at a sum rate
\begin{align}
\label{eq:Rsum_square_matrices_large_m}
\RAHsum({\bf A}_F,\ {\bf B}_F)+\RSWsum({\bf A}_{\Delta},\ {\bf B}_{\Delta}\,\vert\, {\rm SI}_F)\leq 2l^2+0=2l^2 \ .
\end{align}

From Lemma~\ref{lem:entropy_general_matrix_product}, $H_q({\bf A}^{\intercal}{\bf B})=l^2\leq lm $ when $m\geq l$. Thus, employing (\ref{eq:Rsum_square_matrices_large_m}) leads to 
\begin{align}
\label{sum_rate_case_1}
R_1+R_2&\leq \RAHsum({\bf A}_F,\ {\bf B}_F)+\RSWsum({\bf A}_{\Delta},\ {\bf B}_{\Delta}\,\vert\, {\rm SI}_F) \nonumber\\
&\leq 2H_q({\bf A}^{\intercal}{\bf B})=2l^2\leq 2lm = H_q({\bf A}, {\bf B})=\RSWsum({\bf A}, {\bf B}) \ .
\end{align}

\underline{{\bf Case 2 ($m<l$).}} 
For the setting of Lemma~\ref{lem:entropy_general_matrix_product}, we evaluate ${\bf A}^{\intercal}{\bf B}$ exploiting (\ref{eq:RAHsum_square_matrices}), which gives
\begin{align}
\label{eq:RAHsum_square_matrices_small_m}
\RAHsum({\bf A},\ {\bf B})&\overset{(a)}{=}2\cdot \frac{lm}{2}+2\max\Big\{\frac{lm}{2}\ ,\ H_q({\bf A}_1^{\intercal}{\bf A}_2\oplus_q ({\bf U}\ominus_q{\bf A}_2)^{\intercal}{\bf B}_2\,\vert\,{\bf A}_1,{\bf B}_2,{\bf U}={\bf A}_2\oplus_q {\bf B}_1)\Big\}\nonumber\\
&\overset{(b)}{\leq} lm+2\max\Big\{\frac{lm}{2}\ , \ H_q({\bf A}_2)\Big\}
=2lm = \RSWsum({\bf A},\ {\bf B}) \ ,
\end{align}
where $(a)$ follows from incorporating ${\bf U}={\bf A}_2\oplus_q {\bf B}_1$, $(b)$ from noting that $H_q({\bf A}_1^{\intercal}{\bf A}_2\oplus_q ({\bf U}\ominus_q{\bf A}_2)^{\intercal}{\bf B}_2\,\vert\,{\bf A}_1,{\bf B}_2,{\bf U}={\bf A}_2\oplus_q {\bf B}_1)\leq H_q({\bf A}_2)=lm/2$, given side information ${\bf A}_1,{\bf B}_2,{\bf U}$. 

From Lemma~\ref{lem:entropy_general_matrix_product}, when $m<l$, it holds that $H_q({\bf A}^{\intercal}{\bf B})=2lm-m^2\geq lm$. 
Hence, employing (\ref{eq:RAHsum_square_matrices_small_m}) leads to the following sum-rate bound:
\begin{align}
\label{sum_rate_case_2}
R_1+R_2\leq\RAHsum({\bf A},\ {\bf B})\leq 2lm
=\RSWsum({\bf A},\ {\bf B})\leq 2H_q({\bf A}^{\intercal}{\bf B})=4lm-2m^2 \ .
\end{align}

The optimality gap follows directly by comparing  (\ref{sum_rate_case_1}) and (\ref{sum_rate_case_2}) with (\ref{cond_matrix_product_entropy_q_infty}) of Lemma~\ref{lem:entropy_general_matrix_product}. 
\end{proof}

Theorem~\ref{theo:achievability_square_matrix_products} considers the regime $q\geq 2$ where ${\bf A}$ and ${\bf B}$ are drawn independently and uniformly from $\mathbb{F}_q^{m\times l}$. Theorem~\ref{theo:strong_converse_general_matrix_product} proves the optimality of Theorem~\ref{theo:achievability_square_matrix_products} by providing the matching converse (see~(\ref{eq:RAHsumratiotoconverse}) in Corollary~\ref{cor:achievable_rate_lem:entropy_general_matrix_product}) for ${\bf A}$ and ${\bf B}$ chosen independently and uniformly from $\mathbb{F}_q^{m\times l}$, in the asymptotic setting of $q\to\infty$. For finite $q$ or for more general source distributions, Lemma~\ref{lem:entropy_general_matrix_product}  provides an upper bound to $\Rf$ through (\ref{matrix_product_entropy_q_infty}), and similarly to $H_q({\bf A}^{\intercal}{\bf B}\,\vert\, {\bf A})$ through (\ref{cond_matrix_product_entropy_q_infty}) (and analogously for $H_q({\bf A}^{\intercal}{\bf B}\,\vert\, {\bf B})$, by modifying the conditioning variable), respectively. 

We next investigate the optimality gap for {\emph{binary symmetric matrix products}} and {\emph{binary square matrix products}}, in Propositions~\ref{prop:multiplicative_gain_for_binary_symmetric_matrix_products} and~\ref{prop:multiplicative_gain_for_binary_square_matrix_products}, respectively.

\begin{prop}
\label{prop:multiplicative_gain_for_binary_symmetric_matrix_products}
{\bf (Optimality gap for binary symmetric matrix products.)}
Consider the matrix multiplication source network, in the symmetric case, for random matrices ${\bf A}$ and ${\bf B}$ drawn independently and uniformly from $\mathbb{F}_2^{m\times l}$ for $m, \ l>1$ such that $m\geq l$, where 
\begin{align}
\label{eq:elementwise_DSBS_generalized}
(a_{ij},\ b_{ij})\sim {\rm DSBS}(p) \  \,\,  \mbox{are i.i.d. across}\,\, i\in[m] \,\, \mbox{and}\,\, j\in[l] \ .
\end{align}
Then, the optimality gap between the hybrid encoding scheme in Proposition~\ref{prop:KM-OR_sum_rate_for_matrix_vector_product} and the sum-rate lower bound $\RHKsum({\bf A},\ {\bf B})$ in Theorem~\ref{theo:converse_general_matrix_product_Han_Kobayashi} is upper bounded as 
\begin{align}
\label{eq:multiplicative_gap_symmetric} 
\frac{\RKMORsum({\bf A}\oplus_2{\bf B})}{\RHKsum({\bf A},\ {\bf B})}\leq \frac{2mh(p)+m-(l-1)/2\cdot \bar{F}(l(l-1)/2+1;ml,p)}{(2m-l+1)h(p)}\ ,
\end{align}
where the right-hand side of (\ref{eq:multiplicative_gap_symmetric}) is upper bounded by $2+1/h(p)$ as $m\to\infty$ for $m\geq l$, 
and by $1+1/(2h(p))$ as $m\to\infty$ for $l=2$, 
demonstrating the tightness of (\ref{eq:sum_rate_KM_OR}) where ${\bf Y}={\bf A}\oplus_2 {\bf B}$. 
\end{prop}

\begin{proof}
The symmetric matrix product ${\bf A}^{\intercal}{\bf B}$ satisfies Conditions~\ref{cond:3.1} and~\ref{cond:3.11} in Lemma~\ref{lem:lemmas_1_2_Han_Kobayashi}. To prove the optimality gap in the symmetric case, for $q=2$ and $m, \ l>1$ such that $m\geq l$, we thus exploit the tight converse bounds in (\ref{eq:converse_general_matrix_product_Han_Kobayashi}) of  Theorem~\ref{theo:converse_general_matrix_product_Han_Kobayashi}. Hence, the minimum rate required from source one, for the distributed computation of ${\bf \mathbfcal{D}}={\bf A}^{\intercal}{\bf B}\in\mathbb{F}_2^{l\times l}$ must satisfy
\begin{align}
\label{eq:converse_DSBS_source_one_symmetric_D}
R_1&\overset{(a)}{\geq} H({\bf A}\,\vert\, {\bf B},\ {\bf \mathbfcal{D}}={\bf \mathbfcal{D}}^{\intercal})
\overset{(b)}{=}H({\bf Y}\,\vert\, {\bf B},\ {\bf Y}^{\intercal}{\bf B}={\bf B}^{\intercal}{\bf Y})\nonumber\\
&\overset{(c)}{=}\sum\limits_{j\in[l]}H\Big(\{y_{kj}\}_{k\in[m]}\,\Big\vert \, \{y_{ki}\}_{k\in[m],\ i\in[j-1]}, \{b_{ki}\}_{k\in[m],\ i\in[l]}, \Big\{\sum\limits_{k\in[m]} y_{ki}b_{kj}=\sum\limits_{k\in[m]} b_{ki}y_{kj}\Big\}_{i,\ j\in[l]}\Big)\nonumber\\
&\overset{(d)}{=}\sum\limits_{j\in[l]}H\Big(\{y_{kj}\}_{k\in[m]}\,\Big\vert \, \{y_{ki}\}_{k\in[m],\ i\in[j-1]}, \{b_{ki}\}_{k\in[m],\ i\in[l]}, \Big\{\sum\limits_{k\in\mathcal{K}_j} y_{ki}=\sum\limits_{k\in\mathcal{K}_i} y_{kj}\Big\}_{i,\ j\in[l]}\Big)\nonumber\\
&\overset{(e)}{\geq}l(m-(l-1)/2)\cdot h(p) \ ,
\end{align}
and similarly for source two, where $(a)$ follows from exploiting the symmetry ${\bf \mathbfcal{D}}={\bf \mathbfcal{D}}^{\intercal}\in\mathbb{F}_2^{l\times l}$, and $(b)$ from the elementwise DSBS model in (\ref{eq:elementwise_DSBS_generalized}), where ${\bf Y}={\bf A}\oplus_2 {\bf B}\in\mathbb{F}_q^{m\times l}$ such that $y_{ki}\sim{\rm Bern}(p)$ for all $k\in[m]$, $i\in[l]$, and the fact that given ${\bf B}$, the relation ${\bf \mathbfcal{D}}={\bf \mathbfcal{D}}^{\intercal}$ is equivalent to ${\bf Y}^{\intercal}{\bf B}={\bf B}^{\intercal}{\bf Y}$. Step $(c)$ follows from employing the chain rule for entropy, and $(d)$ from setting $\mathcal{K}_i=\{k:\ b_{ki}=1,\ k\in[m] \}$ for a given value of $i\in[l]$. Step $(e)$ follows from noting the following. For the diagonal entries of ${\bf \mathbfcal{D}}$, the relation $\sum\nolimits_{k\in[m]} y_{ki}b_{kj}=\sum\nolimits_{k\in[m]} b_{ki}y_{kj}=\sum\nolimits_{k\in\mathcal{K}_i} y_{ki}$ for $\{i=j\in[l]\}$ holds by definition and therefore provides no reduction in the entropy of $\{y_{kj}\}_{k\in[m]}$. For the non-diagonal entries of ${\bf \mathbfcal{D}}$, the corresponding linear side-information terms $\sum\nolimits_{k\in[m]} y_{ki}b_{kj}=\sum\nolimits_{k\in[m]} b_{ki}y_{kj}$ for $\{i\neq j\in[l]\}$ yield at most $l(l-1)/2$ linearly independent equations in ${\bf Y}$ when $m\gg l$. Consequently, the entropy of ${\bf Y}$ is reduced from $mlh(p)$ by up to $(l(l-1)/2)h(p)$. Thus, (\ref{eq:converse_DSBS_source_one_symmetric_D}) yields the following tight fundamental limit on $\RHKsum$: 
\begin{align}
\label{eq:ach_converse_comparison_symmetric_square}
\RHKsum({\bf A},\ {\bf B})\geq l(2m-l+1)h(p)\ .
\end{align}

From Proposition~\ref{prop:KM-OR_sum_rate_for_matrix_vector_product}, employing (\ref{eq:sum_rate_KM_OR}), and letting ${\bf Y}=(y_{ij})_{i\in[m],\ j\in[l]}={\bf A}\oplus_2 {\bf B}$ where ${\bf A}\independent{\bf Y}$ due to (\ref{eq:elementwise_DSBS_generalized}), the following sum rate is achievable for the receiver to successfully recover ${\bf \mathbfcal{D}}$:
\begin{align}
\label{eq:ach_symmetric_square_binary}
\RKMORsum({\bf Y})&\overset{(a)}{=}2H({\bf Y}\ \vert \ {\bf A}^{\intercal}{\bf B}={\bf B}^{\intercal}{\bf A})+H_{G_{\bf A}}({\bf A}\,\vert\, {\bf Y},\ {\bf A}^{\intercal}{\bf Y}={\bf Y}^{\intercal}{\bf A})\nonumber\\
&\overset{(b)}{\leq} 2mlh(p)+ml-l(l-1)/2\cdot \bar{F}(l(l-1)/2+1;ml,p) \ ,
\end{align}
where $(a)$ is derived from rewriting ${\bf A}^{\intercal}{\bf B}={\bf B}^{\intercal}{\bf A}$ by using ${\bf B}={\bf A}\oplus_2 {\bf Y}$, and $(b)$ by employing (\ref{eq:elementwise_DSBS_generalized}), which yields  
$y_{ij}\sim {\rm Bern}(p)$ for all $i\in[m],\ j\in[l]$, and observing that ${\bf A}^{\intercal}{\bf Y}={\bf Y}^{\intercal}{\bf A}$ yields at most $l(l-1)/2$ linearly independent equations in ${\bf A}$, which holds when ${\bf Y}$ has more than $l(l-1)/2$ nonzero entries, and which occurs with probability $\bar{F}(l(l-1)/2+1;ml,p)=\sum\nolimits_{j\in[l(l-1)/2+1,ml]} {ml\choose j}p^j(1-p)^{ml-j}$. Thus, employing (\ref{eq:ach_converse_comparison_symmetric_square}) and (\ref{eq:ach_symmetric_square_binary}) yields   (\ref{eq:multiplicative_gap_symmetric}).
\end{proof}

\begin{prop}
\label{prop:multiplicative_gain_for_binary_square_matrix_products}
{\bf (Optimality gap for binary square matrix products.)}
For distributed computing of square matrix products, under the elementwise DSBS model of (\ref{eq:elementwise_DSBS_generalized}), it holds that
\begin{align}
\label{rate_ratio_square_matrix_product_v2}
\frac{\RAHrfdsum({\bf A},\ {\bf B})}{\RHKsum({\bf A},\ {\bf B})}\leq 
\begin{cases}
\frac{1+3h(p)}{4h(p)}\  ,\quad h(p)\geq \frac{1}{2}\ , \\
\frac{5}{8h(p)}\  ,\quad h(p)< \frac{1}{2}\ ,
\end{cases}
\end{align}
where $\RAHrfdsum$ provides a refinement over (\ref{eq:RAHsum_square_matrices}) in Theorem~\ref{theo:achievability_square_matrix_products}, as we detail below, and the upper bound on the right-hand side of (\ref{rate_ratio_square_matrix_product_v2}) tends to $1$ as $p\to 1/2$. 
\end{prop}

\begin{proof}
The square matrix product ${\bf A}^{\intercal}{\bf B}$ satisfies Conditions~\ref{cond:3.1} and~\ref{cond:3.11} in Lemma~\ref{lem:lemmas_1_2_Han_Kobayashi}. To prove the optimality gap in the square case, for $q=2$, we thus exploit the tight converse bounds in (\ref{eq:converse_general_matrix_product_Han_Kobayashi}) of  Theorem~\ref{theo:converse_general_matrix_product_Han_Kobayashi}. Hence, the minimum sum rate for the distributed computation of a matrix product ${\bf \mathbfcal{D}}={\bf A}^{\intercal}{\bf B}
\in\mathbb{F}_2^{l\times l}$ must satisfy
\begin{align}
\label{converse_square_matrix_product}
\RHKsum({\bf A},\ {\bf B})\geq H({\bf B}\,\vert\ {\bf A})+H({\bf A}\,\vert\ {\bf B})
=2mlh(p)
=2H({\bf A}\oplus_2{\bf B})\ ,
\end{align}
which follows from the elementwise ${\rm DSBS}(p)$ model, noting that ${\bf A}\oplus_2{\bf B}$ is independent of ${\bf A}$ and ${\bf B}$, and employing $H({\bf A}\ \vert\ {\bf B})=H({\bf B}\ \vert\ {\bf A})=mlh(p)$.

From Theorem~\ref{theo:achievability_square_matrix_products}, employing (\ref{eq:RAHsum_square_matrices}), using the elementwise DSBS model for the matrices ${\bf A},\ {\bf B}\in\mathbb{F}_2^{m\times l}$, where $(a_{ij},\ b_{ij})\sim {\rm DSBS}(p)$, $i\in[m]$, $j\in[l]$, and letting ${\bf Y}_1={\bf A}_1\oplus_2{\bf B}_1\in\mathbb{F}_2^{m/2\times l}$, and 
${\bf Y}_2={\bf A}_2\oplus_2{\bf B}_2\in\mathbb{F}_2^{m/2\times l}$, we rewrite the achievable sum rate for computing ${\bf \mathbfcal{D}}$ as
\begin{align}
\label{eq:RAHsum_square_matrices_DSBS}
\RAHsum({\bf A},\ {\bf B})&=H({\bf A}_1, {\bf B}_2)+2\max\Big\{H(({\bf B}_{2}\oplus_2 {\bf Y}_2)\oplus_2({\bf A}_{1}\oplus_2 {\bf Y}_1)\,\vert\, {\bf A}_1, {\bf B}_2) , \ \nonumber\\
&\hspace{4.35cm}H({\bf A}_{1}^{\intercal}({\bf B}_{2}\oplus_2 {\bf Y}_2)\oplus_2({\bf A}_{1}\oplus_2 {\bf Y}_1)^{\intercal}{\bf B}_{2}\,\vert\, {\bf A}_1, {\bf B}_2, {\bf A}_{2}\oplus_2{\bf B}_{1}) \Big\} \nonumber\\
&\overset{(a)}{=}2\cdot \frac{ml}{2}+2\max\{H({\bf Y}_1\oplus_2 {\bf Y}_2) \ , \ H({\bf A}_{1}^{\intercal}{\bf Y}_2\oplus_2 {\bf Y}_1^{\intercal}{\bf B}_{2}\,\vert\, {\bf A}_1, {\bf B}_2, {\bf Y}_{1}\oplus_2{\bf Y}_{2})\}\nonumber\\
&\overset{(b)}{=}ml+2\max\Big\{\frac{ml}{2}\cdot h(2p(1-p)) \ , \ 
H({\bf A}_{1}^{\intercal}({\bf Y}_s\oplus_2{\bf Y}_1)\oplus_2 {\bf Y}_1^{\intercal}{\bf B}_{2}\,\vert\, {\bf A}_1, {\bf B}_2, {\bf Y}_s)\Big\}\nonumber\\
&=ml+\max\{ml\cdot h(2p(1-p)) \ , \ 2H({\bf A}_{1}^{\intercal}{\bf Y}_1\oplus_2 {\bf Y}_1^{\intercal}{\bf B}_{2}\,\vert\, {\bf A}_1, {\bf B}_2, {\bf Y}_s)\}\nonumber\\
&\overset{(c)}{=} ml+\max\{ml\cdot h(2p(1-p)) \ , \ 2H({\bf Y}_1)\}\nonumber\\
&=ml(1+h(2p(1-p)))\ ,
\end{align}
where  
$(a)$ follows from ${\bf Y}_1\oplus_2 {\bf Y}_2=({\bf A}_1\oplus_2 {\bf B}_2)\oplus_2({\bf A}_{2}\oplus_2{\bf B}_{1})$, with ${\bf A}_1$ and ${\bf B}_2$ given,   
$(b)$ by letting ${\bf Y}_s={\bf Y}_{1}\oplus_2{\bf Y}_{2}$, under the elementwise ${\rm DSBS}(2p(1-p))$ model for ${\bf Y}_s$, which follows directly from (\ref{eq:elementwise_DSBS_generalized}), 
and 
$(c)$ by noting that $H({\bf A}_{1}^{\intercal}{\bf Y}_1\oplus_2 {\bf Y}_1^{\intercal}{\bf B}_{2}\,\vert\, {\bf A}_1, {\bf B}_2, {\bf Y}_s)\leq H({\bf Y}_1)=(ml/2)\cdot h(p)\overset{(d)}{\leq} (ml/2)\cdot h(2p(1-p))$, where $(d)$ is due to the Schur concavity of $h(\cdot)$. Hence, $\max\{ml\cdot h(2p(1-p)) \ , \ 2H({\bf A}_{1}^{\intercal}{\bf Y}_1\oplus_2 {\bf Y}_1^{\intercal}{\bf B}_{2}\,\vert\, {\bf A}_1, {\bf B}_2, {\bf Y}_s)\}=ml\cdot h(2p(1-p))$. Note that (\ref{eq:RAHsum_square_matrices_DSBS}) remains valid even when $\Rf=H({\bf \mathbfcal{D}})>mlh(2p(1-p))$, and it further implies that   
\begin{align}
\label{eq:elemenwise_DSBS_AH_vs_SW}
\RAHsum({\bf A},\ {\bf B})=ml(1+h(2p(1-p))) 
\overset{(a)}{\geq} \RSWsum({\bf A},\ {\bf B})=H({\bf A} \ , {\bf B})=ml(1+h(p)) \ ,
\end{align}
where $(a)$ follows from the Schur concavity of $h(\cdot)$. The relation (\ref{eq:elemenwise_DSBS_AH_vs_SW}) indicates that the computation of ${\bf \mathbfcal{D}}$ is ensured at a higher rate compared to~\cite{SlepWolf1973}, while our scheme that achieves the sum rate $\RAHsum({\bf A},\ {\bf B})$ in (\ref{eq:RAHsum_square_matrices_DSBS}) does not guarantee the recovery of $({\bf A},{\bf B})$.

Motivated by the encoding framework in~\cite{korner1979encode} and its extension to additions over $\mathbb{F}_q$ in~\cite[Lemma~5]{han1987dichotomy}, we proceed to examine the achievability of $\RAHsum\leq 2\Rf$ for the setting of Theorem~\ref{theo:strong_converse_general_matrix_product}. To that end, as $q\to\infty$, for $m\geq l$, $2\Rf=2l^2\leq 2lm=\RSWsum$, and from (\ref{eq:elemenwise_DSBS_AH_vs_SW}) $\RAHsum\leq 2ml$, meaning $\RAHsum\leq 2\Rf$ only when $l=m$. For $m<l$, $2\Rf=4lm-2m^2\geq \RSWsum=2lm\geq \RAHsum$.

Vertically concatenating the columns of the source matrices ${\bf A}$ and ${\bf B}$, we obtain ${\bf X}_1=\begin{bmatrix}
{\bf A}(:,1);
{\bf A}(:,2);
\hdots;
{\bf A}(:,l)
\end{bmatrix}\in\mathbb{F}_2^{ml}$, and ${\bf X}_2=\begin{bmatrix}
{\bf B}(:,1);
{\bf B}(:,2);
\hdots;
{\bf B}(:,l)
\end{bmatrix}\in\mathbb{F}_2^{ml}$, respectively. Following the steps of Lemma~\ref{Elias_lemma} and Proof of Proposition~\ref{prop:KW_sum_rate_for_symmetric_matrix_product}, we let ${\bf Z}={\bf X}_1\oplus_2{\bf X}_2\in \mathbb{F}_2^{ml}$. 
For fixed $\epsilon>0$, $\delta>0$, and for sufficiently large $n$, we choose a binary encoding matrix ${\bf \mathbfcal{C}}$ drawn independently and uniformly from $\mathbb{F}_2^{\kappa_j\times n}$. Then, there exists a decoding function $\psi_j:\mathbb{F}_2^{\kappa_j}\to \mathbb{F}_2^{n}$, for each $j\in[ml]$, that satisfies~\cite{ahlswede1983source}:
\begin{align}
{\bf \hat{Z}}^n(j) \triangleq \phi_j({\bf \mathbfcal{C}}{\bf X}_1^n(j),{\bf \mathbfcal{C}}{\bf X}_2^n(j)) 
\triangleq \psi_j({\bf \mathbfcal{C}}{\bf X}_1^n(j))\oplus_2 {\bf \mathbfcal{C}}{\bf X}_2^n(j)))
\end{align}
such that i) $\kappa_j<n(H({\bf Z}(j)\,\vert\, \{{\bf Z}(j')\}_{j'<j})+\epsilon)$, where $H({\bf Z}(j))=h(p)$ for all $ j\in[ml]$, under (\ref{eq:elementwise_DSBS_generalized}), and ii) $\mathbb{P}(\psi_j({\bf \mathbfcal{C}}{\bf Z}^n(j)))\neq {\bf Z}^n(j))<\delta$. 
Hence, the application of Lemmas~\ref{lem:increasing_k} and~\ref{Elias_lemma_vectors} yields that given a linear encoding matrix ${\bf \mathbfcal{C}}\in\mathbb{F}_q^{\kappa\times n}$ with $\kappa=\sum\nolimits_{j\in [ml]} \kappa_j$, the pair $({\bf \mathbfcal{C}},{\bf \mathbfcal{C}})$ is an $(n,\epsilon,\delta)$-coding scheme~\cite{han1987dichotomy}, and the following rate per source can be achieved for computing ${\bf Z}$:
\begin{align}
\label{eq:rate_vector_encoding_for_A_plus_B}
\frac{\kappa}{n}<H({\bf Z}(j),\,j\in[ml])+\epsilon =mlh(p)+\epsilon \ .
\end{align}
Thus, the achievable sum rate for computing ${\bf \mathbfcal{D}}$ becomes (providing a refinement over (\ref{eq:RAHsum_square_matrices_DSBS})):
\begin{align}
\label{eq:RAHsum_square_matrices_DSBS_v2}
\RAHrfdsum({\bf A},{\bf B})&\overset{(a)}{=}\Big(H({\bf A}_1)-\frac{\kappa}{2n}\Big)+2\max\Big\{H({\bf A}\oplus_2 {\bf B}) , \ \nonumber\\
&\hspace{4.75cm}H({\bf A}_{1}^{\intercal}{\bf A}_{2}\oplus_2({\bf A}_{1}\oplus_2 {\bf Y}_1)^{\intercal}({\bf A}_{2}\oplus_2{\bf Y}_2)\,\vert\, {\bf A}\oplus_2{\bf B},\ {\bf A}_1) \Big\} \nonumber\\
&=\Big(\frac{ml}{2}-\frac{\kappa}{2n}\Big)+2\max\Big\{mlh(p) , \ H({\bf A}_{1}^{\intercal}{\bf Y}_2\oplus_2 {\bf Y}_1^{\intercal}{\bf A}_{2}\,\vert\, 
{\bf Y}_1,\ {\bf Y}_2,\ {\bf A}_1) \Big\} \nonumber\\
&\overset{(b)}{\leq}\Big(\frac{ml}{2}-\frac{\kappa}{2n}\Big)+2\max\Big\{mlh(p) , \ H({\bf A}_{2}\,\vert\, 
{\bf Y}_1,\ {\bf Y}_2,\ {\bf A}_1) \Big\} \nonumber\\
&\overset{(c)}{=}\Big(\frac{ml}{2}-\frac{\kappa}{2n}\Big)+2\max\Big\{mlh(p) , \ \frac{ml}{2} \Big\} \nonumber\\
&\overset{(d)}{=}\begin{cases}
\frac{ml}{2}(1+3h(p))\  ,\quad h(p)\geq \frac{1}{2}\ ,\\
\frac{5ml}{4}\  ,\quad h(p)< \frac{1}{2} \ ,
\end{cases} \nonumber\\
&\leq 2ml
\end{align}
where $(a)$ follows from several observations.  
We begin by linearly encoding ${\bf A}$ using a matrix ${\bf \mathbfcal{C}}\in\mathbb{F}_2^{\kappa\times n}$, and applying an $(n,\epsilon,\delta)$-coding scheme to compute both ${\bf Y}={\bf A}\oplus_2{\bf B}$ and ${\bf A}_1^{\intercal}{\bf A}_2\oplus_2{\bf B}_1^{\intercal}{\bf B}_2$. This results in an encoding rate of $\kappa/n$ for each of ${\bf A}$ and ${\bf B}$. The codewords used to compute ${\bf Y}$ (with $H({\bf Y})=mlh(p)$ from (\ref{eq:elementwise_DSBS_generalized})) additionally enable recovery of the mixed terms ${\bf A}_2\oplus_2{\bf B}_1$ and ${\bf A}_1\oplus_2{\bf B}_2$. Given access to these intermediate terms, fully recovering ${\bf A}_1$ requires an extra rate of $H({\bf A}_1)-\kappa/(2n)$. With this additional information, the receiver can reconstruct ${\bf A}_1, {\bf B}_2$, ${\bf A}_2\oplus_q {\bf B}_1$, and ${\bf A}_{1}^{\intercal}{\bf A}_{2}\oplus_q{\bf B}_{1}^{\intercal}{\bf B}_{2}$, from which ${\bf A}^{\intercal}{\bf B}$ can be fully recovered using the identity in (\ref{eq:sq_matrix_product_representation_2}). Step $(b)$ follows by conditioning, (c) from that ${\bf Y}\independent {\bf A}$ and ${\bf Y}\independent {\bf B}$, and ${\bf A}_1\independent {\bf A}_2$, and $(d)$ by choosing $\kappa/n\leq ml\cdot\max\{h(p), 1/2\}$ from (\ref{eq:rate_vector_encoding_for_A_plus_B}). From (\ref{eq:RAHsum_square_matrices_DSBS_v2}), it holds that $\RAHrfdsum\leq \RSWsum=ml(1+h(p))$, where $\RAHrfdsum$ can be further reduced below (\ref{eq:RAHsum_square_matrices_DSBS}) by setting $\kappa/n\leq l\cdot\max\{mh(p),l\}$.

Given the elementwise DSBS model for matrices ${\bf A},\ {\bf B}\in\mathbb{F}_2^{m\times l}$, we have 
\begin{align}
\label{elementwise_DSBS_matrix_product}
\Rf
&
\overset{(a)}{=}H({\bf A}^{\intercal}({\bf A}\oplus_2{\bf Y})) \nonumber\\
&=H\Big(\big\{\sum\limits_{k\in[m]} a_{ki}(a_{ki}\oplus_2 y_{ki})\big\}_{i\in[l]},\ \big\{\sum\limits_{k\in[m]} a_{ki}(a_{kj}\oplus_2 y_{kj})\big\}_{\substack{i,\ j\in[l]\\ i\neq j}}\Big)\nonumber\\
&\overset{(b)}{=}H\Big(\big\{\sum\limits_{k\in\mathcal{K}_i}  (1\oplus_2 y_{ki})\big\}_{i\in[l]}\Big)+H\Big(\big\{\sum\limits_{k\in\mathcal{K}_i} a_{kj}\oplus_2 y_{kj}\big\}_{\substack{i,\ j\in[l]\\ i\neq j}}\,\vert\, \big\{\sum\limits_{k\in\mathcal{K}_i}  (1\oplus_2 y_{ki})\big\}_{i\in[l]}\Big)\nonumber\\
&\overset{(c)}{=}H\Big(\big\{\sum\limits_{k\in\mathcal{K}_i}  (1\oplus_2 y_{ki})\big\}_{i\in[l]}\Big)+H\Big(\big\{\sum\limits_{k\in\mathcal{K}_i} b_{kj}\big\}_{\substack{i,\ j\in[l]\\ i\neq j}}\,\vert\, \big\{\sum\limits_{k\in\mathcal{K}_i}  (1\oplus_2 y_{ki})\big\}_{i\in[l]}\Big)\nonumber\\
&\overset{(d)}{=}H\Big(\big\{\sum\limits_{k\in\mathcal{K}_i}  (1\oplus_2 y_{ki})\big\}_{i\in[l]}\Big)+H\Big(\big\{\sum\limits_{k\in\mathcal{K}_i} b_{kj}\big\}_{\substack{i,\ j\in[l]\\ i\neq j}}\Big)\nonumber\\
&\overset{(e)}{=}\Big(1-\frac{1}{2^m}\Big) l h((1-p)^{(|\mathcal{K}_i|}))+ \Big(1-\frac{1}{2^m}\Big) l 
(l-1) \nonumber\\
&\overset{(f)}{\geq} \Big(1-\frac{1}{2^m}\Big)l(h(p)+l-1)\ ,
\end{align}
where $(a)$ follows from ${\bf Y}={\bf A}\oplus_2 {\bf B}$, where $\{y_{kj}\}_{k\in [m],\;j\in [l]}$ are i.i.d., and $\{a_{ki}\}_{k\in [m],\;i\in [l]}$ are independently and uniformly distributed, and ${\bf A}\independent{\bf Y}$, $(b)$ from letting $\mathcal{K}_i=\{k:\ a_{ki}=1,\ k\in[m]\}$ for a given $i$, $(c)$ from using $b_{kj}=a_{kj}\oplus_2 y_{kj}\sim{\rm Bern}(1/2)$,  $(d)$ is because $b_{kj}\independent b_{ki}$ for any $j\neq i$ and $b_{kj}\independent \{y_{kj},\ y_{ki}\}$, and $(e)$ from using $1\oplus_2 y_{ki}\sim{\rm Bern}(1-p)$ and letting $p^{(k)}=p^{(k-1)}(1-p)+(1-p^{(k-1)})p$, for $k\geq 2$, setting $p^{(1)}=p$, by substituting $1-p$ for $p$, and noting that $\sum\nolimits_{k\in\mathcal{K}_i}  (1\oplus_2 y_{ki})\sim{\rm Bern}((1-p)^{(|\mathcal{K}_i|)})$ for $|\mathcal{K}_i|\geq 1$, and  $\sum\nolimits_{k\in\mathcal{K}_i} b_{kj}\sim{\rm Bern}(1/2)$,  based on the uniform and i.i.d. nature of the components of ${\bf B}$, provided  $|\mathcal{K}_i|\geq 1$. Step $(f)$ follows from the Schur concavity of $h(\cdot)$. Contrasting (\ref{elementwise_DSBS_matrix_product}) with (\ref{eq:RAHsum_square_matrices_DSBS_v2}), the necessary condition for $\RAHrfdsum\leq 2\Rf$ in the special case of $p=1/2$ is $l\geq {m}\big/{\big(1-2^{-m}\big)}$.

Using $\RAHrfdsum$ in (\ref{eq:RAHsum_square_matrices_DSBS_v2}) and the converse $\RHKsum$ in (\ref{converse_square_matrix_product}) yields the optimality gap in (\ref{rate_ratio_square_matrix_product_v2}).
\end{proof}

\section{Conclusions}
\label{sec:conclusions}
We tackle the well-known open problem of distributed computing of bilinear functions, including dot and matrix products, which form an important class of non-linear functions. Our key contribution is a class of {\emph{structured source codes}} over finite fields for distributed computing of dot and matrix products. These codes underpin our achievability schemes. To this end, we introduced a non-linear source transformation-based design principle by leveraging the linear coding scheme of K\"orner-Marton, and for general square matrix products, a careful calibration of the Ahlswede-Han approach demonstrated optimality as $q\to\infty$ for matrices drawn independently and uniformly from $\mathbb{F}_q^{m\times l}$. Our scheme for $q=2$ is within a constant factor of optimal, under the elementwise DSBS model. Our designs also include a hybrid encoding scheme that combines K\"orner's characteristic graphs with the Orlitsky-Roche approach, as well as additional constructions based on recursive and nested applications of distributed dot products. The proposed schemes can surpass the performance of the state of the art, providing savings in the communication cost. Our schemes are complemented by calibrated converse bounds derived via the Han-Kobayashi approach and the strong converse theorem of Ahlswede-G{\'a}cs-K\"orner, yielding relatively tight sum-rate converses. Furthermore, the exact optimality of our achievability result in Theorem~\ref{theo:achievability_square_matrix_products} is guaranteed for the distributed computation of products of independently and uniformly drawn source matrices from $\mathbb{F}_q^{m\times l}$ as $q \to \infty$, as stated in Corollary~\ref{cor:achievable_rate_lem:entropy_general_matrix_product}.

Further research is required to extend our coding constructions to general source classes. While our results hold asymptotically, practical coding schemes may rely on zero-error adaptations for exact recovery~\cite{korner1998zero,charpenay2023optimal} or on one-shot models for finite-length and approximate representations~\cite{torabilossy,torabi2016distributed}. One can also explore how computational costs, measured in terms of the required number of multiplications, can be reduced by leveraging a class of algorithms like the Strassen-type algorithms, e.g.,~\cite{strassen1969gaussian,fawzi2022discovering}, for matrix multiplication. Our future directions include expanding the proposed design to a wider range of non-linear functions, such as general bilinear maps, including tensor products, chain matrix products, sorting or classification functions, and non-linearly separable functions, and addressing the problems of distributed rank computation, matrix decomposition, and low-rank matrix  factorization~\cite{khalesi2025tessellated}. An interesting research direction can involve security and privacy-related ramifications. In particular, one may explore adapting our schemes to the coded distributed matrix multiplication framework by devising resilient polynomial codes with security and privacy guarantees while keeping communication and computation overheads minimal. Our design and analysis tools are also relevant to coded distributed computing scenarios over real numbers~\cite{moradi2025general}, and to deep neural network and large language model literature, where large-scale matrix multiplication is a key component, and hardware memory is the main bottleneck~\cite{ordentlich2024optimal}. Thus, approximate matrix multiplication via quantization becomes the natural solution, and can be achieved using lossy generalizations of the K\"orner-Marton scheme~\cite{krithivasan2009lattices}, to allow a receiver to estimate the product, and analyze the approximation error as a function of the quantization rate.

\section*{Acknowledgment}
The author acknowledges the constructive discussions with Petros Elia and Arun Padakandla, and thanks the anonymous reviewers for their thorough and constructive feedback, which substantially improved the manuscript, and the editors for their careful handling of the manuscript.

\begin{spacing}{1}
\bibliographystyle{IEEEtran}
\bibliography{Derya}
\end{spacing}
\end{document}